\documentclass[a4paper,fleqn]{cas-sc}

\usepackage{amssymb}
\usepackage{subfig}
\usepackage{amsmath}
\usepackage{multirow}
\usepackage[dvipsnames]{xcolor}
\usepackage[authoryear]{natbib}
\usepackage{lineno}
\usepackage{hyperref}

\def\tsc#1{\csdef{#1}{\textsc{\lowercase{#1}}\xspace}}
\tsc{WGM}
\tsc{QE}

\begin{document}
\let\WriteBookmarks\relax
\def\floatpagepagefraction{1}
\def\textpagefraction{.001}

\shorttitle{Long-term dust dynamics in Didymos and Dimorphos system}
\shortauthors{Madeira et al.}
\title [mode = title]{Long-term dust dynamics in Didymos and Dimorphos system: production, stability, and transport}

%

\author[1]{Gustavo Madeira}[
orcid=0000-0001-5138-230X,
]
\ead{madeira@ipgp.fr}

\affiliation[1]{organization={Université Paris Cité, Institut de Physique du Globe de Paris, CNRS},
            city={Paris},
            postcode={F-75005}, 
            country={France}}
\author[1]{Sebastien Charnoz}

\author[2]{Nicolas Rambaux}
\author[2]{Philippe Robutel}

\affiliation[2]{organization={IMCCE, Paris Observatory, Univ. PSL, Sorbonne Université, CNRS},
            city={Paris},
            postcode={F-75014}, 
            country={France}}
\cortext[1]{Corresponding author}

\nonumnote{}
 
\begin{abstract}
Target of NASA's DART mission, the system of Didymos and Dimorphos will once again be visited by a space mission -- ESA's Hera mission, scheduled to be launch in 2024. Hera will arrive in the system approximately 4 years after the DART impact, a long period compared to Dimorphos' orbital period ($\simeq 12$~hours). It is therefore imperative to understand the dynamics of material in this environment on a long timescale. Here, we explore the long-term dynamics of the binary system (65038) Didymos, in the context of the perturbed, planar, circular and restricted 3-body problem. We design an analytical description for a symmetrical top-shaped object, the shape assumed for the Didymos, while the Dimorphos is considered an ellipsoid. In the absence of external effects, we identify seven stable equatorial regions where particles persist for more than a decade. However, in the presence of the solar radiation effect, the lifetime of small particles ($\lesssim$mm) is in the order of days, being unlikely that Hera spacecraft will encounter clusters of millimetre and sub-millimetre particles in stable equatorial orbits. Nonetheless, large objects may reside in the region for some years, particularly in quasi-satellite orbits, the most stable orbits in the system. Additionally, interplanetary dust impacts onto Didymos populate the region, extending up to a distance of approximately 1500 meters from the primary center, with young dust. These impacts are responsible for a transfer of dust mainly from Didymos to Dimorphos. If the interplanetary dust impacts generate metric-sized boulders, they may persist in the system for years, in first sort orbits around Didymos.
\end{abstract}

\begin{highlights}
\item Didymos environment is essentially chaotic.
\item There are seven stable families in which meter-sized particles can survive for more than a decade.
\item Millimeter-sized particles and smaller are lost within a few days due to the solar radiation effect.
\item Interplanetary dust impacts onto Didymos populate the primary's vicinity region with dust.
\end{highlights}

\begin{keywords}
asteroids, dynamics \sep satellites, dynamics \sep planetary rings
\end{keywords}

\maketitle

\section{Introduction}
On 26 September 2022, NASA's DART mission was responsible for the first astronomical-scale demonstration of the kinetic impact deflection technology \citep{Cheng2018,Rivkin2021,Thomas2023}: DART's probe impacted Dimorphos, the secondary member of the binary system (65803) Didymos. The DART mission is part of the AIDA (Asteroid Impact Deflection Assessment) project, an international cooperation with the objective of developing technologies to deflect the trajectory of objects on a collision course with Earth. The counterpart of the AIDA project is the ESA's Hera mission, scheduled to be launched in October 2024 and which will perform a post-impact characterisation of the system \citep{Michel2018,Michel2022}. 

Boosted by AIDA project, several works have explored the Didymos and Dimorphos environment in recent years, revealing a system of interesting and intricate dynamics. The primary Didymos has an average equatorial radius of $<390$~m meters and was classified as a top-shaped asteroid according to its radar-based shape model \citep{Naidu2020}. However, images from the DART mission showed that the object's shape is actually closer to a flattened shape with an equatorial ridge \citep{Daly2023}. Dimorphos, the secondary, is an oblate ellipsoidal object with volume-equivalent radius of $\sim75$~m \citep{Daly2023}. The system has radial location ranging from $1.01$~AU to $2.27$~AU relative to the Sun (pre-impact orbital pericenter and apocenter, respectively).

It is observed that Didymos rotates close to its critical spin, which may have induced mass shedding on the object's surface, resulting in landslides and ejections \citep{Yu2019}. Mass shedding has been suggested as the origin of the equatorial bulge observed in Didymos \citep{Hyodo2022} and as the source of the material that ultimately gave rise to Dimorphos. The origin of secondary asteroid from a primary has been previously explored by \cite{Walsh2008,Jacobson2011}, which demonstrated that mass shedding \citep{Walsh2008} and rotational fission \citep{Jacobson2011} of the primary asteroid, induced by its fast spin, can lead to the formation of a secondary asteroid. According to a recent model for Dimorphos formation, a ring forms from Didymos' material, viscously spreading and creating aggregates that eventually lead to the formation of Dimorphos \citep{Madeira2023,MadeiraCharnoz2023}. The most common result of direct formation from a ring is prolate satellites, with \cite{MadeiraCharnoz2023} suggesting that the oblate shape of Dimorphos was acquired due to low-velocity collisions of similar-sized objects. However, the process of Dimorphos' formation remains a subject that requires further understanding.

The mass shedding phenomena on Didymos, in turn, have been explored by different works, such as \cite{Yu2018,Yu2019,Zhang2021,Ferrari2022b,Hirabayashi2022,Trogolo2023}. Depending on the (not-yet-known) physical properties of Didymos, it turns out that shedded material can indeed be put into orbit. \cite{Trogolo2023} obtains that most of material put into orbit is accreted back to Didymos, possibly feeding the mass shedding process. \cite{Yu2019}, however, find that most of the material is transferred to Dimorphos, leading to a cumulative growth of the object. Regardless of what the fate of this material is, these works point to the fact that Didymos has hosted orbiting particles in the past, making the study of stability in the system a topic of attention. In addition, stable trajectories are of interest to space missions as they may correspond to possible trajectories for the probe itself. At the same time, these orbits may also host material that can potentially damage the spacecraft, depending on the conditions of the system and the probe. 

Modeling the dynamics of the Didymos system is tricky, since it corresponds to a complex dynamic environment in which particles are expected to move in non-Keplerian motion around one or both of the objects \citep{Rossi2022,Richardson2022}. Some specific factors affecting particle evolution include the proximity and relatively large mass ratio between Didymos and Dimorphos, their irregular shapes, and the system proximity to the Sun. Throughout the years, different studies have investigated the stability on the system with the aim of identifying stable trajectories for the Hera mission \citep{DellElce2017,Oliveira2020,Ferrari2021trajectory,Fodde2023,Raffa2023}. In addition to the Hera mother probe, which will perform hyperbolic arcs within the system, the mission is also made up of two Cube-Sats designed for close orbit the system.

Using the perturbed restricted full three-body problem, \cite{DellElce2017} search for stables trajectories in the system and evaluate the robustness of their solutions in the face of uncertainties in the initial conditions of the system. 
"Perturbed" means that they considered perturbations to the system (solar gravity and radiation force). The shape of the asteroids is taken into account, assuming a polyhedron model for Didymos and considering Dimorphos as an ellipsoid. \cite{DellElce2017} identifies five sets of stable trajectories: two interior and two exterior to Dimorphos (one prograde and another retrograde), and a family of circum-secondary retrograde orbits, referred to as Family A in our work. However, considering the uncertainties of the system, they conclude that the safest solutions for a spacecraft are inner retrograde orbits, where the probe can maintain stable motion for at least 30 days. They also obtain the terminator orbit to be a possible stable trajectory for the probe.

These same conclusions were drawn by \cite{Fodde2023}. They solved the equations of motion, employing spherical harmonics acceleration up to second order and degree to model Didymos' gravitational force, while treating Dimorphos as a point mass. The system was also analysed by \cite{Raffa2023} using a perturbed planar bi-elliptic restricted four-body problem. Their focus was on identifying stable trajectories lasting at least 5 days, which could correspond to possible arcs of the probe's trajectory. In the unperturbed case, a set of different stable trajectories is found,  however, the solar radiation is responsible for destabilizing most of them. One of the only survivors is the circum-secondary retrograde orbits. Here, we also investigate the stable regions in Didymos environment and explore the possibility of the system hosting material.

The dynamics of the system have also been indirectly investigated when analyzing the evolution of DART impact ejecta. Using the polyhedron method, \cite{Rossi2022} discovered particles in stable motion for two years in regions near Didymos (satellite orbits) and in orbits around the binary (circumbinary orbits). However, it is worth noting that their work did not involve a systematic search for stable regions, leaving the extent and potential existence of other stable regions uncertain. When considering the solar radiation effects, \cite{Rossi2022} find that particles in circumbinary orbits collide with Didymos and Dimorphos in a few orbits, while particles around Didymos exhibit longer survival periods.

In an effort to determine whether the material generated by the DART impact ejecta could be trapped in stable regions, \cite{Rossi2022} carry out simulations with particles ejected from Dimorphos, finding that small particles exhibit chaotic motion and are lost within a few days. In contrast, particles with 5-10 cm in size can survive for almost two years. Other works have also analysed the outcome of the DART impact \citep{Yu2017,Fahnestock2022,Ferrari2022a,Moreno2022,Moreno2023}, predicting the formation of an ejecta cloud with a variety of velocities and sizes. Micrometer-sized particles or those with high ejection velocities tend to escape the system due to solar radiation force, while larger, slower particles remain in orbit around the binary until they collide with Didymos or Dimorphos after a few days. In fact, Hubble Space Telescope observations detected the activation of Dimorphos and the formation of an ejecta tail \citep{Li2023,Moreno2023}.

The Hera mission is scheduled to arrive in the system over four years after the DART impact. This can be considered a very long period of time, since it represents more than $3100$~orbits of Dimorphos around Didymos.  Therefore, it is necessary to understand the long-term dynamics of the material in the system in order to anticipate the most dynamically populated regions when the Hera spacecraft arrives in the system (around December 2026).

In this work, we focus on stability in the vicinity of Didymos and Dimorphos before the DART impact, since different studies suggest the possibility of material having existed in the vicinity of the binary. Didymos is thought to have ejected material in the past due to its fast spin \citep{Yu2018,Yu2019,Hirabayashi2022,Trogolo2023}, while it is expected that Didymos has undergone energetic impacts with interplanetary large ongoing objects. Didymos and Dimorphos are expected to be constantly bombarded by interplanetary projectiles, inducing ejections \citep{Janches2021}. Additionally, HST observations have evidenced the ejection of cm to m-size boulders \citep{Jewitt2023}, which are little affected by solar radiation. Therefore, it is plausible that material still remains in the system (in regions of stability), justifying our interest in this topic. The effects that the change in the orbital period of Dimorphos\footnote{The impactor may also have caused changes to the axis of rotation and the shape of Dimorphos \citep{Agrusa2021,Raducan2022,Nakano2022}.} \citep{Thomas2023} due to the impact of DART may have had on the stable regions we leave as a topic for future investigations.

The study of the stability in any system requires several simulations with long timespan ($\sim10^{4}$ orbital period of Dimorphos), which is not feasible when considering high-resolution models for the gravitational potential. Given this, we use the perturbed planar, circular, restricted three-body problem (PPCR3BP). In this model, the shape of objects is taken into account through their inertia moments, which can be determined analytically if we assume Didymos and Dimorphos as non-spherical symmetric bodies \citep[NSSBs,][]{Madeira2022a,Ribeiro2023}. For the calculation to be tractable, we cannot use an over-detailed shape model of the two bodies. Instead, we will adopt a simple but non-trivial shape in order to model the gravitational potential of the system. This will be useful for conducting the calculation in the long term and will allow us to go far beyond the standard point-approximation. We assume Didymos as a top-shaped symmetric object with a circular base, while Dimorphos corresponds to an ellipsoidal body. 

The dynamical model and equations of motions are described in Section~\ref{sec_motion}, while in Section~\ref{sec_PM} we analyse the dynamics of the system assuming Didymos and Dimorphos as spherical bodies. In Section~\ref{sec_SM}, we take this a step further, taking into account the shape of Didymos and Dimorphos. The effects of the solar gravity and radiation force (SRF) are included in Section~\ref{sec_srf}. In Section~\ref{sec_ejected}, we analyse the production and ejection of material from Didymos and Dimorphos due to interplanetary impactors. Our discussion and conclusions are addressed in Section~\ref{sec_discussion}.

\section{Dynamic model} \label{sec_motion}
\begin{figure}
\centering
\includegraphics[width=0.7\columnwidth,trim={0 0 0 0},clip]{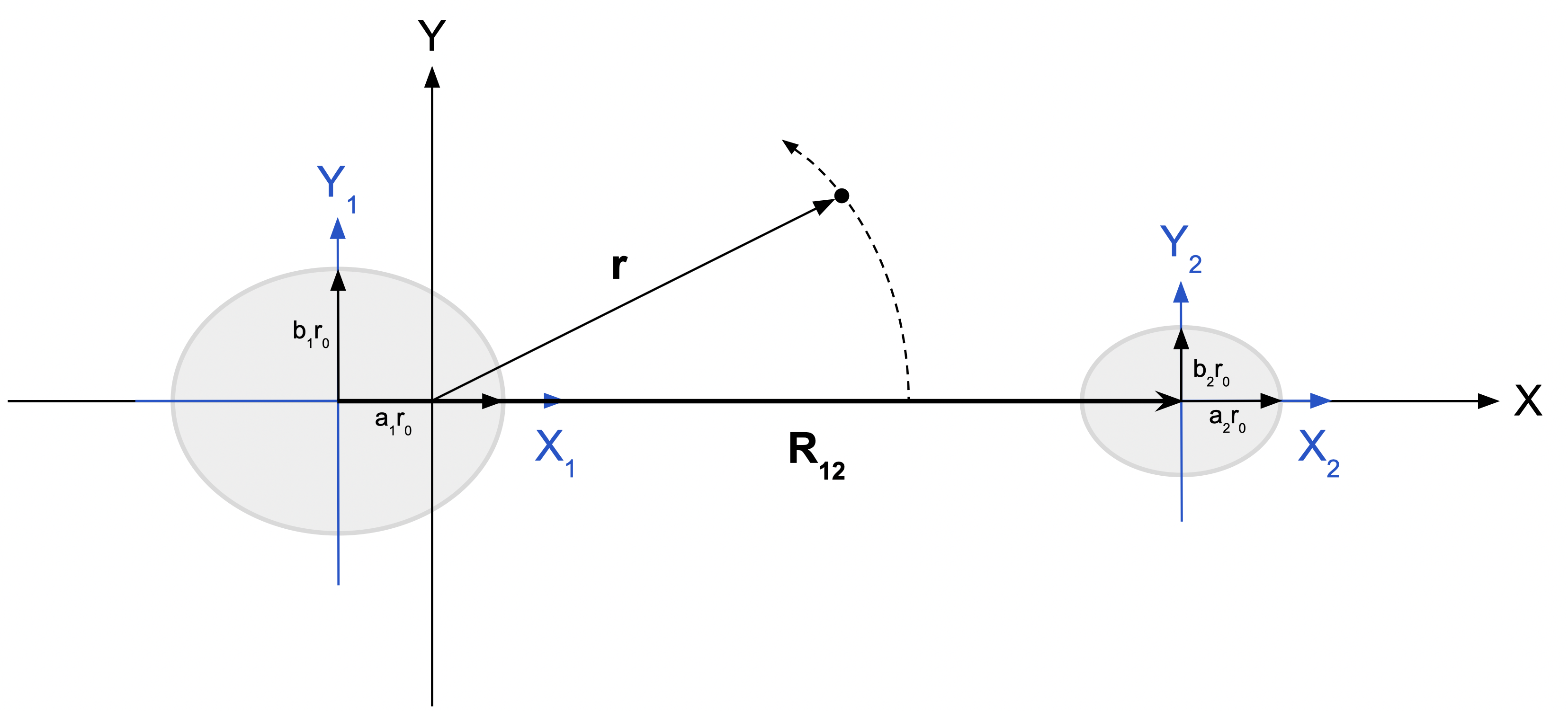}
\caption{Schematic diagram of the Didymos and Dimorphos system. The motion of the particle is given in the frame that rotates with the rotation frequency of the binary (XY, defined in the text), while the inertia moments of Didymos and Dimorphos are calculated in the frames $X_1Y_1$ and $X_2Y_2$, respectively. The figure is not to scale.} 
\label{general}
\end{figure}

In this section, we present the equations responsible for describing the motion of a massless particle in the vicinity of Didymos and Dimorphos. Although our system corresponds to a restricted three-body problem, the large mass ratio of the objects and their non-spherical shape move the system away from the more classical solutions of the three-body problem, such as the Sun-Jupiter-particle system \citep{Poincare1895,Jefferys1971,Winter1994a,Winter1994b}. To evaluate the consequences of each particularity of the system, we will first assume Didymos and Dimorphos as points of mass (CPR3BP, Section~\ref{sec_PM}), which will allow us to analyze how the physical configuration of the bodies modulates the motion of the particles. Next, the shape of the objects will be considered (PCR3BP, Section~\ref{sec_SM}), and we will analyze how they affect the dynamics of the region. By increasing the complexity of the system in stages, we may have an elucidation of how the dynamical environment can be affected if even more accurate models for the shapes of Didymos and Dimorphos are considered. 

In our dynamical model, Didymos is the primary object (body-$1$), Dimorphos is the secondary (body-$2$), and the massless particle is the third body (Figure~\ref{general}). The motion of the particle is calculated in a rotating frame $XY$ centered on the barycenter of the system. The $X$-axis is aligned to the vector ${\mathbf R_{12}}$ connecting the center of Didymos and Dimorphos and the $Y$-axis is the perpendicular axis responsible for defining the equatorial plane of Didymos as the plane of motion. We also define a fixed frame $X_iY_i$ at the physical centre of each body $i$. Finally, we assume Dimorphos in a planar, circular, spin-tidally locked orbit\footnote{See \cite{Agrusa2022,Richardson2022} for a detailed discussion of this assumption.}, which implies that the $X$ and $X_i$ axes are always parallel. 

For simplicity, we normalize the system by setting $G=1$, $M_1+M_2=1$ and $\omega=1$, where $M_i$ is the mass of the body $i$ and $\omega$ is the angular frequency. Consequently, Dimorphos period will correspond to $2\pi$ in normalized units. Assuming that the objects are points of mass, the distance between Didymos and Dimorphos ($R_{12}=|{\mathbf R_{12}}|=1$) can be obtained simply using Kepler's third law, $R_{12}=(G(M_1+M_2)/\omega^2)^{1/3}\sim1200$~m. Now, considering the shape of the objects, the distance and angular frequency are related by the equation \citep{Chandra1942}:
\begin{equation}
\omega^2-\left.\frac{1}{r}\frac{dU}{dr}\right|_{r=R_{12}}=0,
\end{equation}
where $U$ is the gravitational potential, while $R_{12}$ can be determined using the Newton-Raphson method \citep[see][]{Press1988}. For our case, $R_{12}\sim1216$~m, which implies that our units of distance will be slightly different when considering the shape of the objects versus when not considering it.

The physical properties of Didymos and Dimorphos are provided in Table~\ref{initial_data}, where the average radius of Didymos and Dimorphos correspond the radius of volume-equivalent sphere of the objects \citep{Daly2023}\footnote{We assume Didymos and Dimorphos to have the same density, which is consistent if they have same origin. A modest evidence of this possibility is provided by spectral observations of DART ejecta, which show that Dimorphos is a S-type asteroid, just like Didymos \citep{Lin2023,Bagnulo2023}.}. Although the DART impact is responsible for changing the orbital period of Dimorphos by $-33.2$~minutes \citep{Thomas2023}, here we assume the pre-impact orbit of Dimorphos.
\begin{table}{}
\caption{Physical properties of Didymos and Dimorphos \citep{Naidu2020,Daly2023,Thomas2023}}
\label{initial_data}
\centering
\begin{tabular}{lcc}
\hline\hline
Parameters & Didymos & Dimorphos \\ \hline
Mass (kg) & $5.5\times 10^{11}$ & $4.3\times 10^{9}$  \\
Bulk density (kg/m$^3$) & $2400$ & $2400$  \\
Average radius (m) & 380.5 & 75.5  \\
Orbital period (h) & -- & $11.92$ \\
\hline
\end{tabular}
\end{table}

Equations of motion in the $XY$ frame are given by \citep{Singh2013,Woo2014}:
\begin{equation}
    \ddot{X}-2\dot{Y}-X=U_x \label{motionx}
\end{equation}
and
\begin{equation}
    \ddot{Y}+2\dot{X}-Y=U_y \label{motiony}
\end{equation}
where $U_x$ and $U_y$ stand for the partial derivatives of the gravitational potential $U$. Their explicit relation are \citep{Woo2014}:
\begin{equation}
\begin{split}
U_x=\frac{\partial U}{\partial X}=&-v\left[\frac{1}{r_{13}^3}+\frac{3\epsilon}{2r_{13}^5}\left(3p_{x1}^2+p_{y1}^2+p_{z1}^2-\frac{5}{r_{13}^2}\left[\left(\frac{1-v}{u}+X\right)^2p_{x1}^2+Y^2p_{y1}^2 \right] \right) \right] \left( \frac{1-v}{u}+X \right)- \\
&(1-v)\left[\frac{1}{r_{23}^3}+\frac{3\epsilon}{2r_{23}^5}\left(3p_{x2}^2+p_{y2}^2+p_{z2}^2-\frac{5}{r_{23}^2}\left[\left(-\frac{v}{u}+X\right)^2p_{x2}^2+Y^2p_{y2}^2 \right] \right) \right] \left(-\frac{v}{u}+X \right),
\end{split}
\end{equation}
and
\begin{equation}
\begin{split}
U_y=\frac{\partial U}{\partial Y}=&-v\left[\frac{1}{r_{13}^3}+\frac{3\epsilon}{2r_{13}^5}\left(p_{x1}^2+3p_{y1}^2+p_{z1}^2-\frac{5}{r_{13}^2}\left[\left(\frac{1-v}{u}+X\right)^2p_{x1}^2+Y^2p_{y1}^2 \right] \right) \right]Y- \\
&(1-v)\left[\frac{1}{r_{23}^3}+\frac{3\epsilon}{2r_{23}^5}\left(p_{x2}^2+3p_{y2}^2+p_{z2}^2-\frac{5}{r_{23}^2}\left[\left(-\frac{v}{u}+X\right)^2p_{x2}^2+Y^2p_{y2}^2 \right] \right) \right]Y
\end{split}
\end{equation}
where $v=M_1/(M_1+M_2)$ is the reduced mass of Didymos, and u is the normalized characteristic length of the mutual orbits. $p_{xi}$, $p_{yi}$ and $p_{zi}$ are the radii of gyration of the bodies, corresponding to the moments of inertia normalized by the mass (calculated later in the section). $\epsilon$ is the normalized average radius of Didymos ($\epsilon=0.359$), while $r_{13}$ and $r_{23}$ are the distance from the particle to Didymos and Dimorphos, respectively:
\begin{equation}
r_{13}=\sqrt{\left(\frac{1-v}{u}+X\right)^2+Y^2}
\end{equation} 
and
\begin{equation}
r_{23}=\sqrt{\left(-\frac{v}{u}+X\right)^2+Y^2}
\end{equation}
The general expression of the gravitational potential $U$ is expressed as \citep{Woo2014}:
\begin{equation}
\begin{split}
U=&v\left[\frac{1}{r_{13}}+\frac{3\epsilon}{2r_{13}^3}\left(\frac{p_{x1}^2+p_{y1}^2+p_{z1}^2}{3}-\frac{\left(\frac{1-v}{u}+X\right)^2p_{x1}^2+Y^2p_{y1}^2}{r_{13}^2} \right) \right]+ \\ 
&(1-v)\left[\frac{1}{r_{23}}+\frac{3\epsilon}{2r_{23}^3}\left(\frac{p_{x2}^2+p_{y2}^2+p_{z2}^2}{3}-\frac{\left(-\frac{v}{u}+X\right)^2p_{x2}^2+Y^2p_{y2}^2}{r_{23}^2} \right) \right] \label{potential}
\end{split}
\end{equation}

Although the total energy of the system is conserved, the energy of a given particle is not constant with the time, and it is useful to define the Jacobi constant $C_J$. This parameter is a constant for the same particle and relates its velocity and position \cite[see][]{Murray1999}. The Jacobi constant is expressed as \citep{Jefferys1971}:
\begin{equation}
C_J=X^2+Y^2+2U-\dot{X}^2-\dot{Y}^2 \label{jacobi}
\end{equation}

We perform our numerical simulations by integrating the equations of motion given in this section using the function \texttt{IVP} from the Python library \texttt{SciPy} \citep{Hunter2007}. This library is composed by a wide range of tools for scientific applications, including ordinary and partial differential equation solvers.

\subsection{The shapes of Didymos and Dimorphos}
In Equation~\ref{potential}, the contribution to the gravitational potential of the objects' mass distribution is computed only by means of the radii of gyration, which means that the formalism given in Section~\ref{sec_motion} can be used for any object, provided that the radii of gyration can be computed. In Section~\ref{sec_PM}, we analyse the system for Didymos and Dimorphos as points of mass, with radii of gyration given by:
\begin{equation}
p_{x}=p_{y}=p_{z}=\sqrt{\frac{2}{5}}=0.6325 \label{eq_spheres}
\end{equation}
Applying Equation~\ref{eq_spheres} in Equation~\ref{potential}, we recover the gravitational potential of the CPR3BP \citep{Murray1999}
\begin{equation}
U=\frac{v}{r_{13}}+\frac{1-v}{r_{23}} 
\end{equation}

\begin{figure}
\centering
\subfloat[Didymos]{\includegraphics[width=0.45\columnwidth]{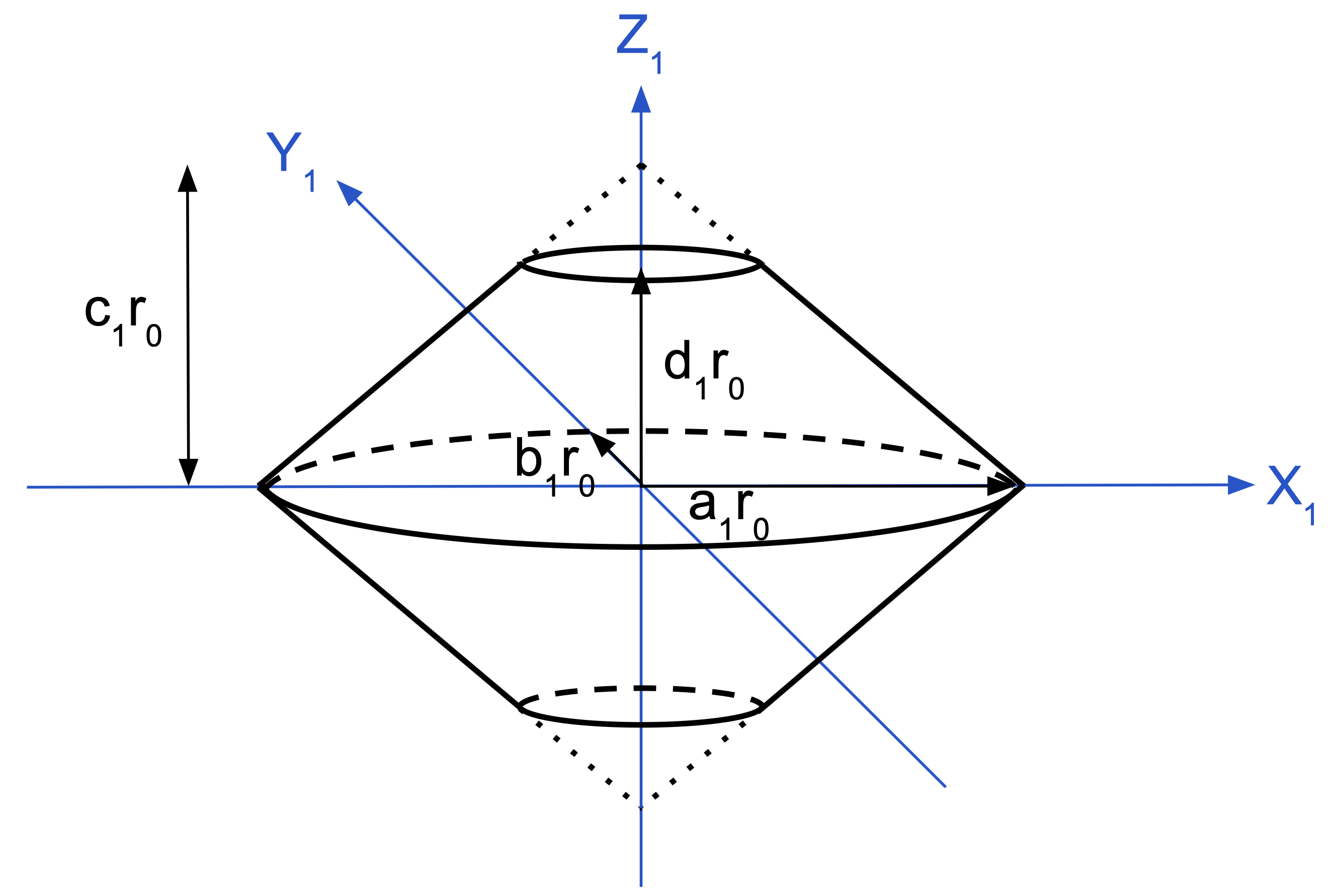}\label{shapesa}}
\quad
\subfloat[Dimorphos]{\includegraphics[width=0.35\columnwidth]{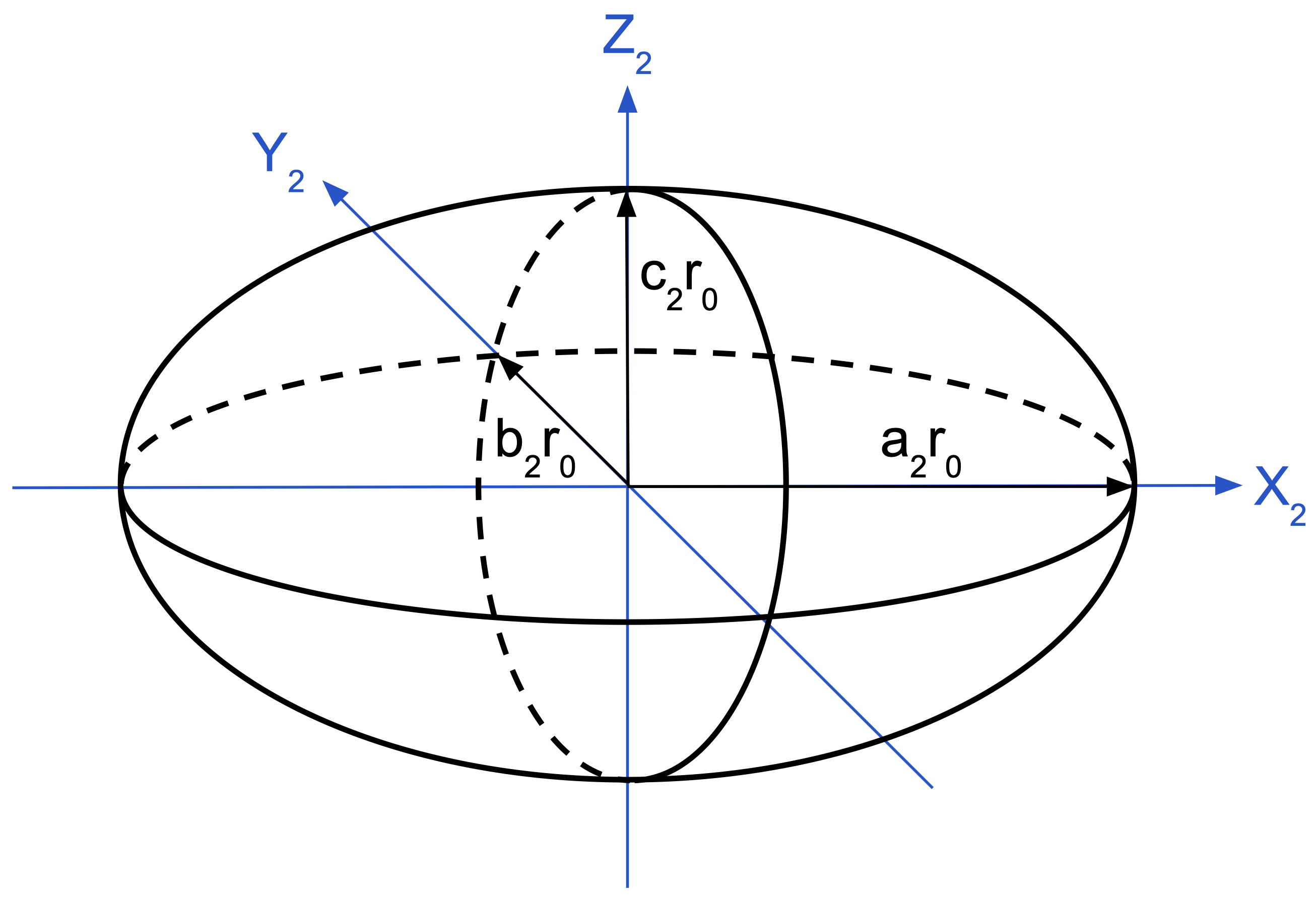}\label{shapesb}}
\caption{The shapes assumed for Didymos and Dimorphos.}
\label{shapes}
\end{figure}

Now, in Section~\ref{sec_SM} we consider Didymos and Dimorphos as objects with non-spherical shapes, and it is necessary to consider the full gravitation potential given in Equation~\ref{potential}. To obtain analytically tractable radii of gyration, we assume Didymos to be a top-shaped symmetric object, constructed by union of two truncated cones, as shown in Figure~\ref{shapesa}. We describe each cone by means of the parameters $a_1$, $b_1$, $c_1$, and $d_1$: $a_1r_0$ and $b_1r_0$ are the semi-axes of the ellipse defining the base of the cone, $c_1r_0$ is the height of the object if it were a complete cone, and $d_1r_0$ is its actual height; $d_1$ defines where the complete cone was ``sliced" \citep[see][]{Oliveira2020}. 

Using cylindrical coordinates ($R_1$,$\theta_1$,$Z_1$), the position of an element of the surface will be given by $\vec{R_1}={\tilde R_1\cos{\theta_1}}\hat{X_1}+{\tilde R_1\sin{\theta_1}}\hat{Y_1}+Z_1\hat{Z_1}$, where $\tilde R_1$ is given by:
\begin{equation}
\tilde R_1=\frac{a_1b_1(c_1r_0-|Z_1|)}{c_1(a_1^2+\sin^2\theta_1+b_1^2+\cos^2\theta_1)^{1/2}}
\end{equation}
Assuming a constant bulk density $\rho$, the mass of the symmetric top-shaped object is:
\begin{equation}
M_1=\rho\int_{-c_1r_0}^{c_1r_0}\int_0^{2\pi}\int_0^{\tilde R_1}R_1dR_1d\theta_1 dZ_1=\frac{2}{3}\pi a_1b_1\rho\left(\frac{c_1^3-(c_1-d_1)^3}{c_1^2}\right)r_0^3.     
\end{equation}
The radius of gyration in an arbitrary axis $\gamma$ is related to the moment of inertia through the relation $I_\gamma=Mp_\gamma^2r_0^2$ \citep{Woo2014}. Using this relation, we compute the radii of gyration of Didymos as:
\begin{equation}
\begin{split}
p_{x1}=&\left[\frac{\rho}{M_1r_0^2}\int_{-c_1r_0}^{c_1r_0}\int_0^{2\pi}\int_0^{\tilde R_1}((R_1\sin\theta_1)^2+Z_1^2)R_1dR_1d\theta_1 dZ_1\right]^{1/2}\\
=&\left[\frac{3b_1^2}{20c_1^2}\frac{c_1^5-(c_1-d_1)^5}{c_1^3-(c_1-d_1)^3}+\frac{d_1^3}{10}\frac{10c_1^2-15c_1d_1+6d_1^2}{c_1^3-(c_1-d_1)^3}\right]^{1/2},  
\end{split}
\end{equation}
\begin{equation}
\begin{split}
p_{y1}=&\left[\frac{\rho}{M_1r_0^2}\int_{-c_1r_0}^{c_1r_0}\int_0^{2\pi}\int_0^{\tilde R_1}((R_1\cos\theta_1)^2+Z_1^2)R_1dR_1d\theta_1 dZ_1\right]^{1/2}\\
=&\left[\frac{3a_1^2}{20c_1^2}\frac{c_1^5-(c_1-d_1)^5}{c_1^3-(c_1-d_1)^3}+\frac{d_1^3}{10}\frac{10c_1^2-15c_1d_1+6d_1^2}{c_1^3-(c_1-d_1)^3}\right]^{1/2},
\end{split}
\end{equation}
and
\begin{equation}
p_{z1}=\left[\frac{\rho}{M_1r_0^2}\int_{-c_1r_0}^{c_1r_0}\int_0^{2\pi}\int_0^{\tilde R_1} R_1^3dR_1d\theta_1 dZ_1\right]^{1/2}=\left[\frac{3(a_1^2+b_1^2)}{20c_1^2}\frac{c_1^5-(c_1-d_1)^5}{c_1^3-(c_1-d_1)^3}\right]^{1/2}.
\end{equation}
Based on the triaxial ellipsoidal shape for Didymos, we set $a_1r_0=b_1r_0=425.0$~m for the base the cone \citep{Daly2023}. We assume the base of the cone to be circular, so the Didymos rotational effect can be neglected in our system. Such an assumption is reasonable, given that the ellipticity of Didymos' triaxial ellipsoidal shape is observed to be very small ($\sim 10^{-3}$) in addition to the fact that the rotation period ($\sim2.26$ hr) of Didymos is much smaller than the timescales considered here. We set $c_1r_0=641.0$~m and $d_1r_0=412.0$~m, these values being obtained to reproduce the mass of Didymos while roughly reproducing the physical extents of the object. For this set of parameters, $p_{x1}=0.6473$, $p_{y1}=0.6473$, and $p_{z1}=0.6244$.

For Dimorphos (Figure~\ref{shapesb}), we assume the oblate ellipsoidal shape reported by \cite{Daly2023}, with semi-axes $a_2r_0=88.5$~m, $b_2r_0=87.0$~m, and $c_2r_0=58.0$~m. Dimorphos mass is computed as:
\begin{equation}
 M_2=\frac{4}{3}\pi a_2b_2c_2\rho r_0^3, 
\end{equation}
while the radii of gyration are:
\begin{equation}
p_{x2}=\left[\frac{1}{5}(b_2^2+c_2^2)\right]^{1/2},
\end{equation}
\begin{equation}
p_{y2}=\left[\frac{1}{5}(a_2^2+c_2^2)\right]^{1/2},
\end{equation}
and
\begin{equation}
p_{z2}=\left[\frac{1}{5}(a_2^2+b_2^2)\right]^{1/2}.
\end{equation}
Using the quantities given in Table~\ref{initial_data}, we obtain the radii of gyration as $p_{x2}=0.1229$, $p_{y2}=0.1243$, and $p_{z2}=0.1459$. In the following sections, we calculate the motion of particles in Didymos-Dimorphos vicinity.

\section{Didymos and Dimorphos as mass points} \label{sec_PM}
\begin{table}
	\centering
	\caption{Location and Jacobi constant of the Lagrangian points of the system in normalized units, considering Didymos and Dimorphos as mass points and as objects with non-spherical shapes.}
	\label{lagrange}
	\begin{tabular}{ccccccc}
		\hline
    & \multicolumn{3}{c}{Points-of-mass} & \multicolumn{3}{c}{NSSBs}  \\       
    & X & Y & $C_J$ & X & Y & $C_J$ \\
		\hline
 $L_1$ & 0.860 & 0.0 & 3.15 & 0.851 & 0.0 & 3.14  \\
 $L_2$ & 1.137 & 0.0 & 3.13 & 1.143 & 0.0 & 3.14  \\
 $L_3$  & -1.003 & 0.0 & 3.01 & -0.999 & 0.0 & 3.00  \\
 $L_4$  & 0.494 & 0.865 & 2.99 & 0.503 & 0.854 & 2.98 \\
 $L_5$ & 0.494 & -0.865 & 2.99 & 0.503 & -0.854 & 2.98  \\
  \hline
	\end{tabular}
	\centering
\end{table}

In this section, the motion in the vicinity of Didymos and Dimorphos is analysed using PCR3BP, which means that we disregard any effects caused by the shape of the objects and consider them as points of mass. We start the investigation by looking for the equilibrium locations of the system, the Lagrangian points. The equilibrium locations correspond to the positions at which the forces acting on a particle are in equilibrium, which implies that the total acceleration of the particle in the rotating frame is zero. If the initial velocity of the particle in the rotating frame is also zero, the particle will remain stationary with respect to the primary and secondary bodies.

We determine the equilibrium locations numerically by setting $\ddot{X}=\ddot{Y}=\dot{X}=\dot{X}=0$ in equations~\ref{motionx} and \ref{motiony}, and searching for the roots of the equations via Newton-Raphson method \citep{Press1988}. We then use equation~\ref{jacobi} to obtain the Jacobi constants associated to the equilibrium points. The location and Jacobi constant of the equilibrium points are given in Table~\ref{lagrange}, along with the values obtained when the shapes of Didymos and Dimorphos are considered.

\begin{figure}
\centering
\includegraphics[width=0.5\columnwidth,trim={0 0 0 0},clip]{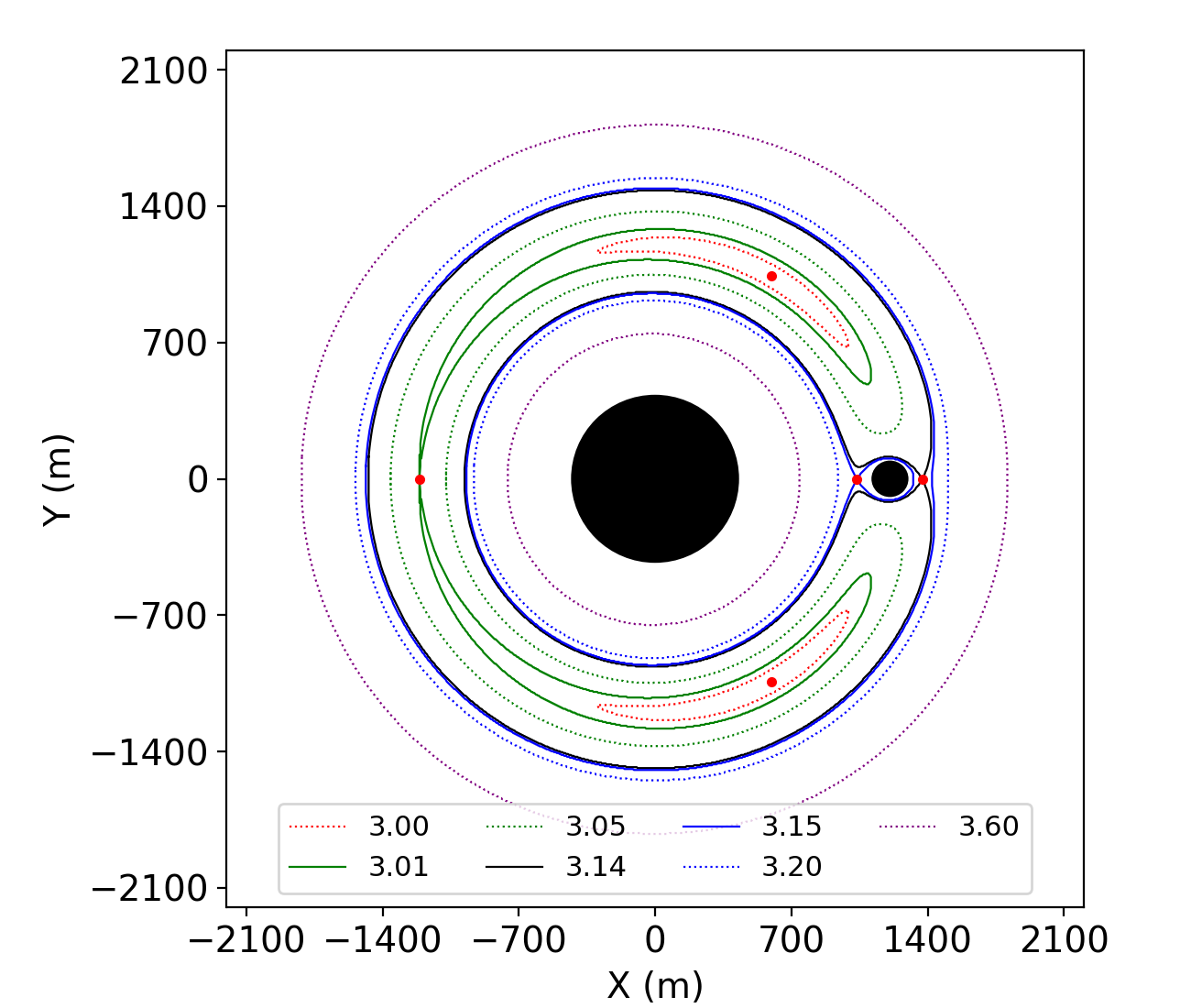}
\caption{Zero-velocity curves for Didymos and Dimorphos as mass points. The blue, black and green solid lines correspond to curves with $L_1$, $L_2$ and $L_3$ values of Jacobi constant, respectively, while the dotted lines are for other values of $C_J$, in normalized units. The Lagrangian points of the system are given by the red dots.}
\label{zero_velocity}
\end{figure}
Figure~\ref{zero_velocity} shows the Lagrangian points of the system and some selected zero-velocity curves. The zero-velocity curves are obtained by setting $\dot{X}=\dot{Y}=0$ in Equation~\ref{jacobi} for a fixed value of $C_J$, and are responsible for providing the boundaries at which a particle cannot cross \citep{Szebehely1967}. In the figure, the blue, black and green solid lines are associated with $L_1$, $L_2$ and $L_3$ values of Jacobi constant, respectively, while the dotted lines correspond to curves related to other selected values of $C_J$. For $C_J=3.00$, we have that particles initially within the regions enclosed by the dotted red lines would have imaginary orbital velocities. Therefore, these regions are forbidden regions, meaning that particles can perform any type of trajectory in the rotating frame (including tadpole orbits around $L_4$ or $L_5$), but can never enter these closed regions.

As we increase $C_J$, the forbidden regions around the triangular points grow, touching at $L_3$ for $C_J=3.01$ and then continue to grow until they reach $L_1$ and $L_2$. At this point, orbits around Lagrangian points are no longer allowed, and the particles can remain in (quasi-)satellite orbits around Didymos or Dimorphos, or around the binary in circumbinary orbits. For even higher values of $C_J$, the forbidden region swallows Dimorphos, and the only possible orbits are orbits very close to the surface of Didymos or very far from the binary. These correspond to extreme cases, for which the particle is close enough to Didymos to feel only a small perturbation of Dimorphos, or far enough away from the binary that it can be approximated to a single body. Such cases are those found by \cite{Rossi2022}.

In the planar case, we have that the motion of a particle is completely described in a phase space of four-dimensions ($X$, $Y$, $\dot{X}$, and $\dot{Y}$). The Poincaré map technique \citep{Poincare1895} consist of fixing two quantities ($Y$ and $\dot{Y}$) so that the trajectories of a group of particles can be presented in a two-dimensional phase map. Here, we will use of this technique by fixing a value of Jacobi constant, $\dot{Y}$ being determined via Equation~\ref{jacobi}. We then integrate the equations of motion, marking on the Poincaré map the values of $X$ and $\dot{X}$ for which $Y=0$. Thinking of a simple circular orbit, we have that the particle crosses the phase plane in two places: one on each side of the primary, with opposite signs of $\dot{Y}$. For this reason, a condition concerning the sign of $\dot{Y}$ is also considered. Here, we will assume the condition $\dot{Y}>0$ and $\dot{Y}<0$ for particles initially inside and outside the Dimorphos orbit, respectively.

For different values of Jacobi constant, we integrate the equations of motion of 50 particles equally distributed on the $X$-line from the surface of Didymos until $X=3500$~m. The particles are initially with $Y=\dot{X}=0$ and the integration time is equal to $10^4$ Dimorphos orbital periods. We consider a particle to be lost when it reaches the surface of Didymos or Dimorphos, or when its orbital radius becomes larger than the mutual Hill radius of the system ($a_{\rm Hill}=111500$~m). For $X>3500$~m, the particles are usually in stable circumbinary orbits (classified as Family F orbits in this article) and do not exhibit particularly interesting dynamics.

\begin{figure}
\subfloat[$C_J=2.60$]{\includegraphics[width=0.5\columnwidth,trim={0 0 0 0},clip]{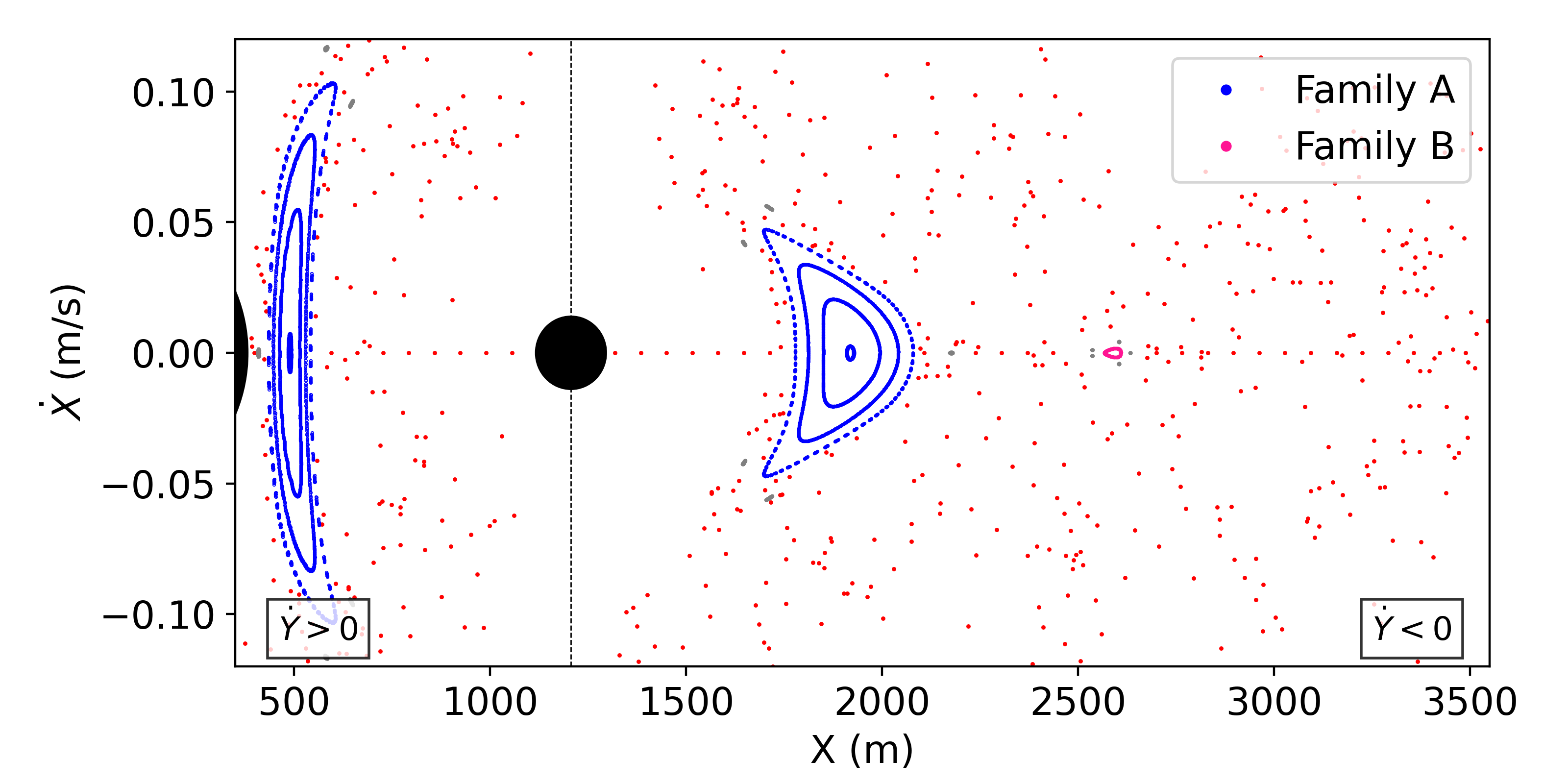}\label{ssp_pma}}
\subfloat[$C_J=2.75$]{\includegraphics[width=0.5\columnwidth,trim={0 0 0 0},clip]{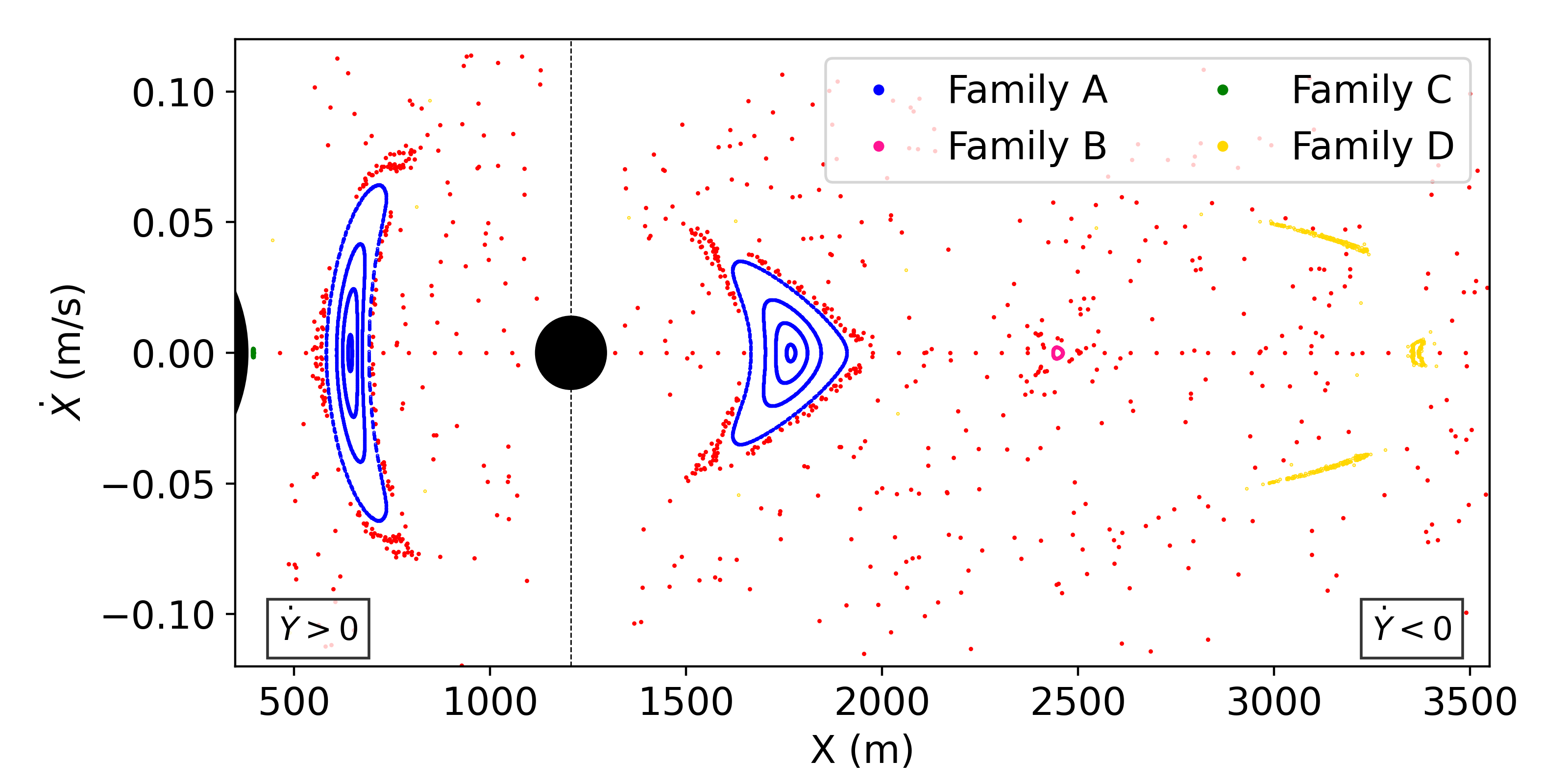}\label{ssp_pmb}}\\
\subfloat[$C_J=2.92$]{\includegraphics[width=0.5\columnwidth,trim={0 0 0 0},clip]{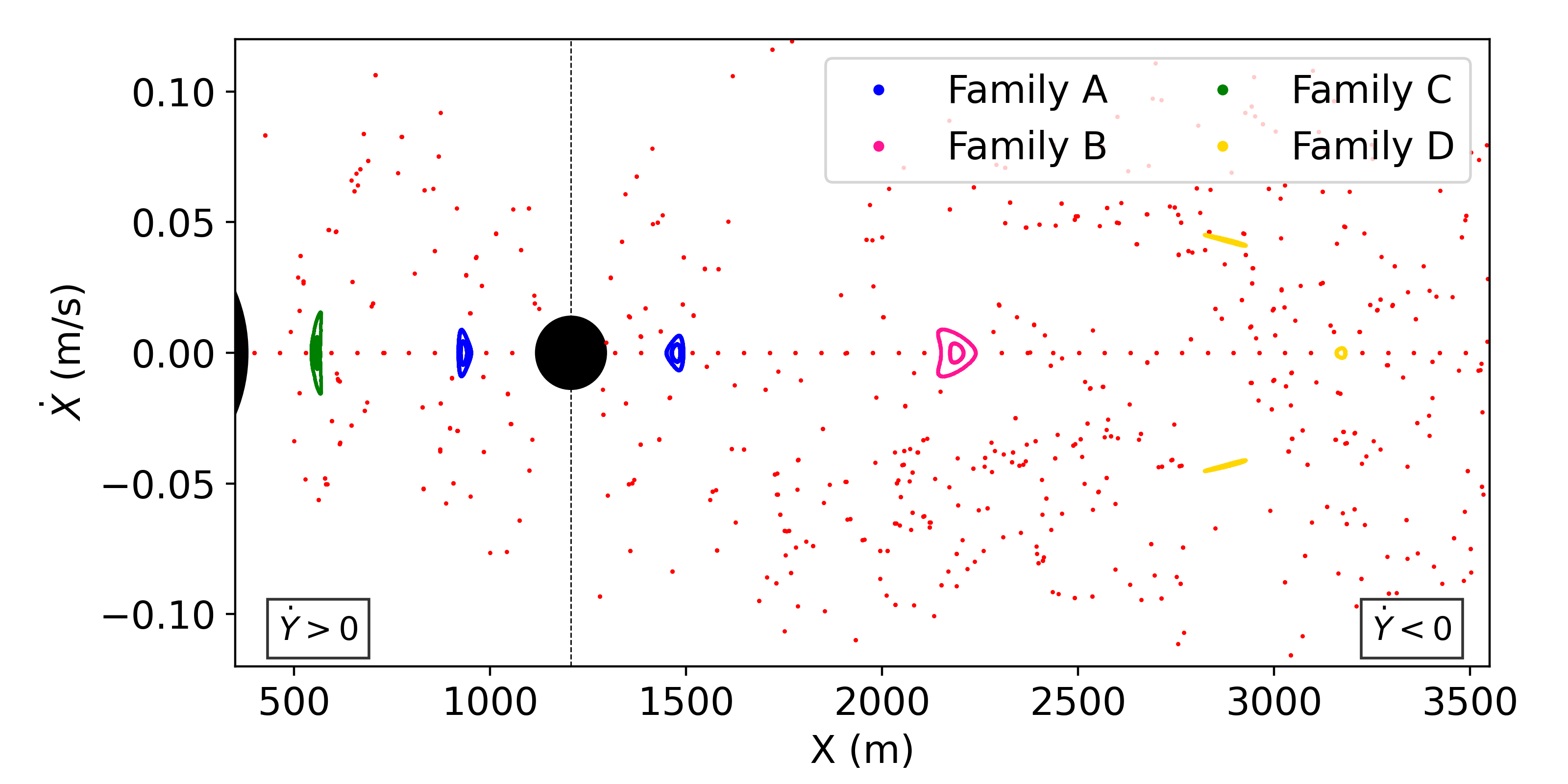}\label{ssp_pmc}}
\subfloat[$C_J=3.00$]{\includegraphics[width=0.5\columnwidth,trim={0 0 0 0},clip]{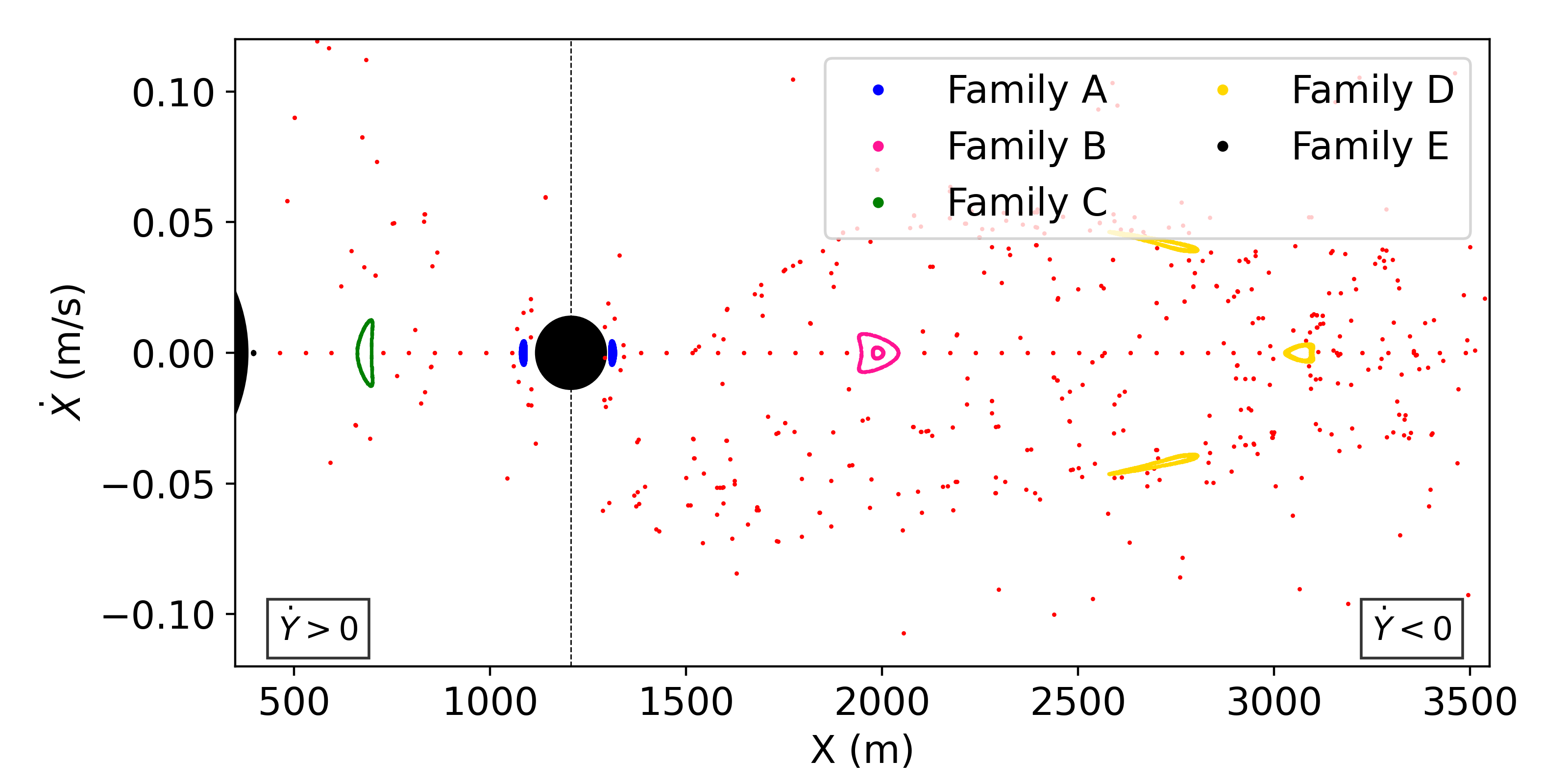}\label{ssp_pmd}}\\
\subfloat[$C_J=3.10$]{\includegraphics[width=0.5\columnwidth,trim={0 0 0 0},clip]{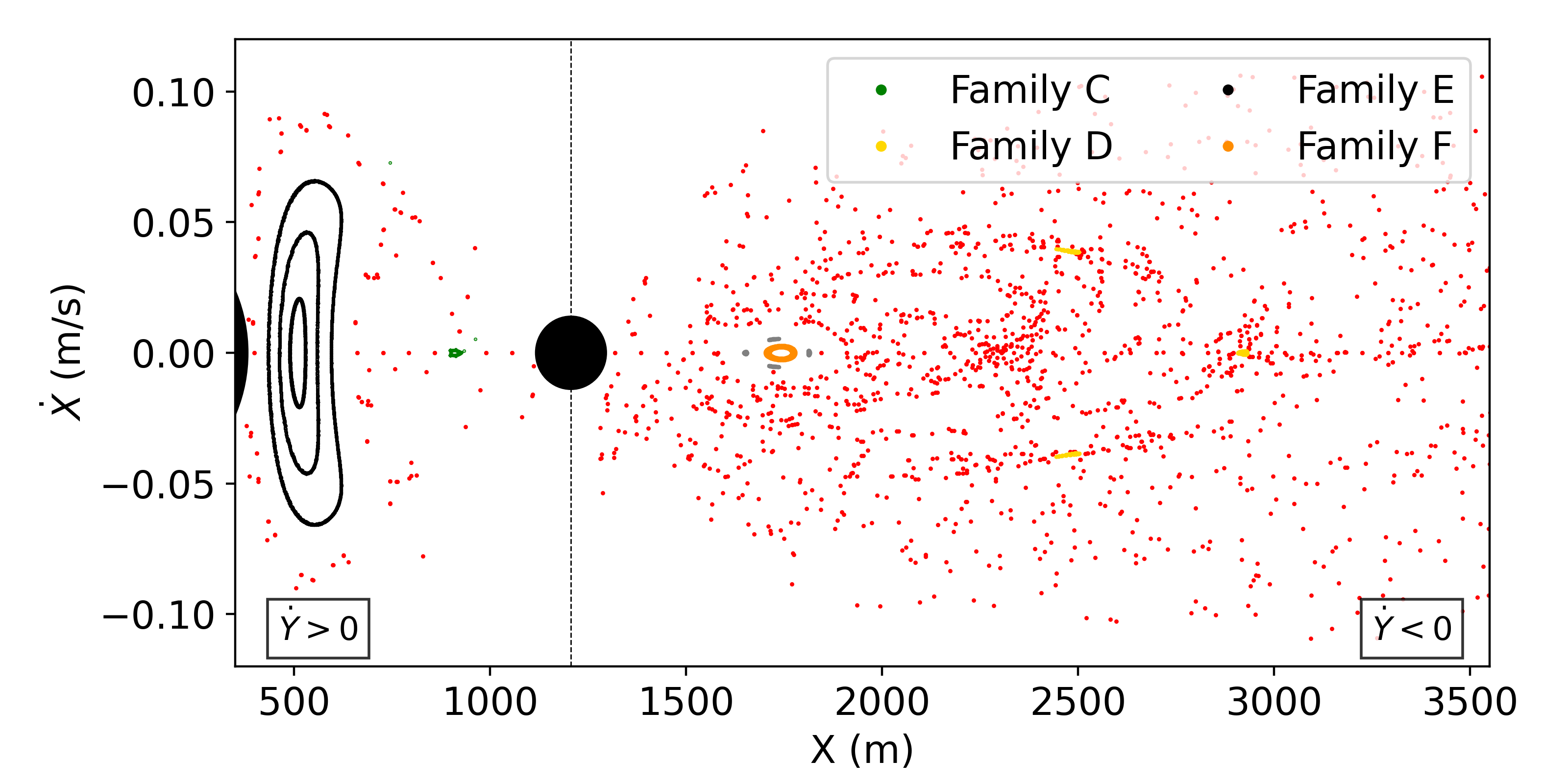}\label{ssp_pme}}
\subfloat[$C_J=3.20$]{\includegraphics[width=0.5\columnwidth,trim={0 0 0 0},clip]{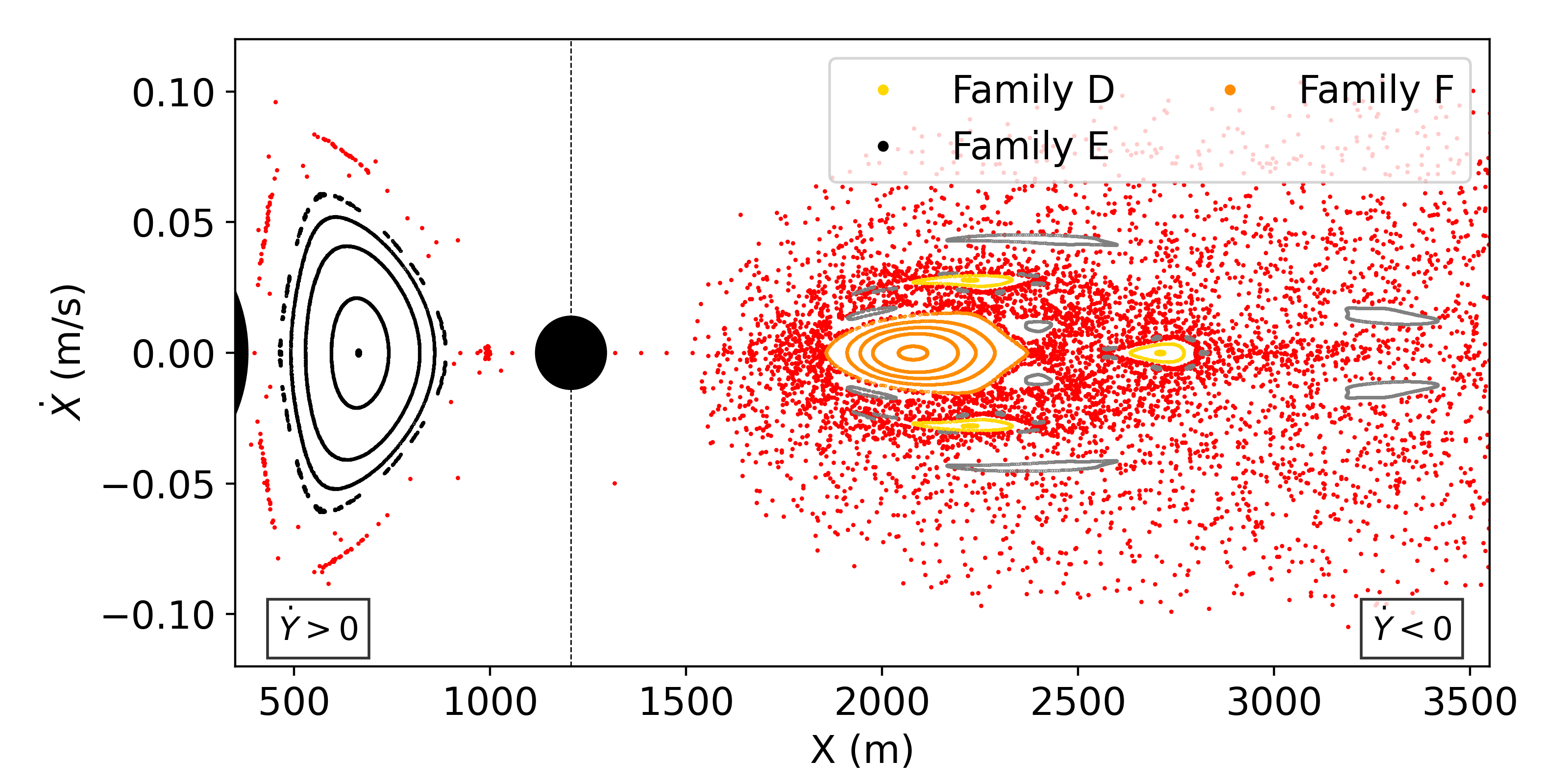}\label{ssp_pmf}}\\
\subfloat[$C_J=3.40$]{\includegraphics[width=0.5\columnwidth,trim={0 0 0 0},clip]{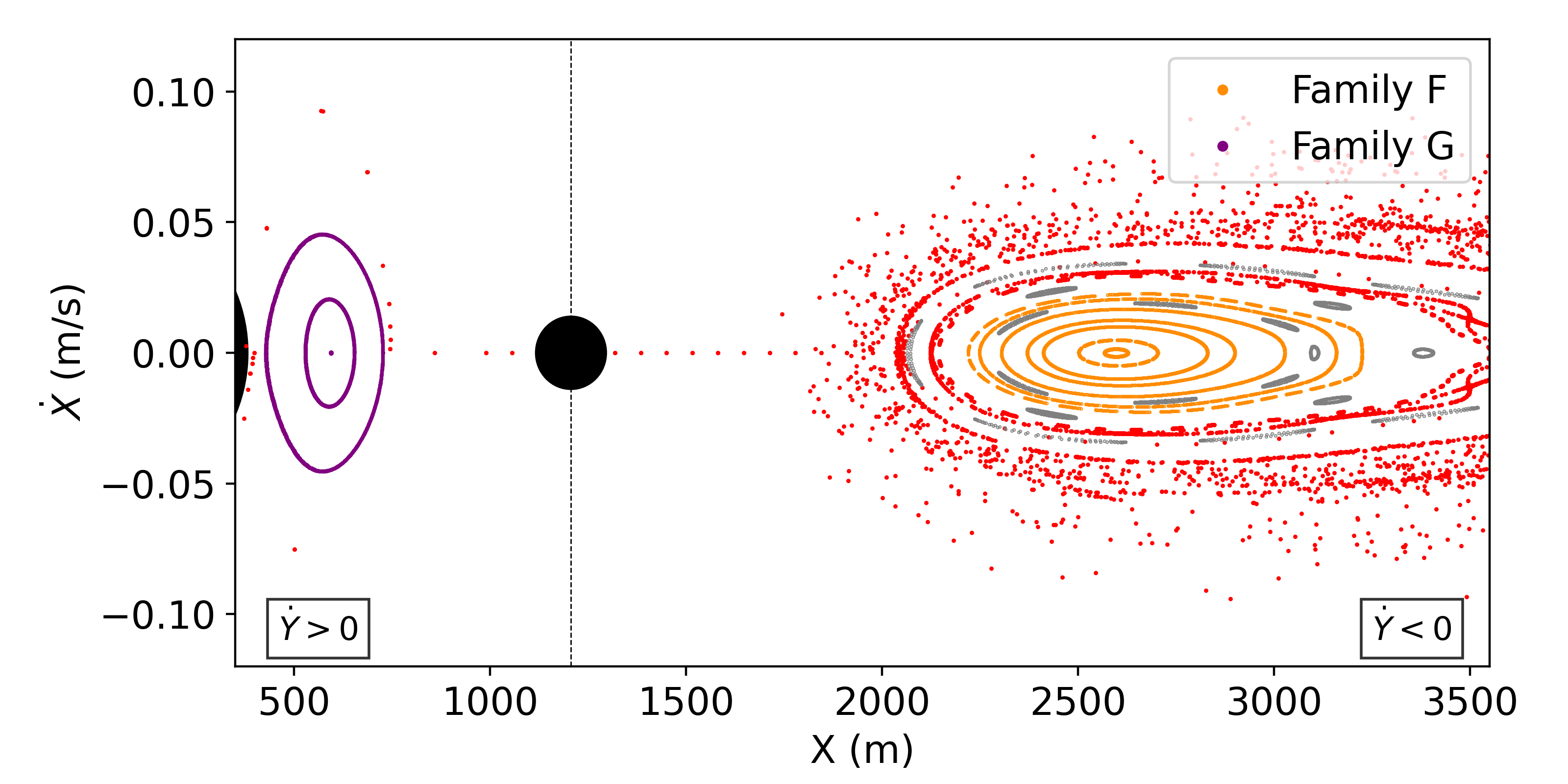}\label{ssp_pmg}}
\subfloat[$C_J=3.80$]{\includegraphics[width=0.5\columnwidth,trim={0 0 0 0},clip]{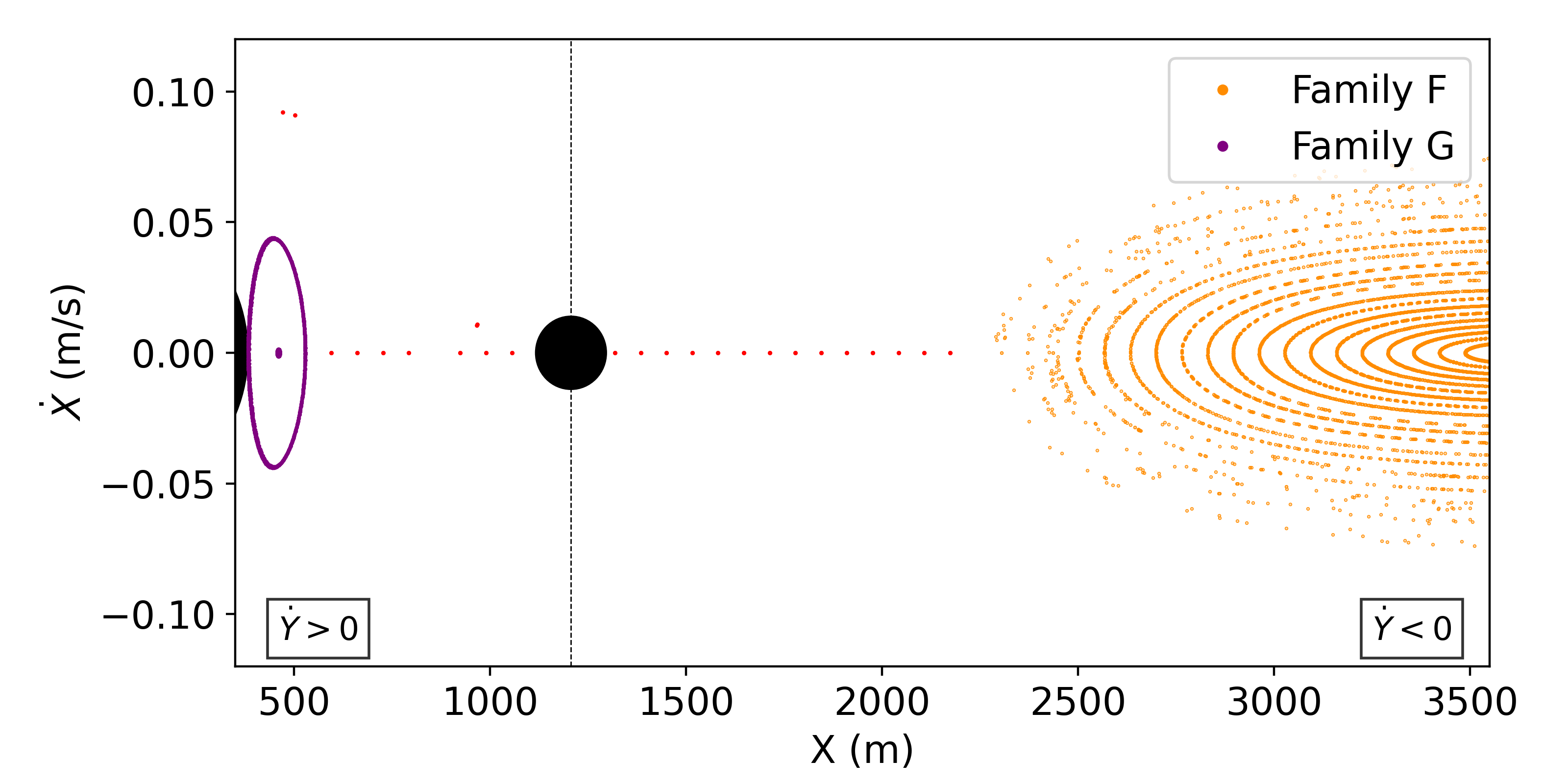}\label{ssp_pmh}}\\
\caption{Poincaré maps for different values of Jacobi constant, indicated in the legend of each panel. For particles initially with $X<R_{12}$, we display only the points with $\dot{Y}>0$, while for $X>R_{12}$, we display the points with $\dot{Y}<0$. The colored islands correspond to stable orbits, while the red dots are chaotic orbits. Here, we are assuming Didymos and Dimorphos as mass points.}
\label{ssp_pm}
\end{figure}

Figure~\ref{ssp_pm} shows a set of Poincaré maps, for $2.60\leq C_J\leq 3.80$. Particles in chaotic orbits are responsible for irregular distributions of points on the maps, coloured red here, while islands in other colors are associated with stable orbits. The closed islands correspond to quasi-periodic stable orbits, and are related to the stable orbit located in the center of a group of closed islands. Here we will name a family of stable orbits (periodic orbit + group of quasi-periodic orbits) by different letters of the Latin alphabetic. The islands in gray correspond to higher-order mean motion resonances (MMRs) that are not of our interest, given that they only exist in a very restricted range of the Jacobi constant. Therefore, we will not label them.

\cite{Poincare1895} classifies periodic orbits into two sorts: the periodic orbit of first sort is the "most circular" orbit obtained in the system for a given $C_J$ and intersects the Poincaré map at only one location, while the periodic orbit of second sort is the orbit at the centre of a $m$:$m-j$ MMR with the secondary body. Each periodic orbit of second sort crosses the Poincaré map at $j$ different locations, $j$ corresponding to the order of the resonance. In addition to the number of locations that cross the map, a simple way to distinguish these two kinds of orbits is by the period: The period of the periodic orbit of the first sort changes with the Jacobi constant, while that of the periodic orbit of the second sort does not change. For systems with mass ratio $M_2/M_1\gtrsim10^{-2}$, we have the emergence of additional types of periodic orbits \citep{Broucke1968} in the system, such as "Family A" that we will obtain in this work.

\begin{figure}
\centering
\subfloat[Family A]{\includegraphics[width=0.33\columnwidth,trim={0 0 0 0},clip]{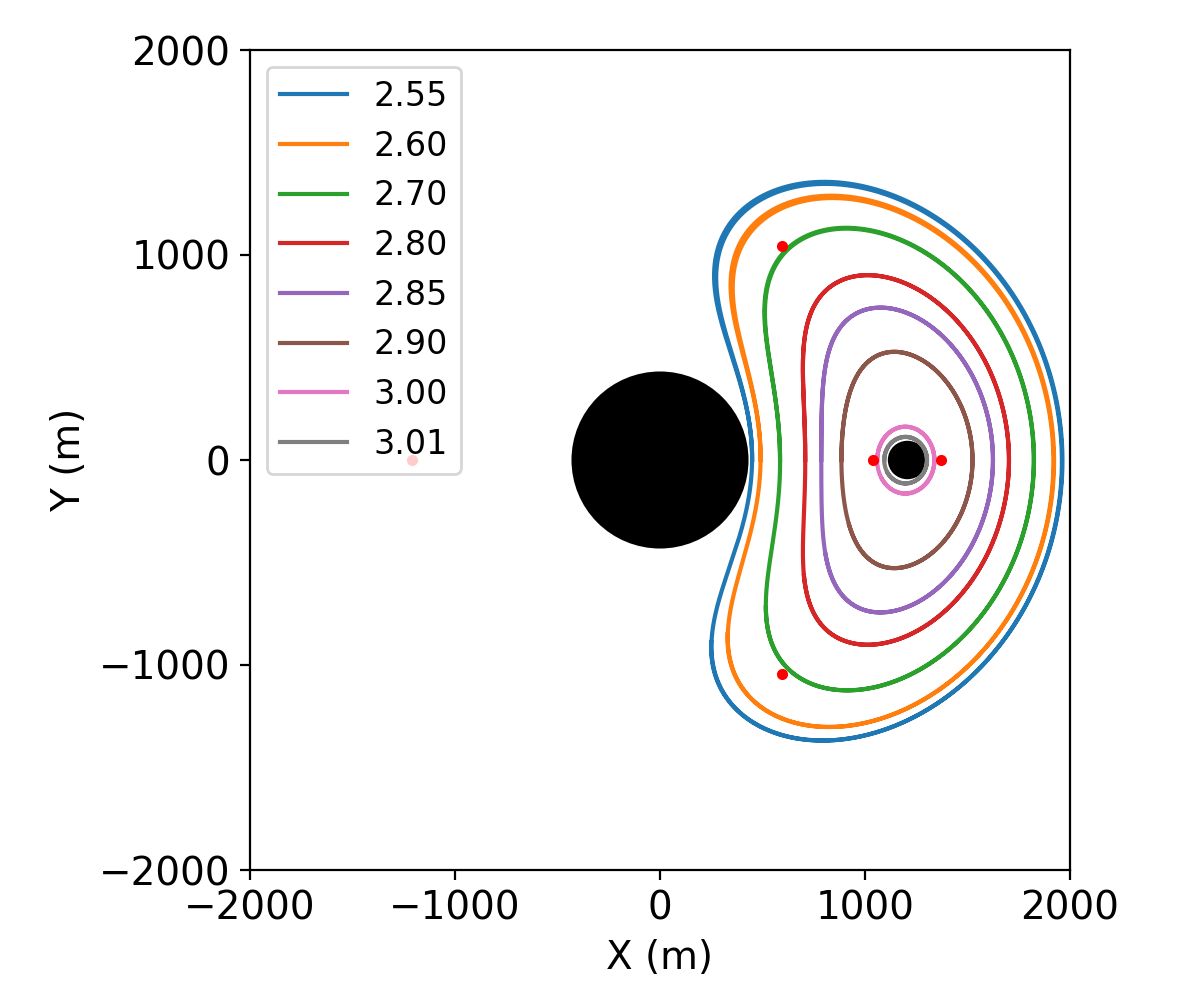}\label{p_orbitsa}}
\subfloat[Family B]{\includegraphics[width=0.33\columnwidth,trim={0 0 0 0},clip]{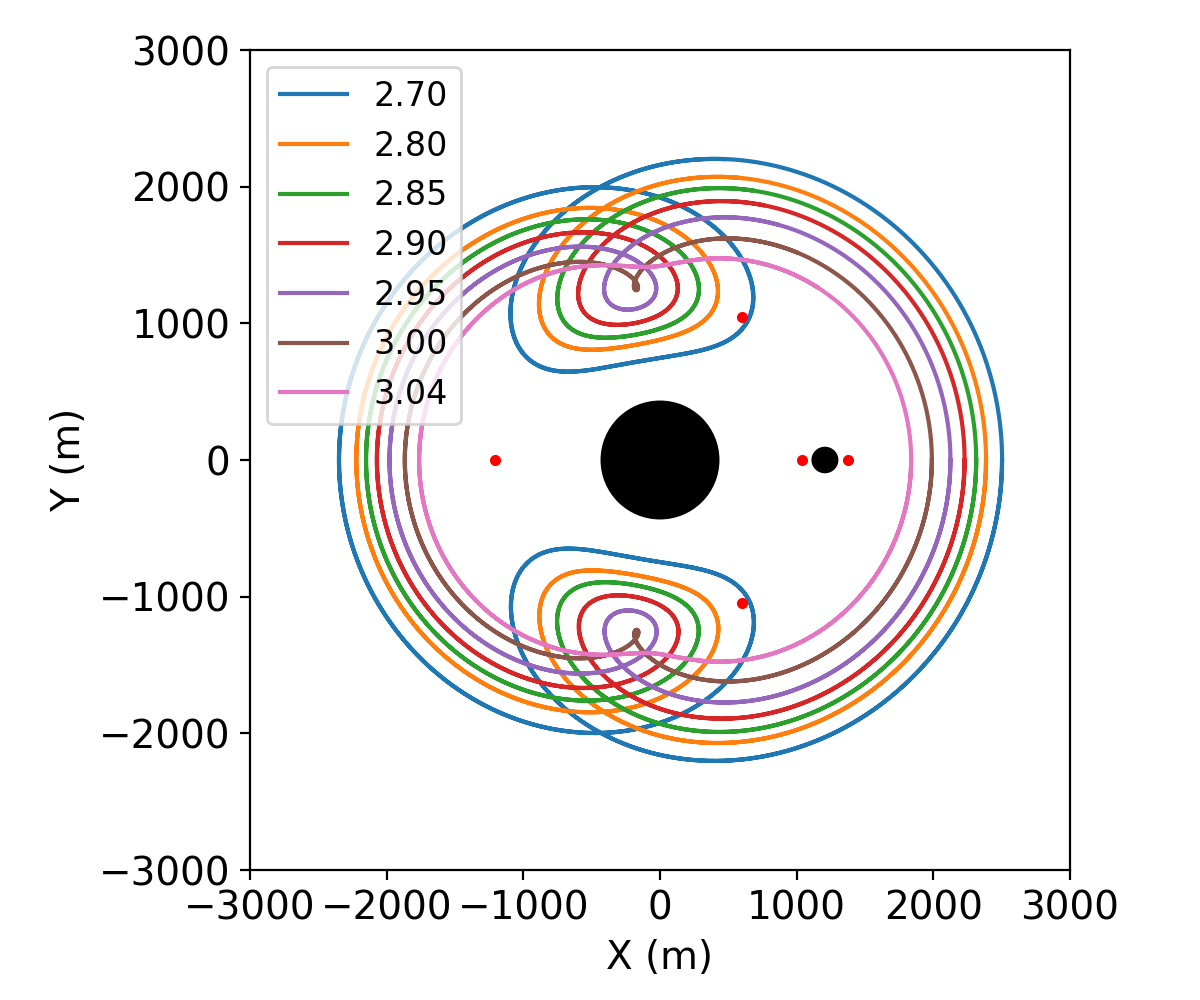}\label{p_orbitsb}}
\subfloat[Family C]{\includegraphics[width=0.33\columnwidth,trim={0 0 0 0},clip]{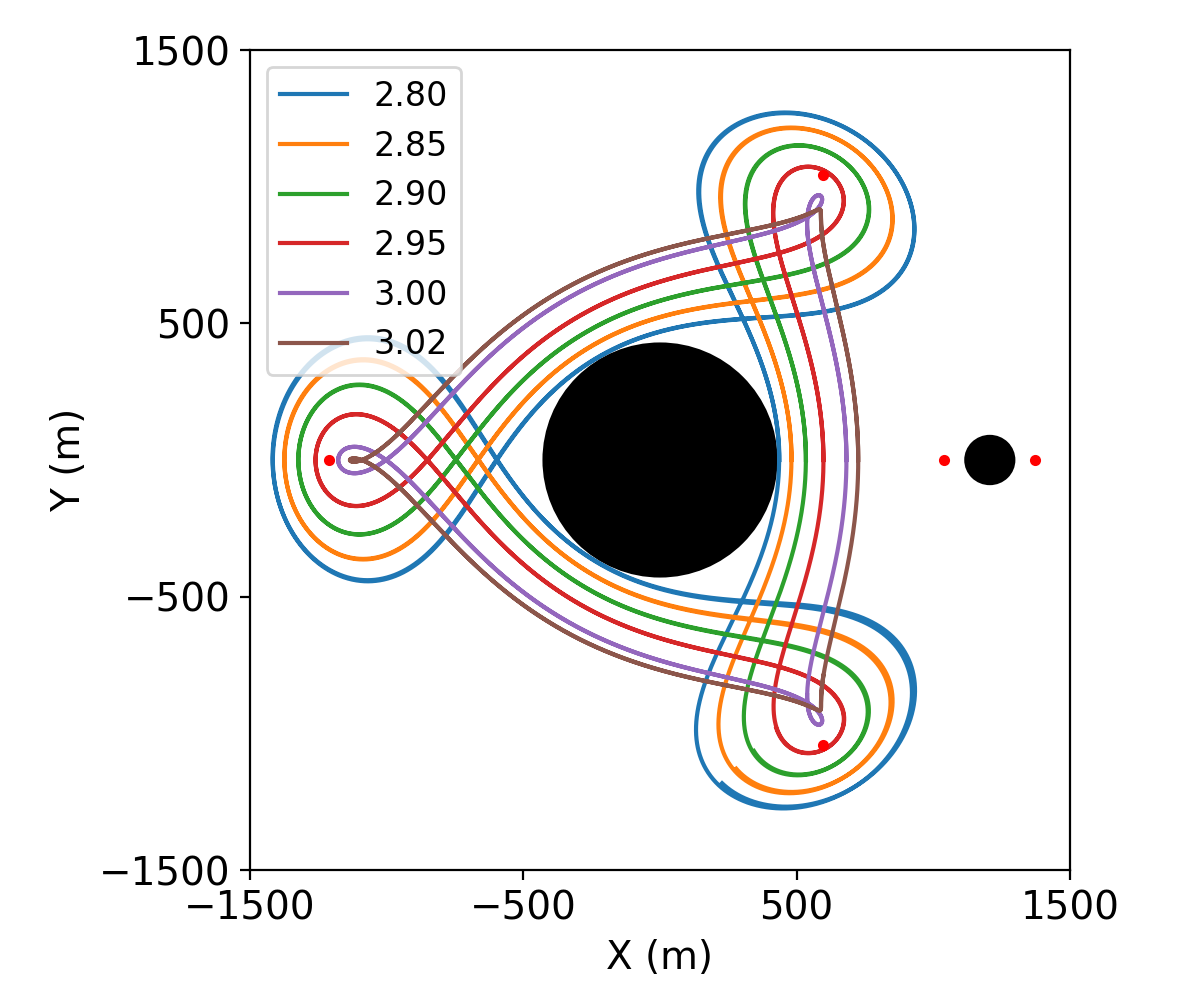}\label{p_orbitsc}}\\
\subfloat[Family D]{\includegraphics[width=0.33\columnwidth,trim={0 0 0 0},clip]{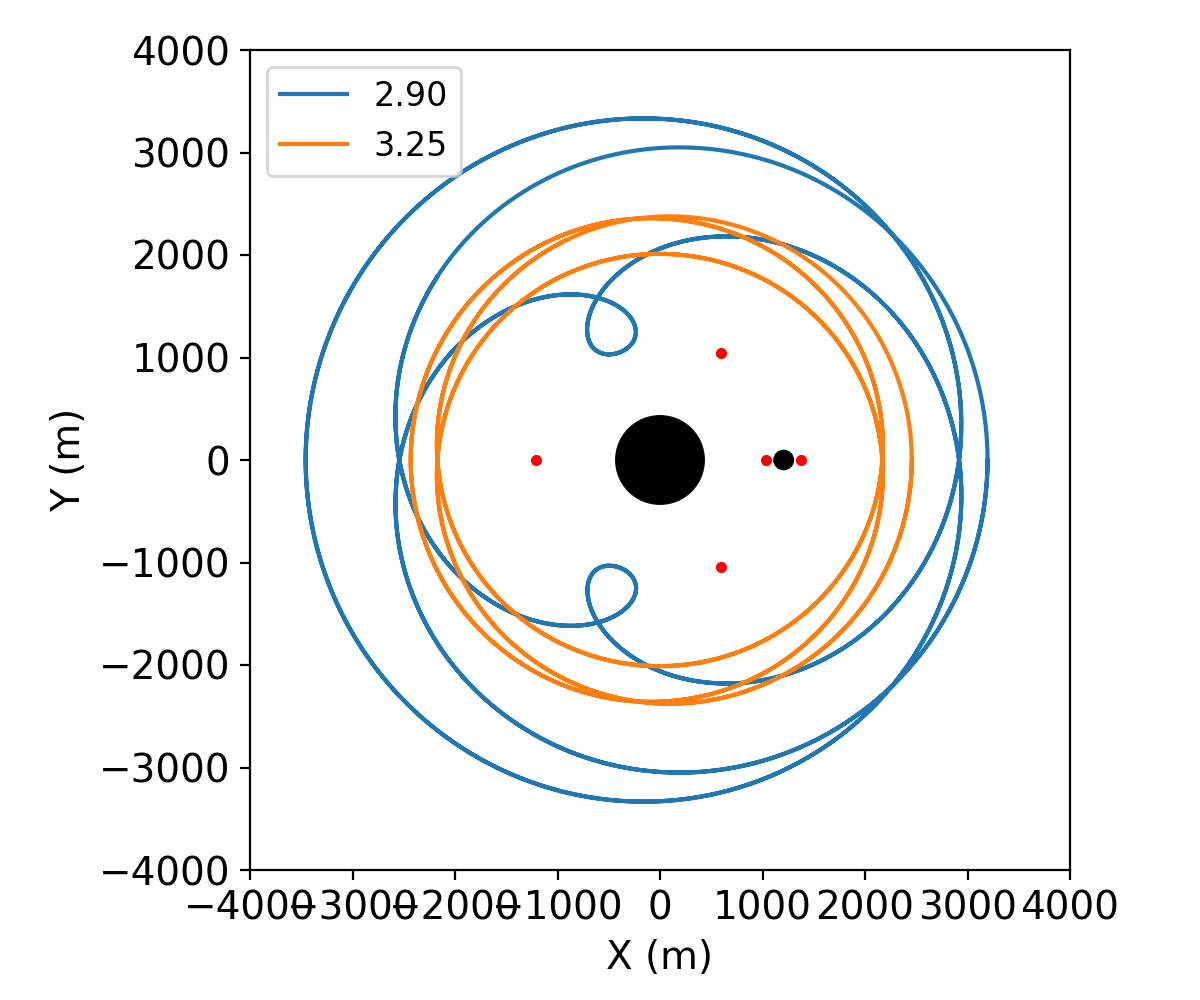}\label{p_orbitsd}}
\subfloat[Family E]{\includegraphics[width=0.33\columnwidth,trim={0 0 0 0},clip]{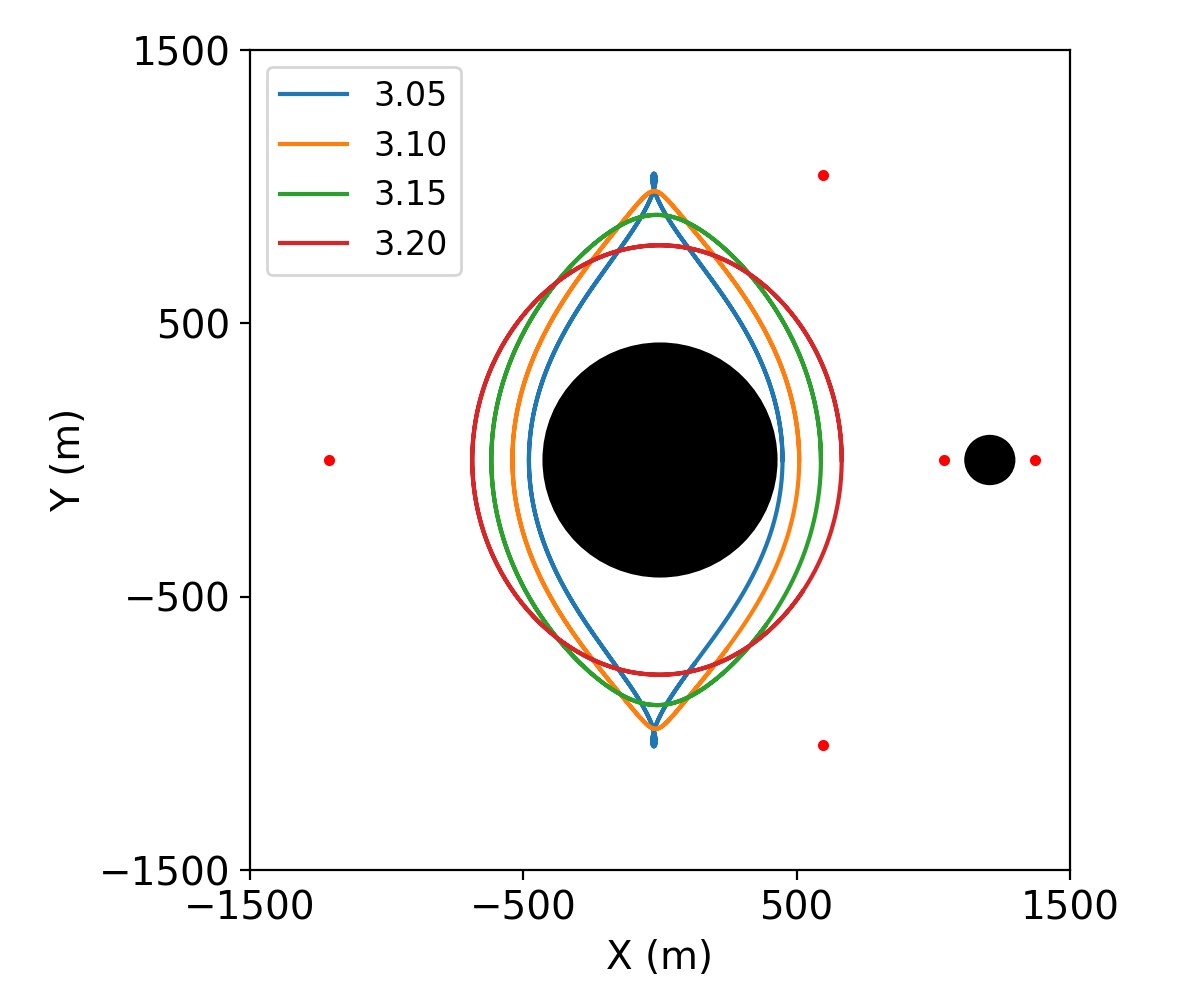}\label{p_orbitsf}}
\subfloat[Family F]{\includegraphics[width=0.33\columnwidth,trim={0 0 0 0},clip]{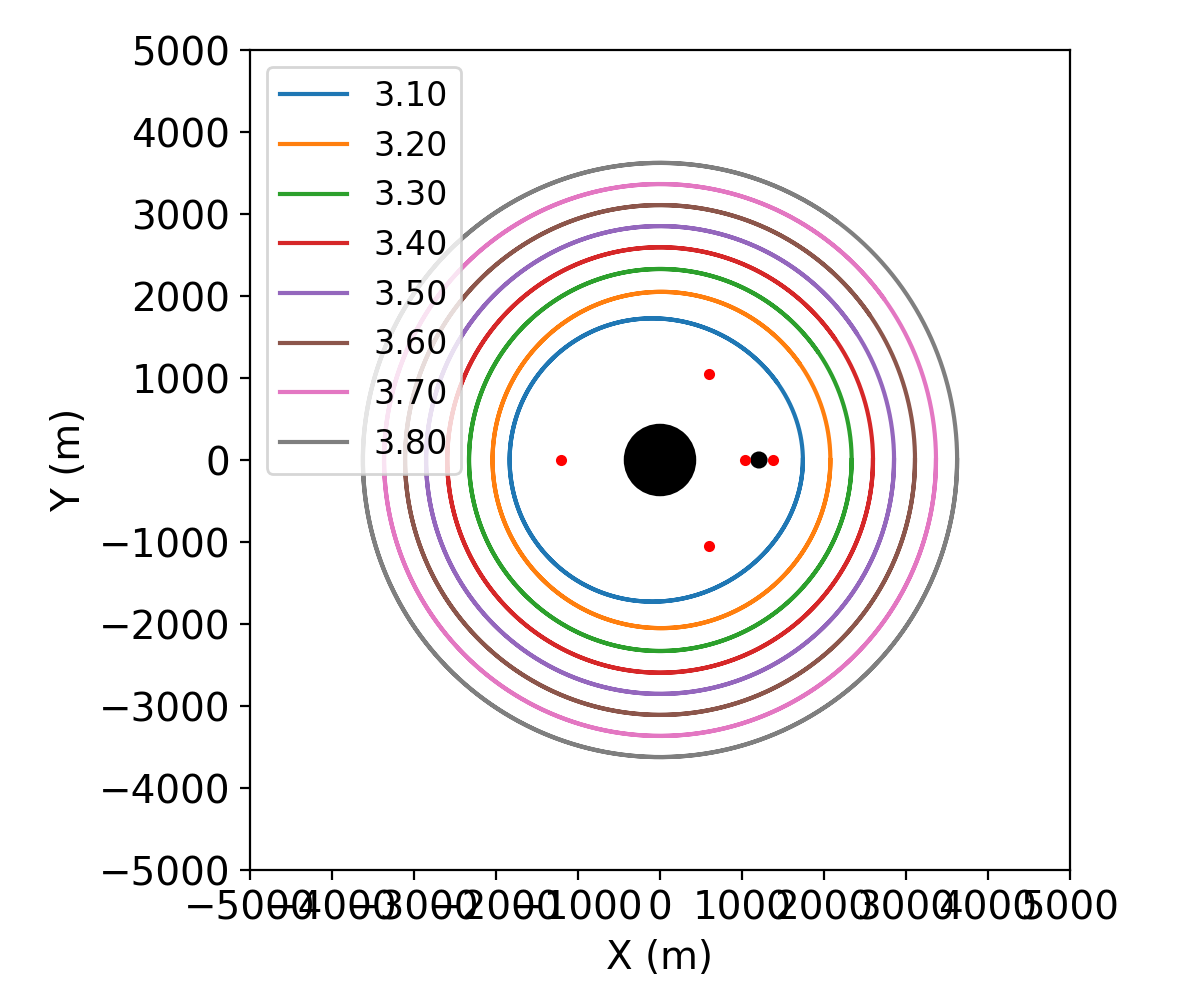}\label{p_orbitsg}}\\
\subfloat[Family G]{\includegraphics[width=0.33\columnwidth,trim={0 0 0 0},clip]{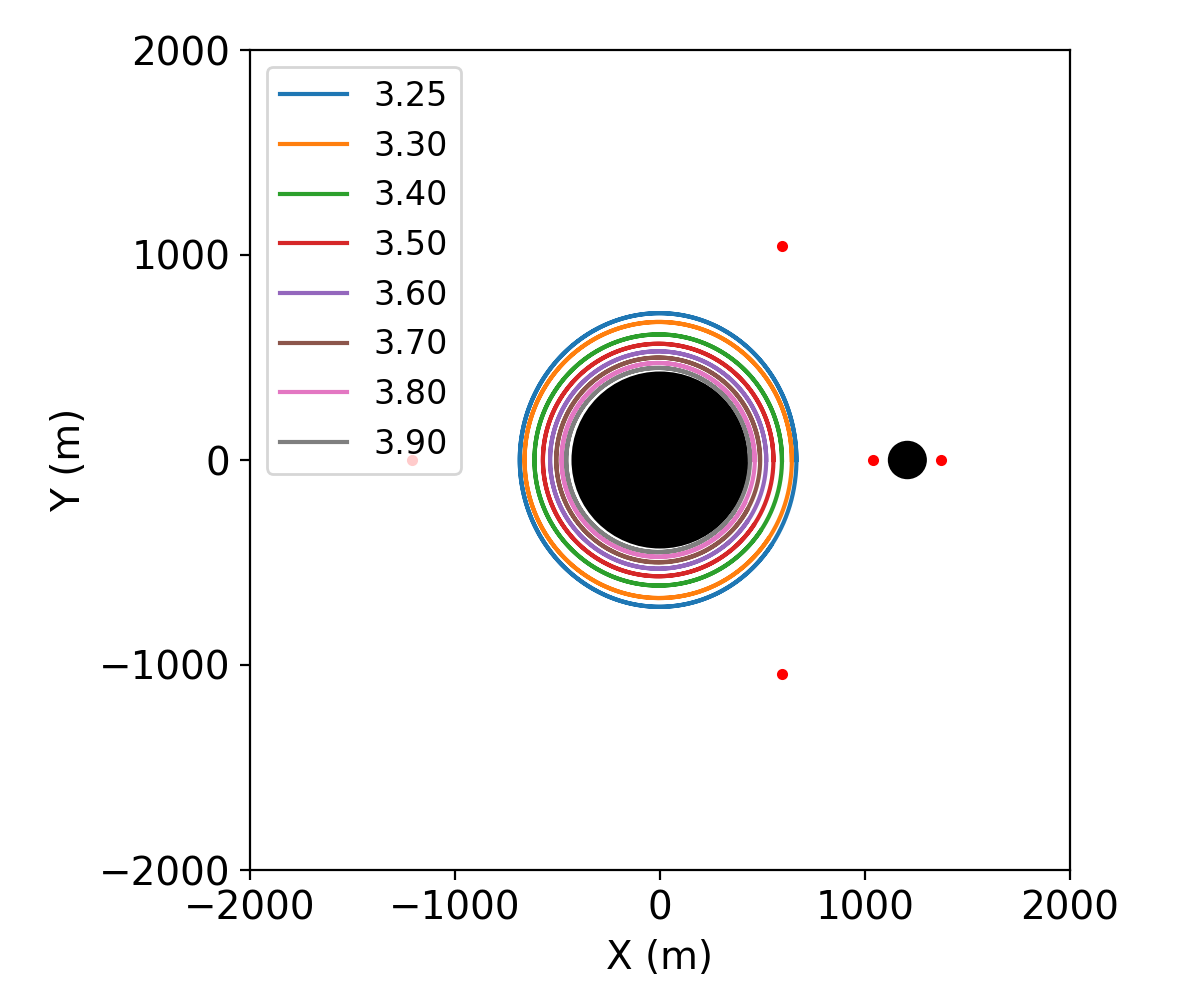}\label{p_orbitsh}}
\caption{Periodic orbits of the families of stable orbits for different values of $C_J$. Didymos and Dimorphos are assumed as mass points.}
\label{p_orbits}
\end{figure}

For $C_J=2.60$ (Figure~\ref{ssp_pma}), we obtain that the motion of the particles is mostly chaotic, which is a general result found in our simulations. The first set of stable orbits we obtain is Family A (in blue), with its periodic orbit crossing the Poincaré map in two locations, one on each side of Dimorphos. This family does not fit the classical classification of \cite{Poincare1895}, but is associated with simple-periodic symmetrical retrograde satellites (or circum-secondary) orbits, according to the classification of \cite{Henon1969b,Henon1970}. The periodic orbits of Family A for different values of $C_J$ are presented in Figure~\ref{p_orbitsa}. For lower values of $C_J$, the particles are in a particular kind of co-orbital resonance, and the periodic orbits librate around Dimorphos, encompassing the $L_1$ and $L_2$ points. These orbits are commonly called quasi-satellite orbits \citep{Mikkola1997,Pousse2017}. With increasing $C_J$, the periodic orbit approaches the Hill sphere of Dimorphos. The orbit is unstable at the Hill sphere, located at $\sim0.14$ from Dimorphos, in normalized units. Increasing $C_J$ even more, Family A becomes stable again, performing retrograde satellite orbits around Dimorphos \citep{Namouni1999,Pousse2017}.

In Figure~\ref{ssp_pma}, one can also identify Family B (in pink), related with the $2$:$3$ MMR with Dimorphos. The trajectories of the stable orbits of this family are shown in Figure~\ref{p_orbitsb}. As a general rule, the trajectory of a particle in $m$:$m-j$ MMR will always exhibit some specific structures in the rotating frame, these being: i) the orbit has $|m|$ identical sectors and ii) each sector has $j-1$ self-crossing points. For the $2$:$3$ MMR, $m=-2$\footnote{$m<0$ corresponds to resonances external to Dimorphos}, and we get orbits with 2 identical sectors. Orbits in first-order resonances tend not to cross themselves, being the loops observed in some orbits caused by the eccentricity. In general, the eccentricity decreases with increasing $C_J$ \citep{Madeira2022a} and we observe that the loops become smaller as we increase $C_J$, disappearing for the largest value of $C_J$, for which the orbit has no self-crossing.

As $C_J$ increases, Family B approaches Dimorphos, initially becoming larger (Figs.~\ref{ssp_pmb},\ref{ssp_pmc}) and then shrinking (Fig.~\ref{ssp_pmd}), which corresponds to a classical pattern of resonant orbits in the Poincaré Map. In Figure~\ref{ssp_pmb}, we have the emergence of Families C (in green) and D (in yellow), related to $3$:$2$ and $2$:$5$ MMRs with Dimorphos. While the $3$:$2$ MMR orbits have no self-crossing for low eccentricities (Figure~\ref{p_orbitsc}), we obtain two self-crossing points in each of the two sectors of the $2$:$5$ MMR orbits (Figure~\ref{p_orbitsd}). For $C_J=3.00$ (Figure~\ref{ssp_pmd}), we have the appearance of Family E (in black), related to the $2$:$1$ MMR with Dimorphos (Figure~\ref{p_orbitsf}).

For $C_J=3.10$ (Figure~\ref{ssp_pme}), we identify a new family of stable orbits, Family F (in orange). The stable orbits of this family are shown in Figure~\ref{p_orbitsg}, corresponding to circumbinary orbits, first sort orbits around the barycenter of the system. The last family is Family G (in purple) which appears in Figure~\ref{ssp_pmg} and is related to first sort orbits around Didymos (Figure~\ref{p_orbitsf}). Unlike the resonant orbits, we have that Families F and G move away from Dimorphos when increasing $C_J$, corresponding to the only families that exist for larger values of Jacobi constant.

Bearing in mind the inverse relationship between eccentricity and Jacobi constant, we conclude that low eccentricity orbits can be stable only when very close to Didymos (Family G) or far from the binary (Family F). This result was expected. However, we also find that particles can reside in eccentric orbits in co-orbital motion (Family A), in MMR regions with Dimorphos (Families B, C, D, and E), or even in retrograde low eccentricity orbits around Dimorphos (also Family A). Next, we assess how these regions are altered when considering the shapes of Didymos and Dimorphos.

\section{Didymos and Dimorphos as NSSBs} \label{sec_SM}
\begin{figure}
\subfloat[$C_J=2.60$]{\includegraphics[width=0.5\columnwidth,trim={0 0 0 0},clip]{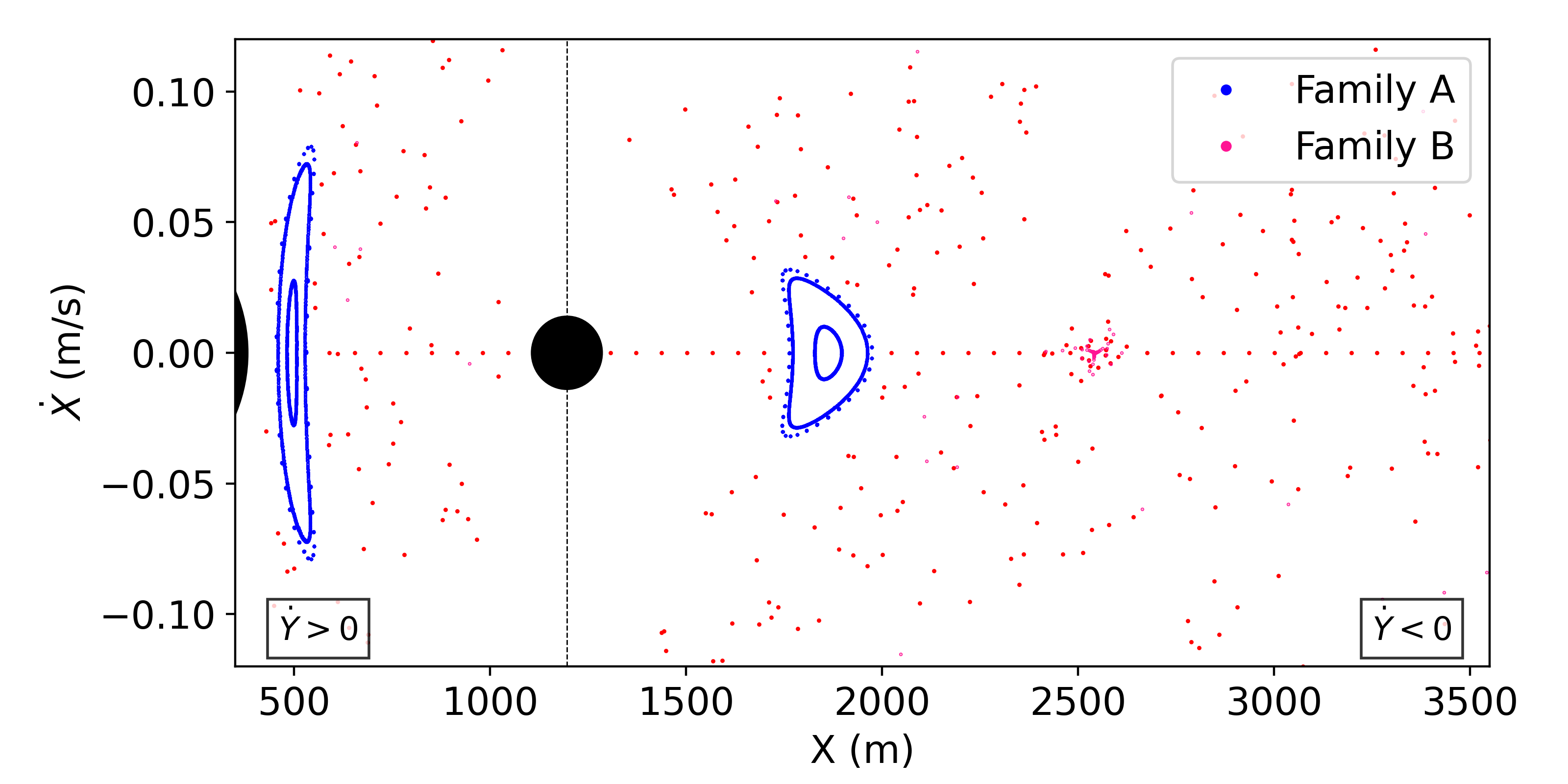}}
\subfloat[$C_J=2.75$]{\includegraphics[width=0.5\columnwidth,trim={0 0 0 0},clip]{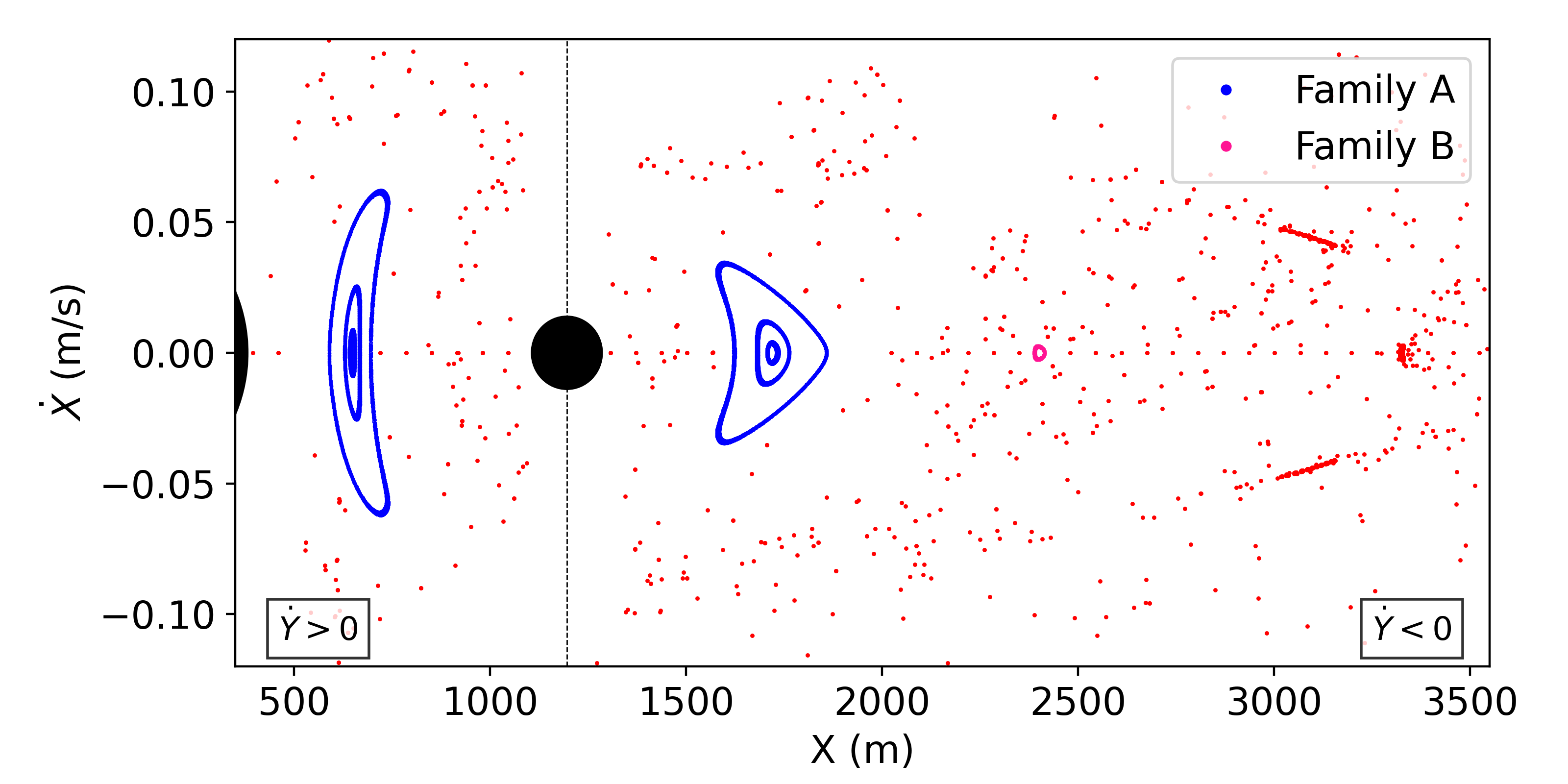}}\\
\subfloat[$C_J=2.92$]{\includegraphics[width=0.5\columnwidth,trim={0 0 0 0},clip]{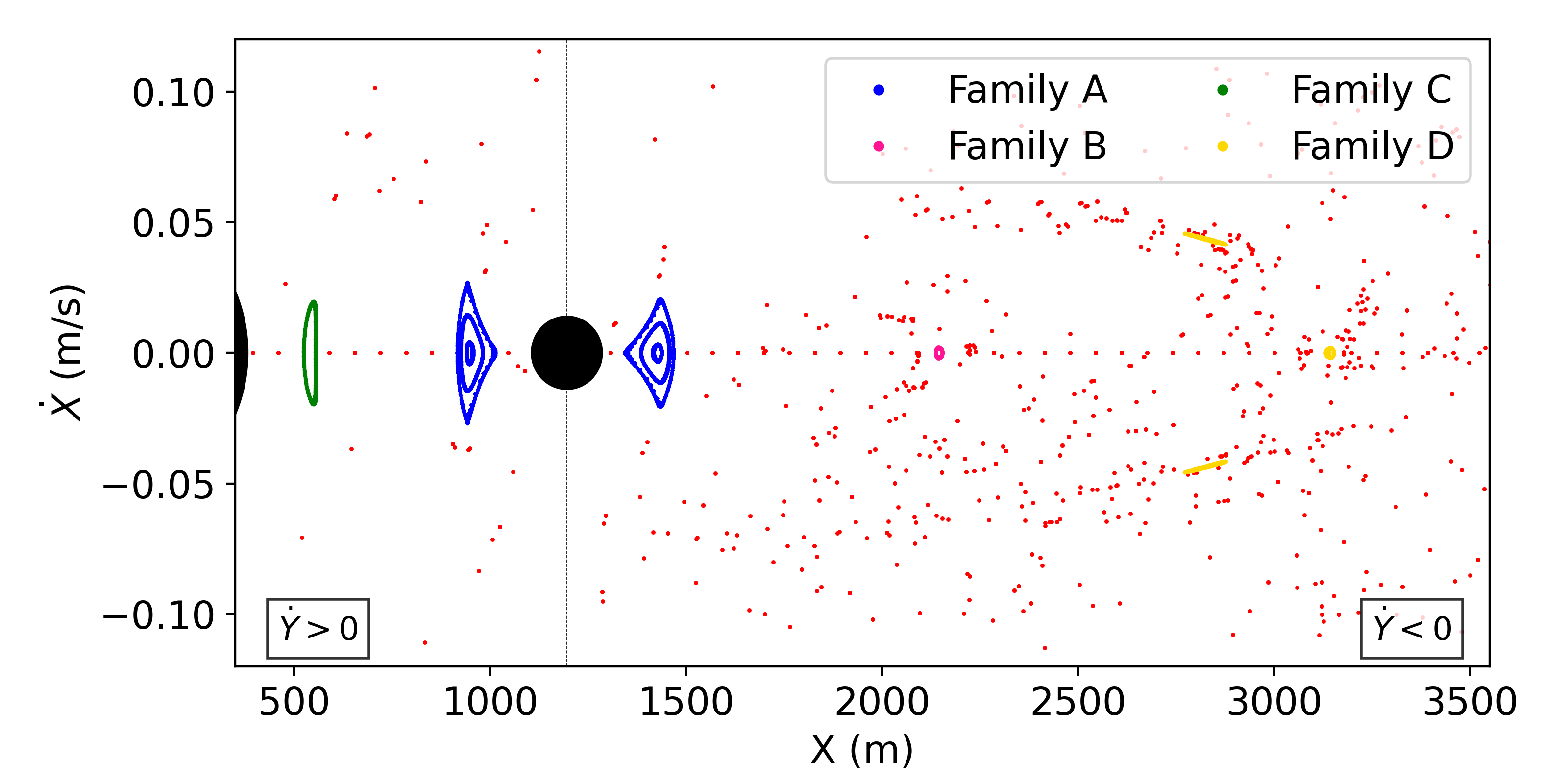}}
\subfloat[$C_J=3.00$]{\includegraphics[width=0.5\columnwidth,trim={0 0 0 0},clip]{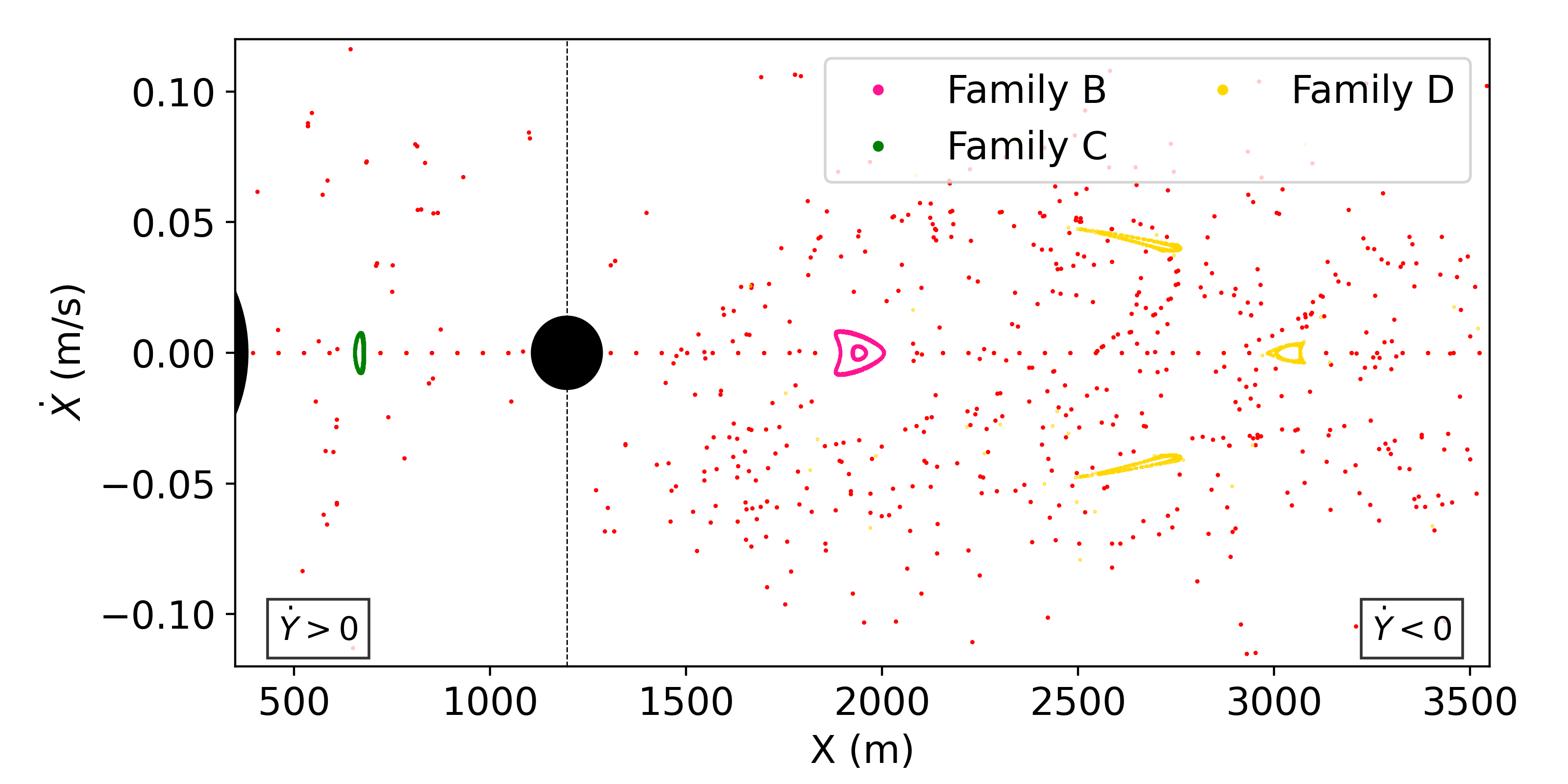}}\\
\subfloat[$C_J=3.10$]{\includegraphics[width=0.5\columnwidth,trim={0 0 0 0},clip]{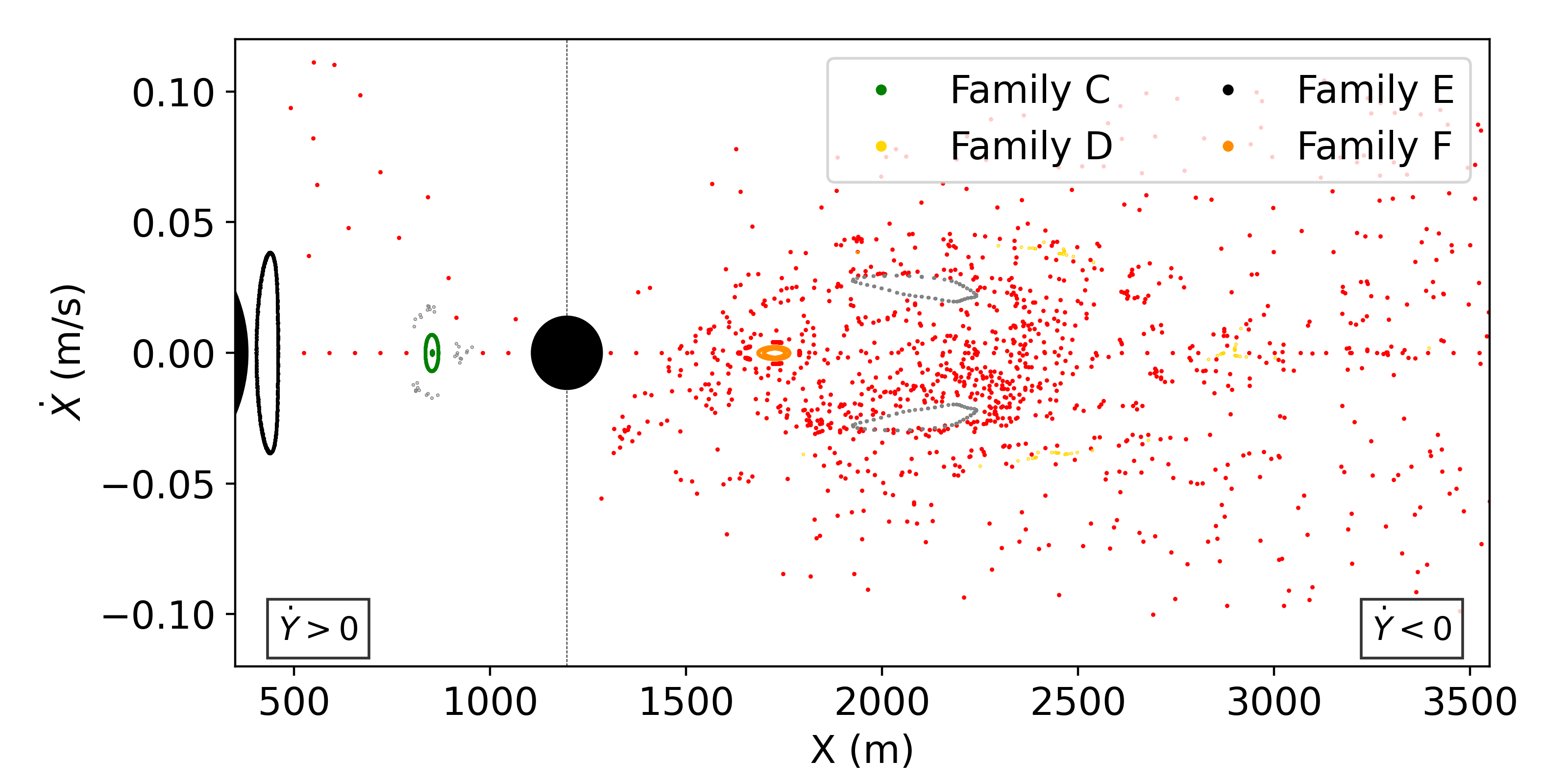}}
\subfloat[$C_J=3.20$]{\includegraphics[width=0.5\columnwidth,trim={0 0 0 0},clip]{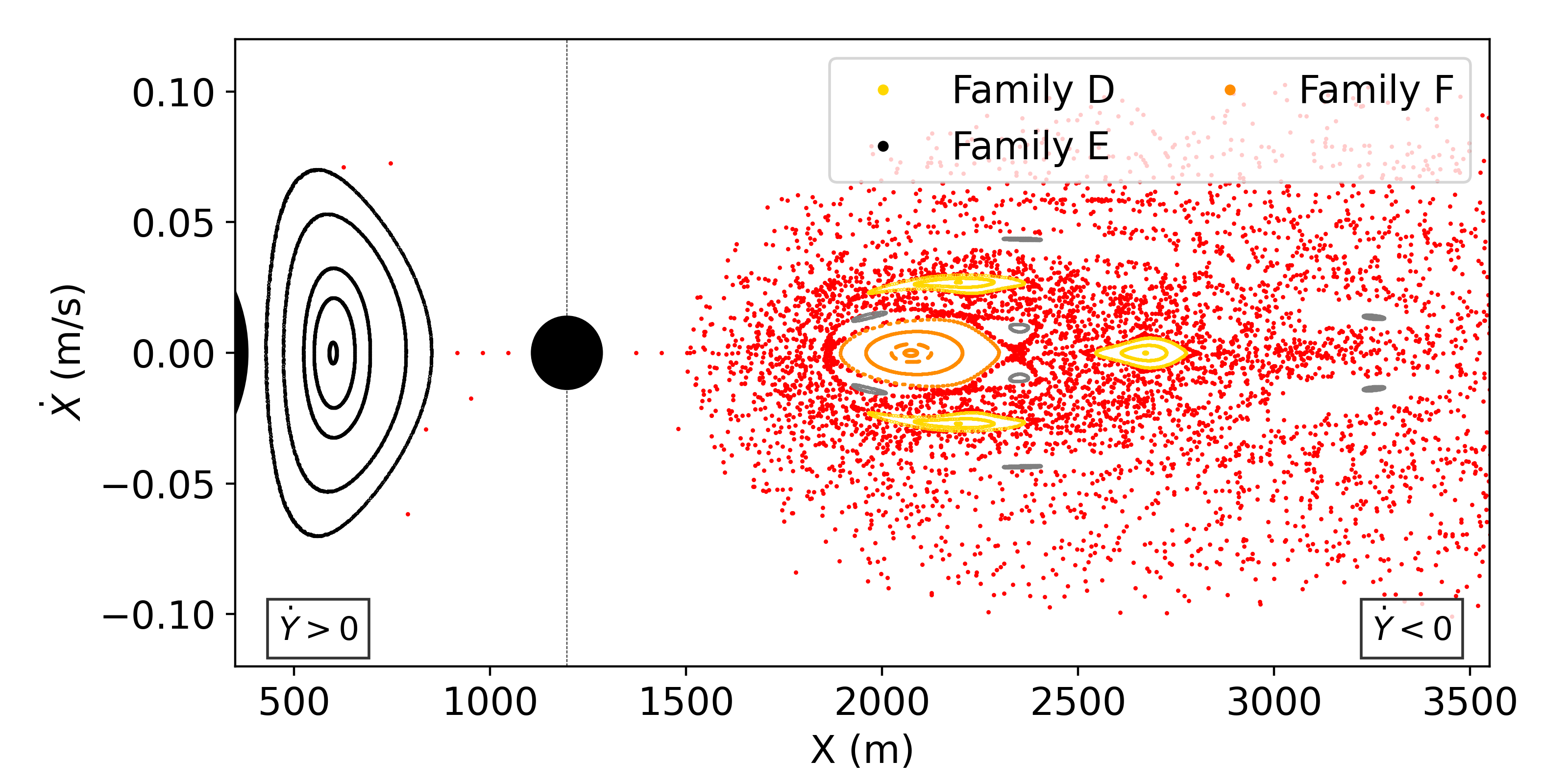}}\\
\subfloat[$C_J=3.40$]{\includegraphics[width=0.5\columnwidth,trim={0 0 0 0},clip]{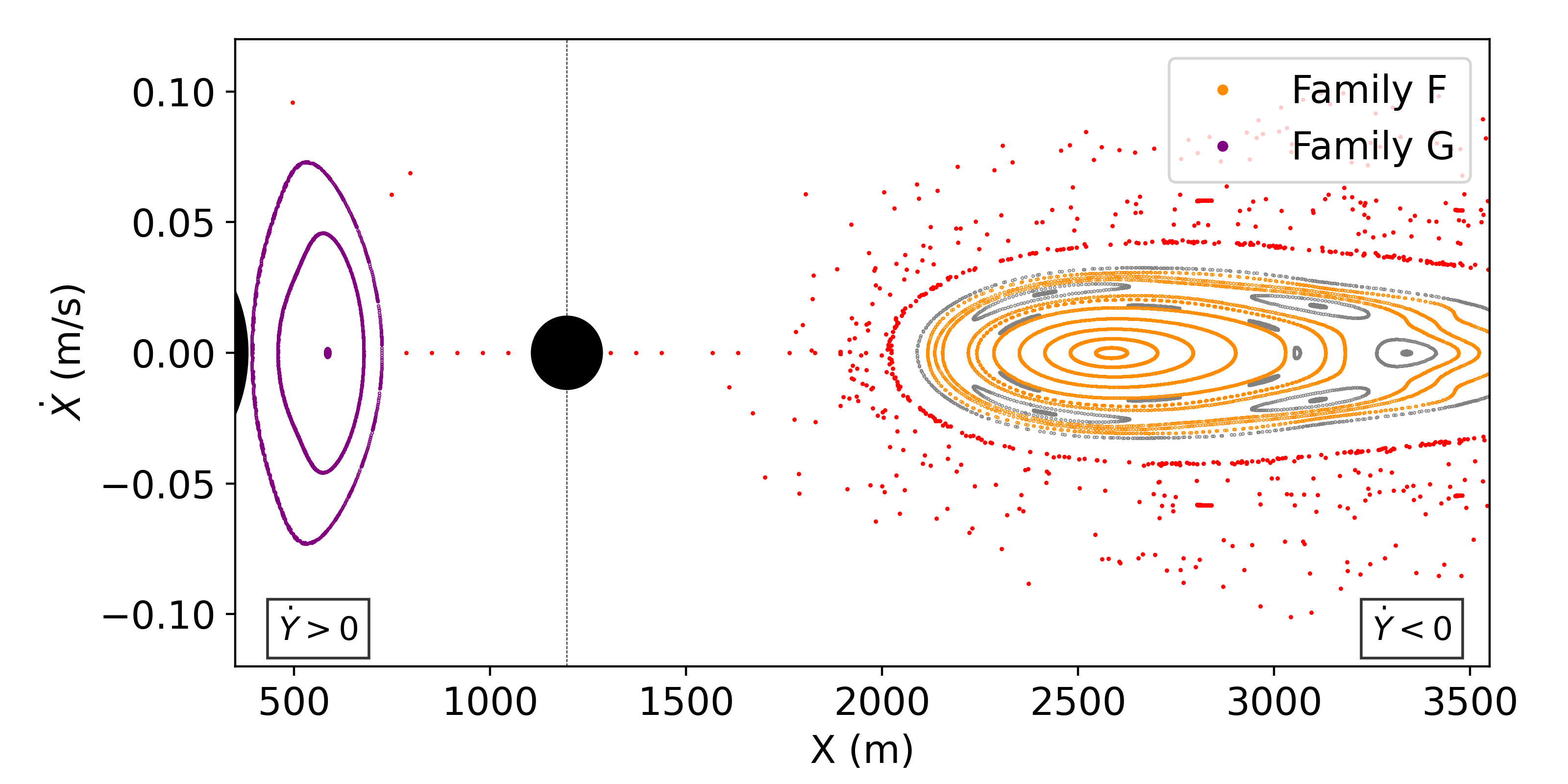}}
\subfloat[$C_J=3.80$]{\includegraphics[width=0.5\columnwidth,trim={0 0 0 0},clip]{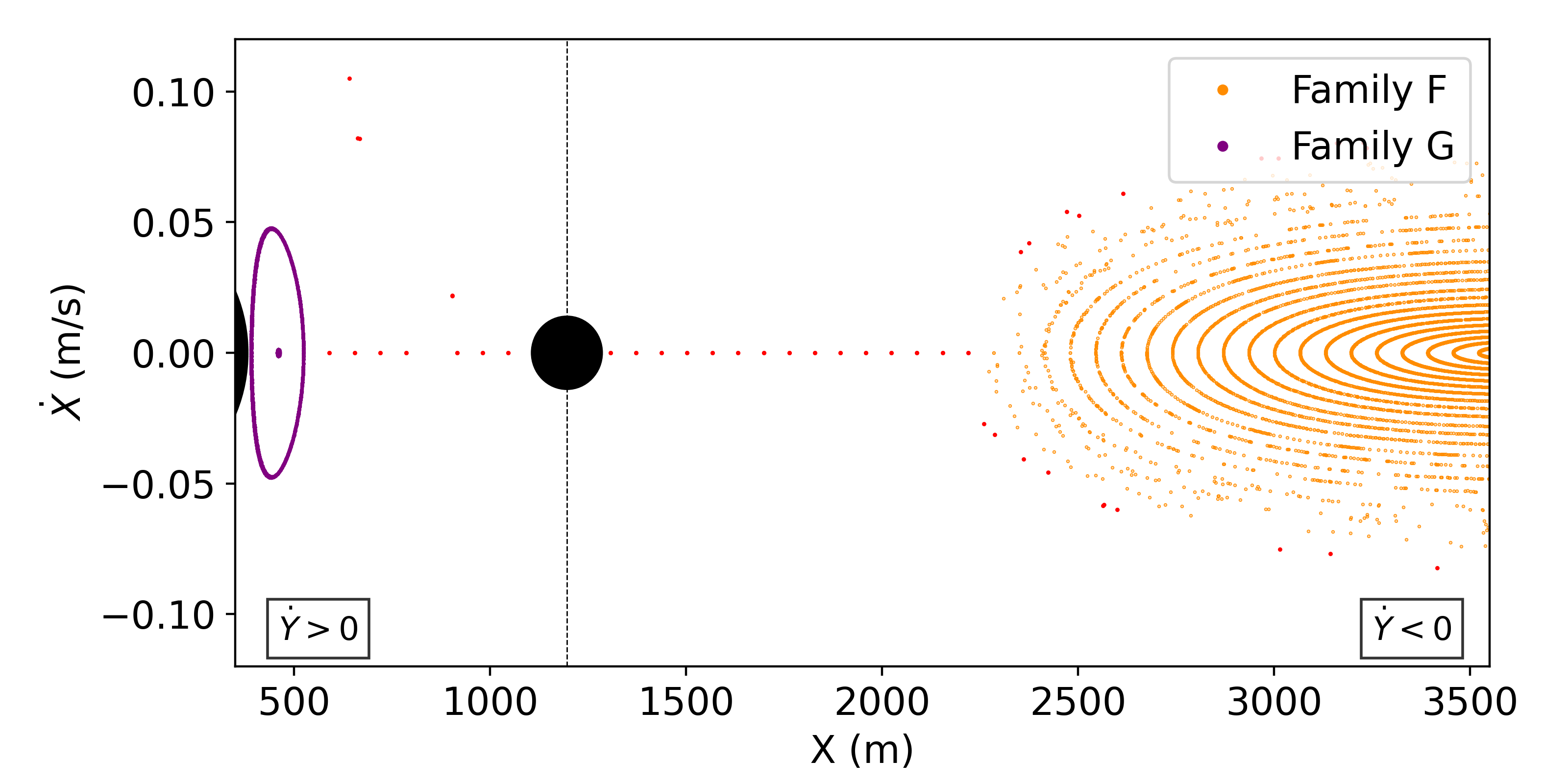}}\\
\caption{Poincaré maps when considering Didymos and Dimorphos as NSSBs. The Jacobi constant considered for each graph is provided in the caption of each panel. The red dots correspond to chaotic orbits, while the dots with different colors represent different families of stable orbits. For particles initially with $X<R_{12}$, we display only the points with $\dot{Y}>0$, while for $X>R_{12}$, we display the points with $\dot{Y}<0$.}
\label{ssp_sm}
\end{figure}

In this section, we re-investigate the dynamics in the vicinity of Didymos and Dimorphos, assuming the objects with the shapes given in Figure~\ref{shapes}. As a first step, we recalculate the location of the Lagrangian points, also given in Table~\ref{lagrange}. In a first approximation, we can say that the distances from $L_1$ and $L_2$ to Dimorphos are equal to the theoretical Hill radius of Dimorphos. However, due to the Didymos tidal effect, $L_1$ (between the asteroids) is, in fact, further from Dimorphos than $L_2$ (beyond Dimorphos), which can be seen even in the mass points case. When considering the shape of the objects, this pattern becomes even more evident, indicating that Didymos mass distribution causes an even more asymmetrical Hill region in Dimorphos. Due to the higher-order terms in the gravitational potential, additional equilibrium points may exist very close to the surface of Didymos and Dimorphos or within them \citep{Yu2019}. We will not consider such possible points in this paper.

\begin{figure}
\centering
\includegraphics[width=0.5\columnwidth,trim={0 0 0 0},clip]{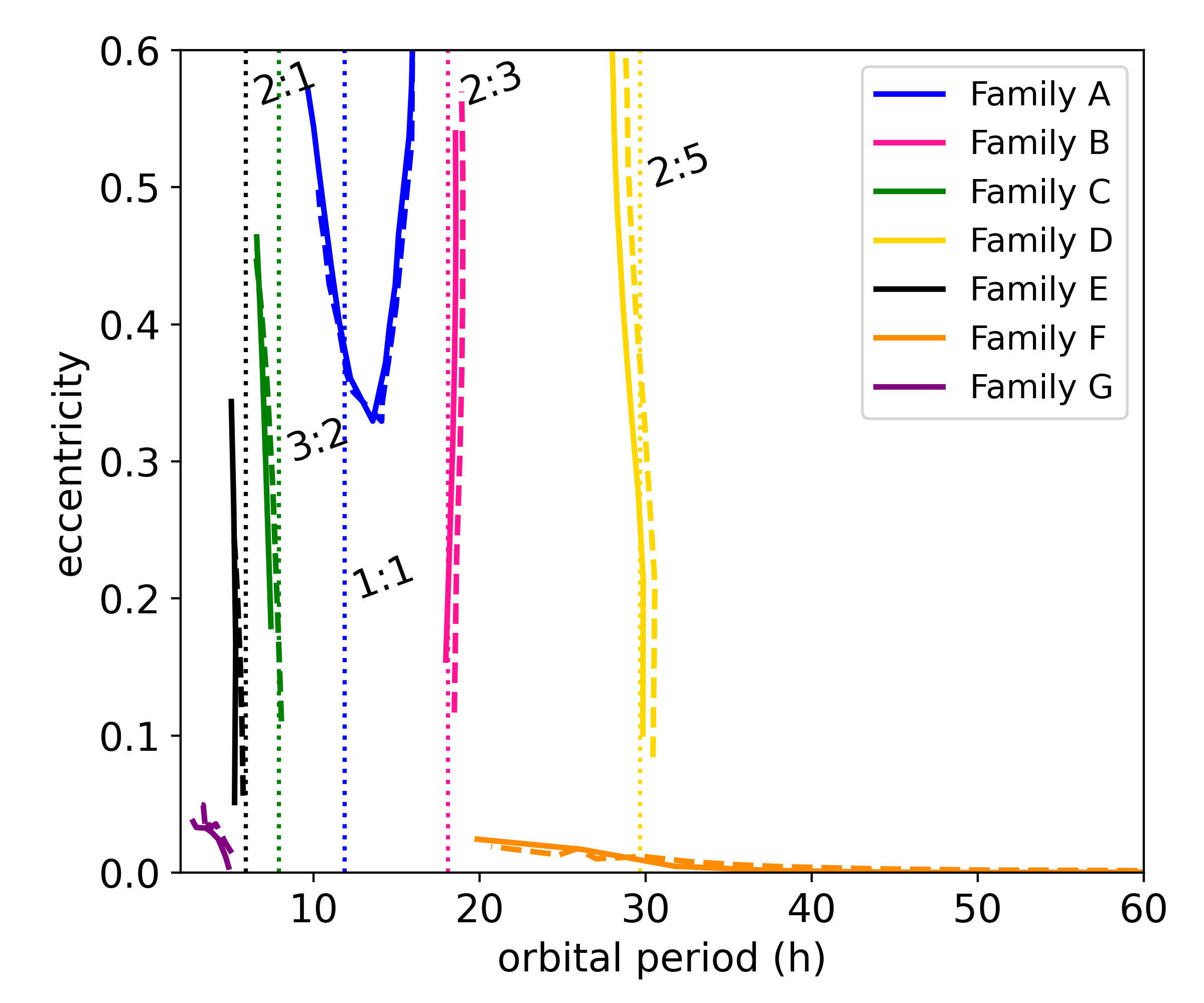}
\caption{Orbital period and eccentricity of the periodic orbits, for Didymos and Dimorphos as mass points (solid lines) and as NSSBs (dashed lines). The families of stable orbits are shown in different colors, while the colored dotted lines indicate the theoretical orbital period of MMRs with Dimorphos.}
\label{families_ae}
\end{figure}

Figure~\ref{ssp_sm} shows the Poincaré maps for Didymos and Dimorphos as NSSBs. Despite some visual differences, the dynamic obtained in Figures~\ref{ssp_pm} and \ref{ssp_sm} are similar, with the same families of stable orbits being obtained in both cases. To compare the results systematically, we extracted from the Poincaré maps, $X$, $\dot{X}$, and $C_J$ of the periodic orbits and used them to calculate the Didymos-centered orbital elements of the orbits.

It is well known that orbital elements of a particle orbiting a non-spherical primary will exhibit short-period variations, associated with an apsidal precession of the orbit \citep{Greenberg1981}. These variations can be reduced by using the geometrical elements, defined to account for the effect of the zonal terms of the primary. Having seen this, we expanded Didymos gravitational potential into spherical harmonics, obtaining the first zonal coefficients as $J_2=2.2\times 10^{-1}$, $J_4=-6.1\times 10^{-3}$, and $J_6=4.0\times 10^{-5}$. Then, the algorithm described in \cite{Renner2006} was used to obtain the orbital elements of the orbits (for the mass points case, we set $J_2=J_4=J_6=0$).

The orbital period and eccentricity of the periodic orbits found by us are shown in Figure~\ref{families_ae}. The different families are shown in different colors, with the solid lines corresponding to the case with Didymos and Dimorphos as mass points, and the dashed lines to the case with both as NSSBs. The dotted vertical lines provide the theoretical location of the MMRs $m$:$m-j$ with Dimorphos ($T=2\pi(m-j)/(m\omega)$). The good agreement between the solid and dashed lines obtained for all families shows that the Didymos shape has only a perturbative effect on the particles and does not act to destabilize the stable families. We have performed a set of test simulations with Didymos as mass point and Dimorphos as an ellipsoid, whereby we conclude that Dimorphos shape has a minor effect on particle evolution.

We draw the reader's attention to the distinction between families that are or are not related to MMRs in Figure~\ref{families_ae}. Resonant orbits cover narrow regions and have high eccentricities, while non-resonant orbits have low eccentricity and cover wider regions. It is interesting to note that Family A corresponds to an intermediate case, with high eccentricities and covering a wide region. We point out, however, that the satellite retrograde orbits of Family A orbit Dimorphos and not Didymos, and therefore the use of Didymos-centric coordinates is not appropriate for this case. In Dimorphos-centered coordinates, the eccentricity of these orbits is $\lesssim 10^{-2}$.

\subsection{Stability Map}

We now evaluate the overall stability in Didymos vicinity by integrating 60,000 particles regularly distributed on a grid of initial position $X$ and eccentricity $e$, with initial position ranging from the Didymos surface until $X=3500$~m and initial eccentricity from $0$ to $1$. In our simulations, the initial eccentricity is used as a parameter to obtain the initial velocity of the particle $\dot{Y}$, and it is assumed that the particles inside the Dimorphos orbit are initially at the pericenter, $\dot{Y}=\sqrt{GM_1(1+e)/X}-\omega X>0$, and those outside the Dimorphos orbit are at the apocenter, $\dot{Y}=\sqrt{GM_1(1-e)/X}-\omega X<0$. We used this approach to obtain a unified stability map that encompasses all the families previously identified in the Poincaré maps. If we started all the particles at their pericenter, we would not capture the stable regions associated with Families B and D in the stability map.

The timespan of the simulation is $10^4$ orbital periods of Dimorphos (13.5~years). This time interval was chosen based on the works of \cite{Holman1999,Winter2001}, which show that $10^4$ orbital period of the secondary is sufficient to determine whether a particle is in a stable orbit or not. A particle can collide with Didymos or Dimorphos upon reaching their surface or be considered ejected if its orbital radius is greater than the Hill radius of the system. In this section, we assume the Hill radius as the fixed value of $a_{\rm Hill}=111500$~m. In the following sections, for which solar gravity will be included in the system, we calculate the instantaneous Hill radius using the equation $a_{\rm Hill}=r_\odot((M_1+M2)/(3M_{\odot}))^{1/3}$, where $r_\odot$ is the distance between the center of the system and the Sun, and $M_{\odot}$ is the solar mass.

\begin{figure}
\centering
\includegraphics[width=0.8\columnwidth,trim={0 0 0 0},clip]{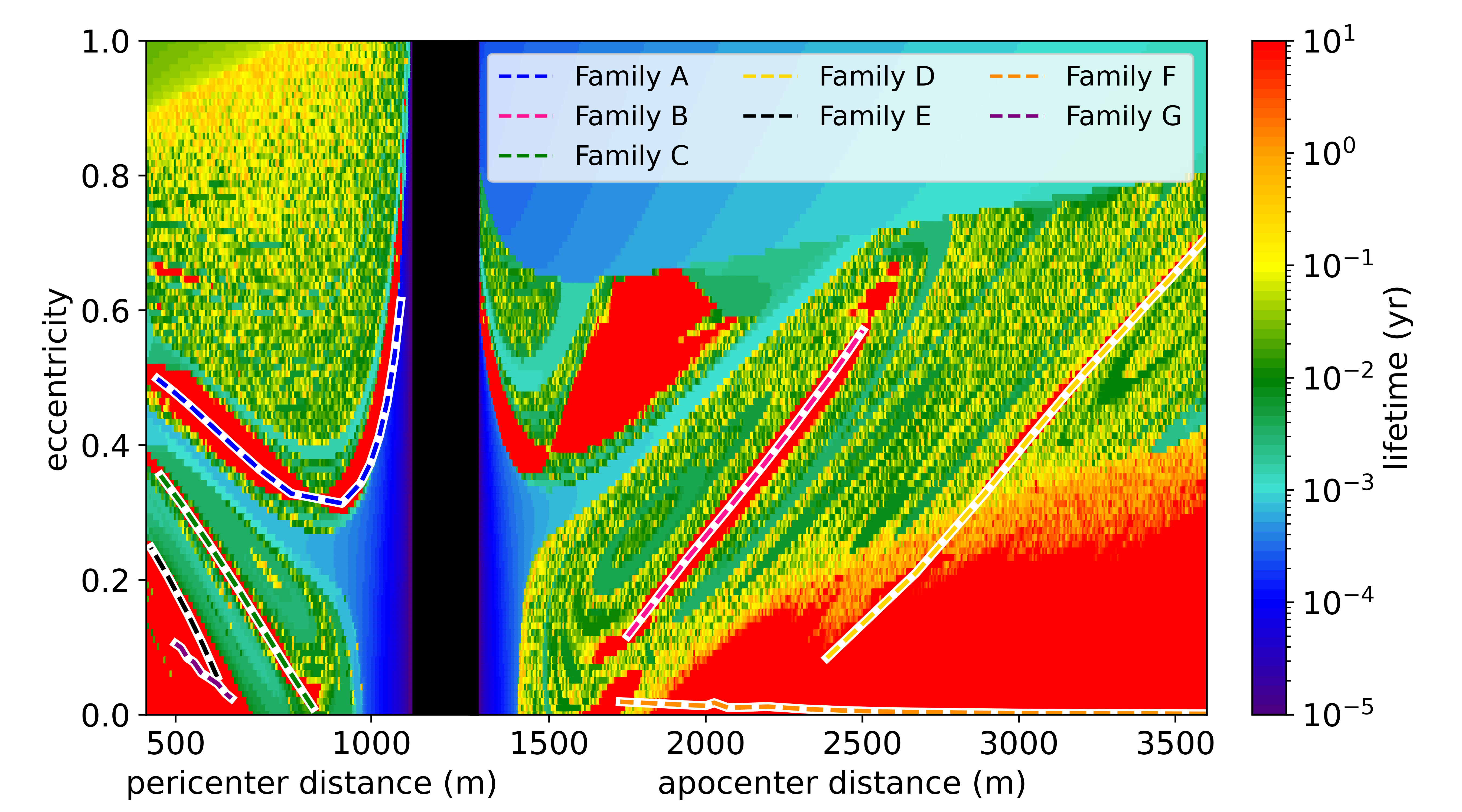}
\caption{Lifetime of particles in the vicinity of Didymos and Dimorphos. Each dot corresponds to a particle with initial distance and eccentricity given on $Ox$ and $Oy$ axis, respectively: particles inside the Dimorphos orbit are initially at the pericenter, while particles outside the Dimorphos orbit are initially at the apocenter. The color represents the lifetime of the particles. The dashed lines give the periodic orbits of the stability families, and the black region places the region inside Dimorphos}
\label{map_SM}
\end{figure}

In Figure~\ref{map_SM}, we show the lifetime of the particles through different colors, while the black region corresponds to the region inside Dimorphos. We also give the elements of the stable orbits shown in Figure~\ref{families_ae}. As can be seen, the stable regions obtained in the stability map (in red) correspond to stable families obtained using the Poincaré map. The particles in the red region to the right of Dimorphos with $0.4<e<0.7$ also belong to Family A. The cusps observed in this family correspond to the orbits of the particles initiated in the Hill sphere of Dimorphos.

It is found that 81\% of the simulated particles are in chaotic motion, while 19\% of the particles reside in stable regions. Chaotic particles initially follow the nearest stable family, but undergo continuous orbit deflections due to close encounters, leading to eventual collision or ejection. 29\% and 43\% of the set of particles are removed due to impacts with Didymos and Dimorphos, respectively. The average lifetime of the particles is of 12~hours (for collisions with Didymos) and 15~hours (for collisions with Dimorphos). Here, the average lifetime represents the time interval for 50\% of the particles to be removed. Ejected particles corresponds to 9\% of the particles, with an average lifetime of 8~days. 90\% of the chaotic particles are removed in $\sim0.23$~years (a few hundreds of Dimorphos orbits). Beyond $X=3500$~m, the particles are mainly in stable trajectories associated with Family F orbits.

\begin{figure}
\subfloat[]{\includegraphics[width=0.5\columnwidth,trim={0 0 0 0},clip]{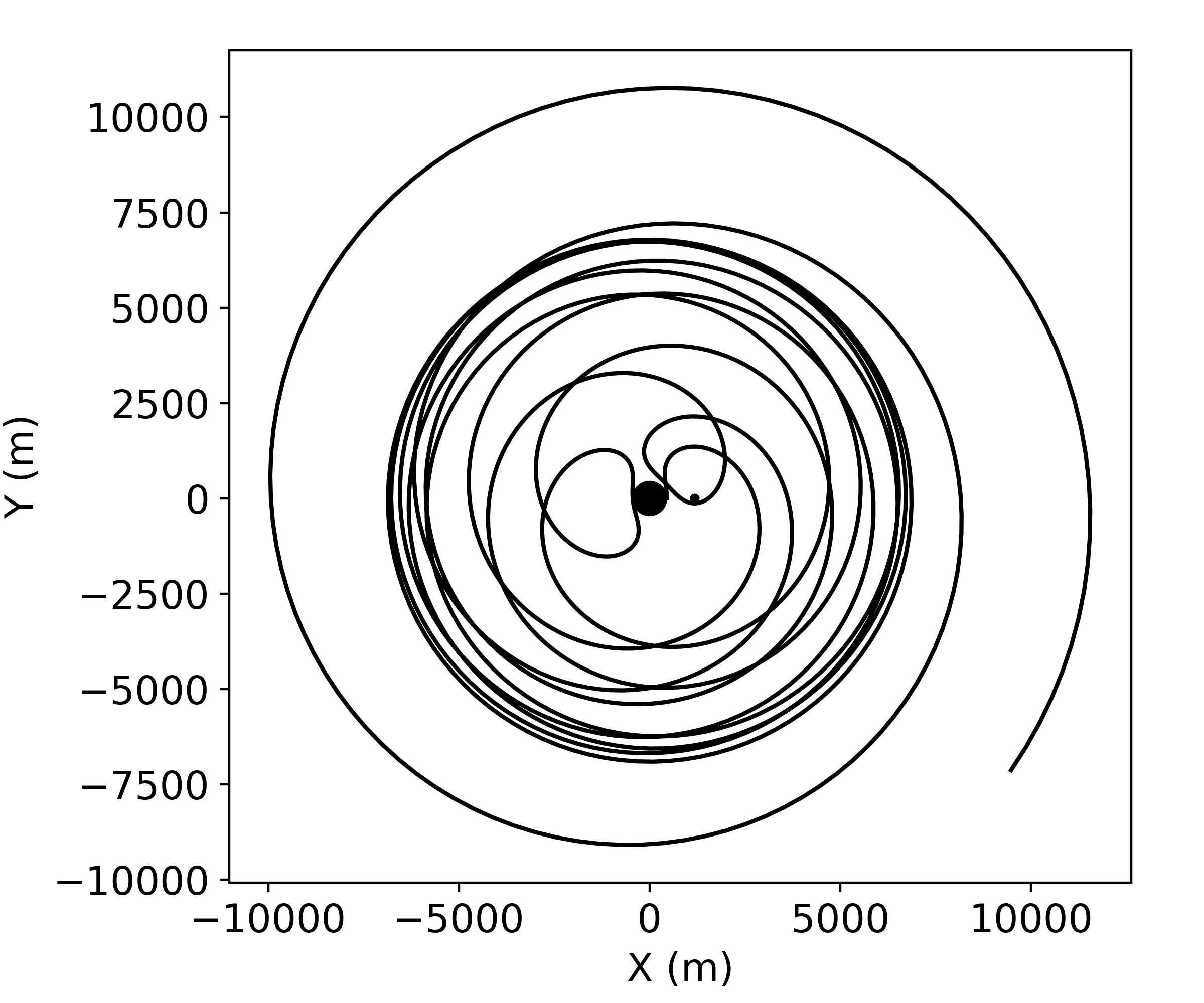}\label{exs_partsa}}
\subfloat[]{\includegraphics[width=0.5\columnwidth,trim={0 0 0 0},clip]{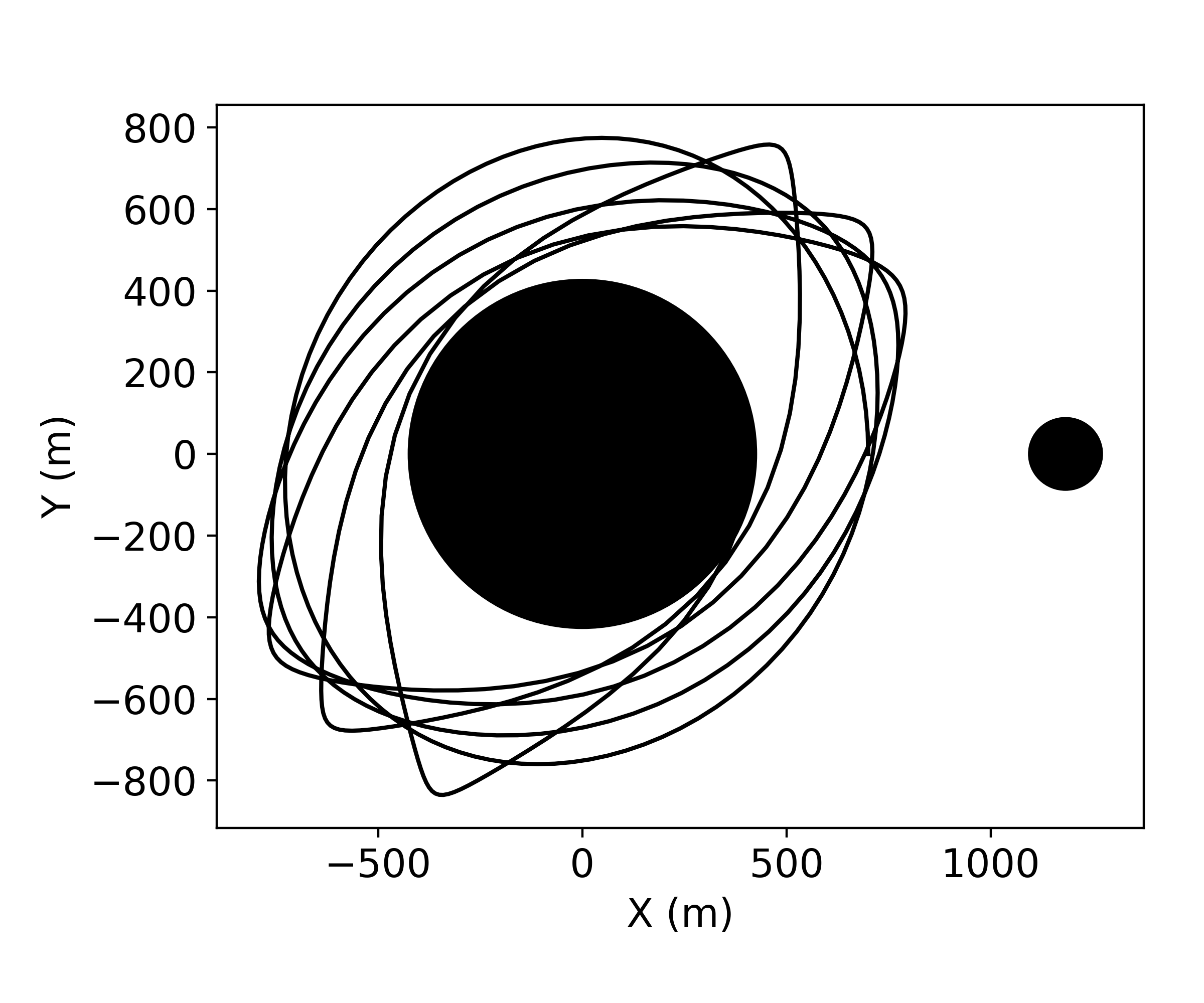}\label{exs_partsb}}
\caption{Particle trajectory initially with a) $q=450$~m and $e=0.75$ and b) $q=700$~m and $e=0$.}
\label{exs_parts}
\end{figure}
Some examples of chaotic trajectories are shown in Figure~\ref{exs_parts}. The particle in Figure~\ref{exs_partsa} initially follows Family A, is deflected into a trajectory of Family B, and then into a spiral orbit, being ejected in $\sim1$~year. The particle in Figure~\ref{exs_partsb}, on the other hand, is initially close to an orbit of the first sort (Family G) and is deflected due to a close encounter with a trajectory of Family E, in precession. The particle collides with Didymos in $\sim0.5$~days. More than half of the stable particles (56\%) belong to the Family F (i.e., they are in circumbinary orbits), which could mean that this is one of the families most likely to harbor material. However, it is important to note that this region is expected to be the most susceptible to solar tides. The effects of the solar tide are discussed in the following section.

\section{Stability of particles under the solar influence} \label{sec_srf}

\subsection{Solar gravity} \label{sec:gravity}

Didymos orbits the Sun in a highly eccentric orbit ($e=0.38$), with a radial distance ranging from 1.01 AU at its pericenter to 2.27 AU at the apocenter. The barycenter orbital period is about 770 days, which means that the system completes more than 5 revolutions around the Sun during the 13.5-year period we consider in our study. Therefore, the gravitational influence of the Sun must be considered.

We include the solar effect on the particles by adding to the equations of motion the gravity gradient between the center of our system (barycenter) and the particle. The force per mass $\vec{f}_\odot$ is given by \citep[e.g.][]{Ferrari2021trajectory}:
\begin{equation}
\vec{f}_\odot=G(M_\odot+M_1+M_2)\frac{\vec{r}_\odot}{|\vec{r}_\odot|^3}-GM_\odot\frac{\vec{r}_\odot-\vec{r}}{|\vec{r}_\odot-\vec{r}|^3}   
\end{equation}
where $\vec{r}$ is the position-vector of the particle, and $\vec{r}_\odot$ corresponds to the position-vector of the Sun. It is assumed that the binary is in a Keplerian orbit around the Sun, with orbital elements extracted from Horizons System\footnote{\url{https://ssd.jpl.nasa.gov/horizons/} [last accessed December 20, 2023].}. To determine the solar position-vector, we apply the frame transformations given by \cite{DellElce2017}, which allow us to transition from the Sun-centered inertial frame to the barycenter-centered rotation frame. Initially, we place the binary system at the apocenter of its heliocentric orbit. Although Didymos being in an heliocentric orbit with an inclination of $i=3.41^{\circ}$, our analysis will focus only on the equatorial motion of the particles. We refer the reader to \cite{Fodde2023} for a more detailed analysis of the inclined stable orbits in the system. 

Particles in periodic orbits near Didymos and Dimorphos are weakly affected by solar tides, remaining in closed periodic trajectories in the rotating frame. The gravitational effects of the Sun become significant only for periodic particles located at distances greater than $\sim$24 km (or 20 in normalized units). In this region, the particles, which are initially in circumbinary orbits (Family F), lose their periodic nature and exhibit erratic motion until they are eventually lost. In our region of interest ($X<3500$ m), we find that solar gravity has a negligible effect on the trajectory of all particles in low eccentricity orbits, a result was somewhat expected, considering our analysis focuses on the evolution of particles in the region close to Didymos, well inside the Hill sphere. These findings align with those of \cite{Ferrari2021trajectory}, which obtain that the solar effect becomes dominant only at distances of tens of kilometers from the barycenter of the Didymos and Dimorphos system.

For families associated with highly eccentric orbits (Families A and D), we observe that solar tides still have a negligible effect on the trajectory of the periodic particles. However, solar influence becomes more prevalent for quasi-periodic particles near the limits of the stable region (red regions in Figure~\ref{map_SM}). In these cases, particles also begin to exhibit erratic motion, being removed mainly by collisions with Dimorphos, but also by ejections, especially when the binary is near its perihelion (at perihelion, the instantaneous Hill sphere of the system is reduced to $\sim$58, in normalized units). Figure~\ref{mapsrff} shows the same 60,000 particles as Figure~\ref{map_SM}, including solar gravity. The fraction of particles in stable orbits is reduced from 19\% in the case without solar tides to 12\% in the case with Sun, with the 7\% of difference being ejections and collisions with Dimorphos. Of these, 5\% belong to Family A and 2\% to Family D. In the other families, the effect of the Sun is negligible. Next, we increase the complexity of our dynamical system by also considering the effects of the solar radiation force. 

\subsection{Solar radiation force} \label{sec:radiation}
The trajectory of the particles is also affected by the radiation emitted by the Sun: the photons of the radiation impact the particles, causing an effect called radiation pressure. At the same time, the particles absorb and re-emit solar radiation, giving rise to the Poynting-Robertson effect. Here, we consider both effects through the force per unit of mass $\vec{f}_{rf}$\footnote{Radiation pressure corresponds to the first term of Equation~\ref{eq_srf}, independent on velocity, while the Poynting-Robertson component corresponds to the terms that depend on velocities} \citep{Wyatt1950,Mignard1984,Yu2017}
\begin{equation}
\vec{f}_{rf}=\beta\frac{GM_\odot}{|\vec{r}-\vec{r}_\odot|^2}\left[\left(1+\frac{\vec{v}}{c}\cdot\frac{\vec{r}-\vec{r}_\odot}{|\vec{r}-\vec{r}_\odot|}\right)\frac{\vec{r}-\vec{r}_\odot}{|\vec{r}-\vec{r}_\odot|}+\frac{\vec{v}}{c} \right] \label{eq_srf}
\end{equation}  
where $\vec{v}$ is the particle velocity-vector relative to the Sun, $c$ is the speed of light and $\beta$ is the ratio between solar radiation pressure and solar gravity \citep{Burns1979}
\begin{equation}
\beta=\frac{3S_\odot AU^2}{4cGM_{\odot}\rho_pr_p},
\end{equation}
being $S_\odot=1.36\times 10^3~{\rm W/m^2}$ the solar radiation per unit area at 1 Astronomical Unit (AU), $\rho_p$ the particle bulk density, and $r_p$ the particle physical radius.

\begin{figure}
\centering
\subfloat[]{\includegraphics[width=0.45\columnwidth,trim={0 0 0 0},clip]{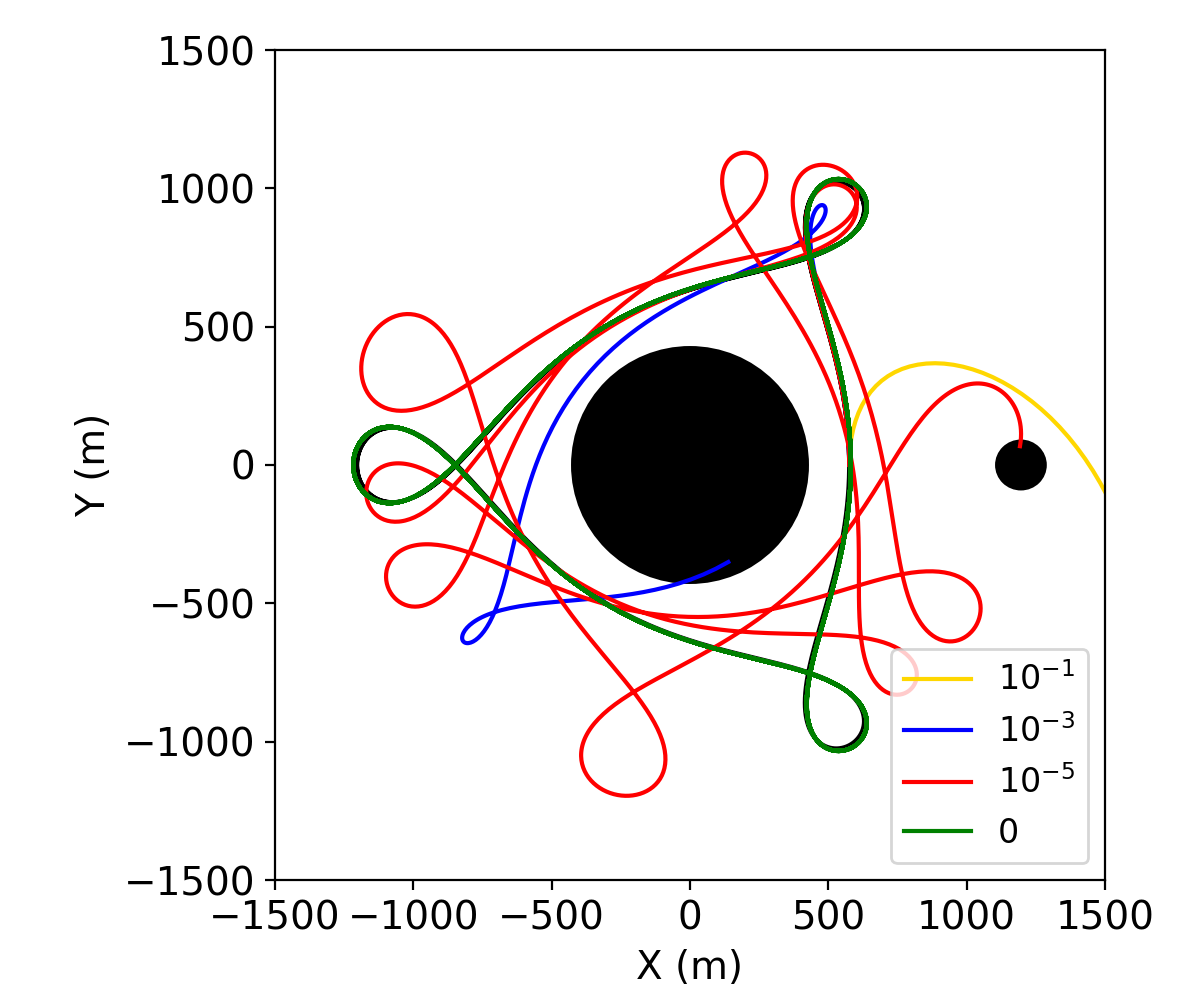}\label{orbitsrfa}}
\quad
\subfloat[]{\includegraphics[width=0.45\columnwidth,trim={0 0 0 0},clip]{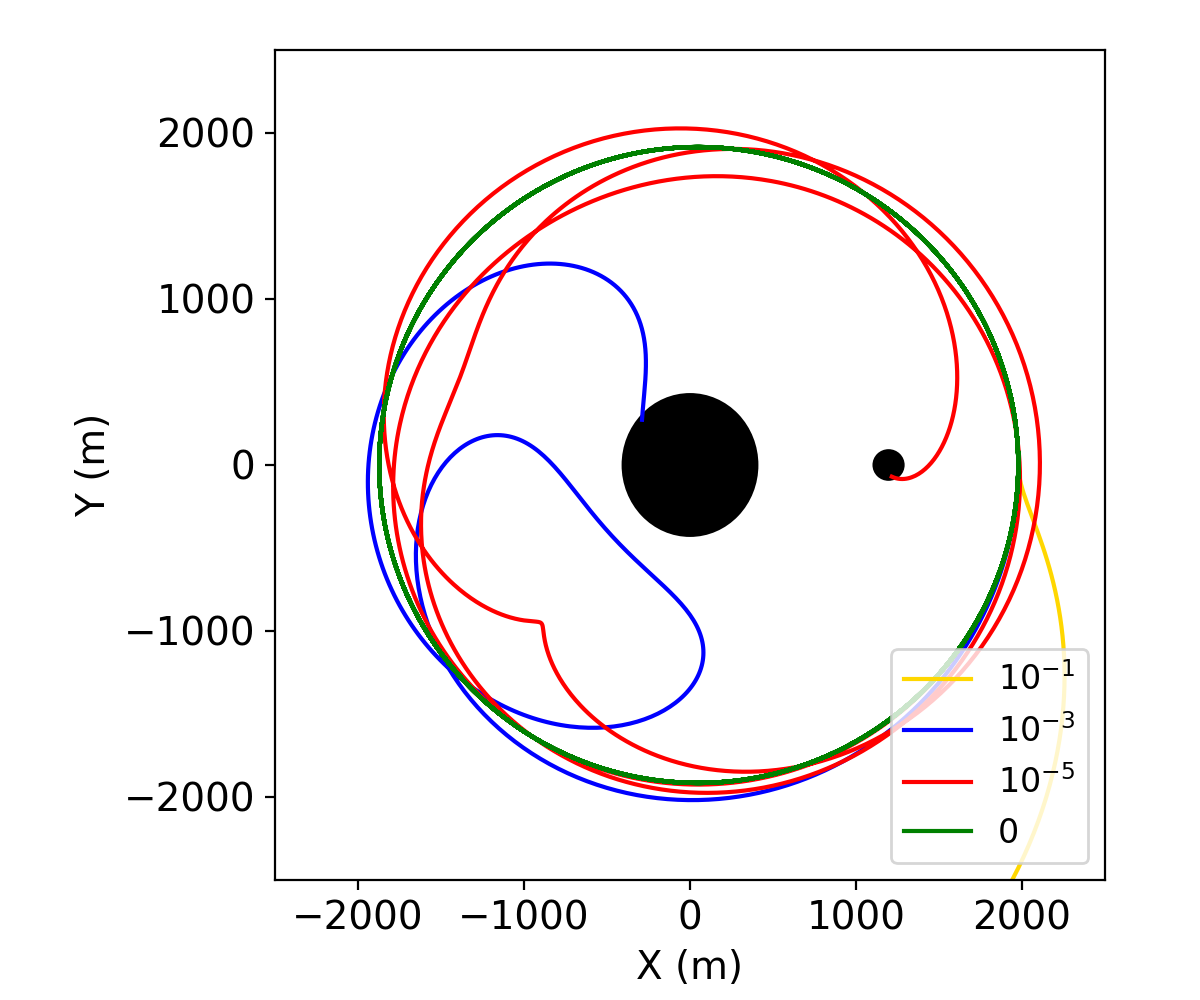}\label{orbitsrfb}}
\caption{Trajectory of representative particles for different values of $\beta$. The particles are initially at a) $a=0.654$ and $e=0.257$ and b) $a=1.600$ and $e=0.043$. $\beta=0$ corresponds to the case with solar gravity but without solar radiation force.}
\label{orbitsrf}
\end{figure}

To demonstrate the effects of SRF, we show in Figure~\ref{orbitsrf} the trajectory of representative particles of families C (Figure~\ref{orbitsrfa}) and F (Figure~\ref{orbitsrfb}), for different values of $\beta$. Under the presence of a weak non-conservative force, the periodic orbits starts to librates similarly to quasi-periodic orbits \citep{Ferraz2003,Beauge2006}, which can be seen in Figure~\ref{orbitsrf} for $\beta\lesssim10^{-5}$ (red lines in Figure~\ref{orbitsrf}). For higher values of $\beta$ ($\beta=10^{-3}$), we obtain that particle trajectories become erratic under the effect of the radiation force and the particles experience close encounters that lead to their losses (blue lines in Figure~\ref{orbitsrf}). For $\beta\gtrsim10^{-1}$, the eccentricity of the particles becomes greater than one, and the particles are blown out from the system (yellow lines in Figure~\ref{orbitsrf}).

We evaluate the effect of the radiation force on the system by reintegrating the particles from Figure~\ref{map_SM}, assuming $\beta=10^{-3}$, $10^{-4}$, $10^{-5}$, $10^{-6}$ and $10^{-7}$. These values correspond to spherical particles with radii of $0.17$~mm, $1.7$~mm, $17$~mm, $17$~cm, and $1.7$~m, respectively, assuming the same bulk density ($\rho_p=3400~{\rm kg/m^3}$) as \cite{Yu2019}. Our results are summarized in Table~\ref{tab_lifetime}, which shows the lifetime of the families for different values of $\beta$. We also present the lifetime of the particles outside the stable regions (Non-family case). Here, we consider the lifetime of a set of particles as the time for $90\%$ of the ensemble to be lost from the system.

\begin{figure}
\centering
\subfloat[$\beta=10^{-3}$]{\includegraphics[width=0.48\columnwidth,trim={0 0 0 0},clip]{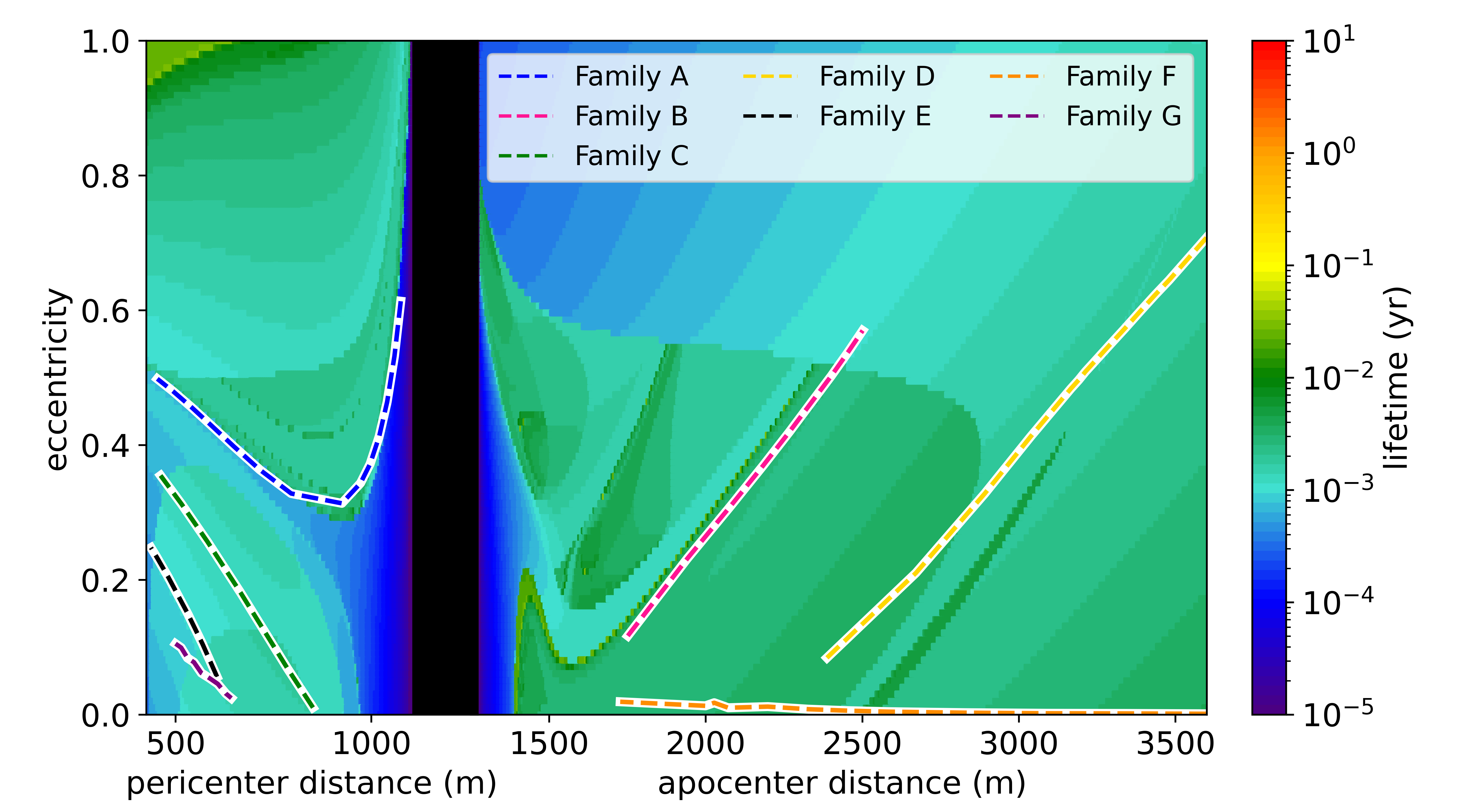}\label{mapsrfa}}
\subfloat[$\beta=10^{-4}$]{\includegraphics[width=0.48\columnwidth,trim={0 0 0 0},clip]{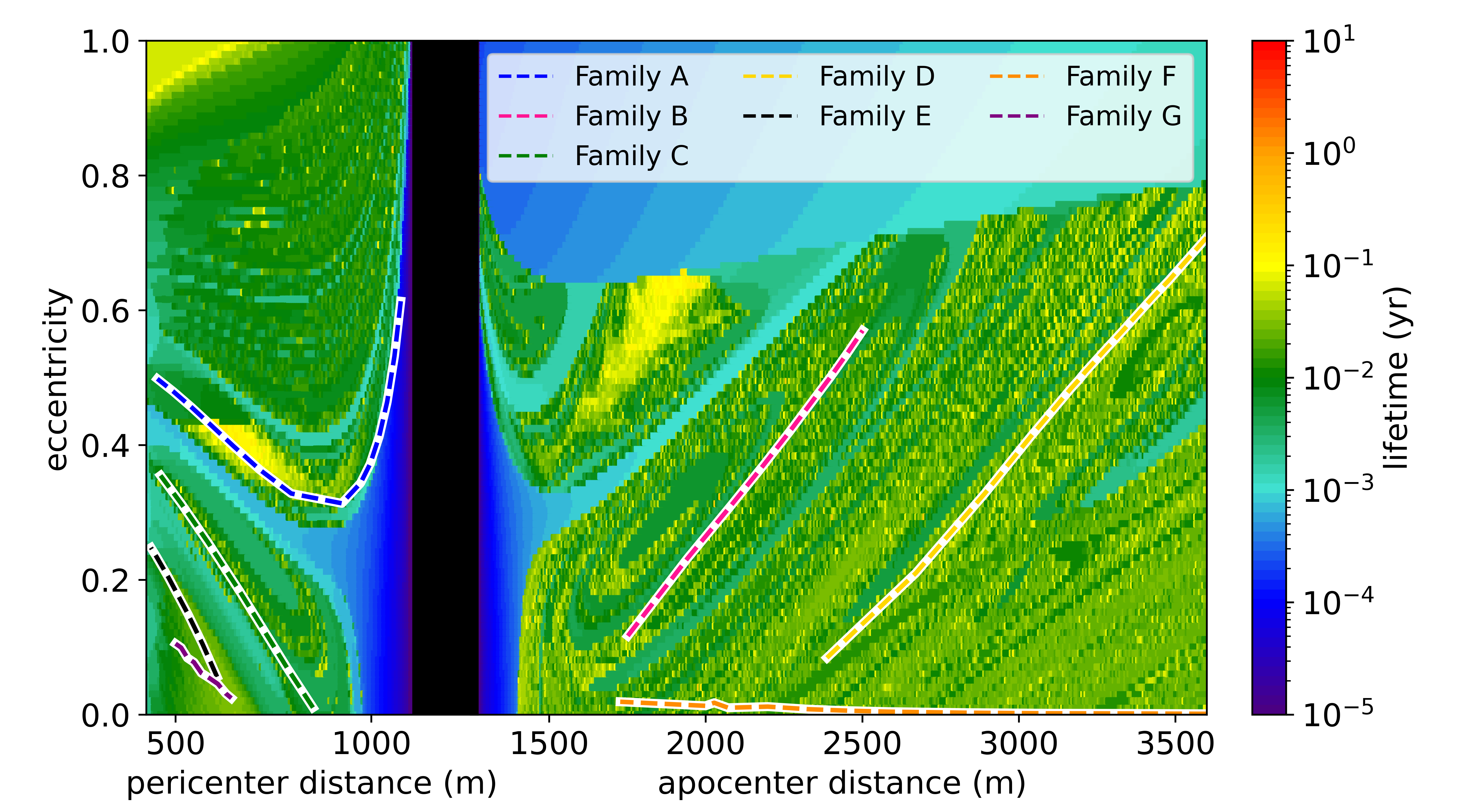}\label{mapsrfb}}\\
\centering
\subfloat[$\beta=10^{-5}$]{\includegraphics[width=0.48\columnwidth,trim={0 0 0 0},clip]{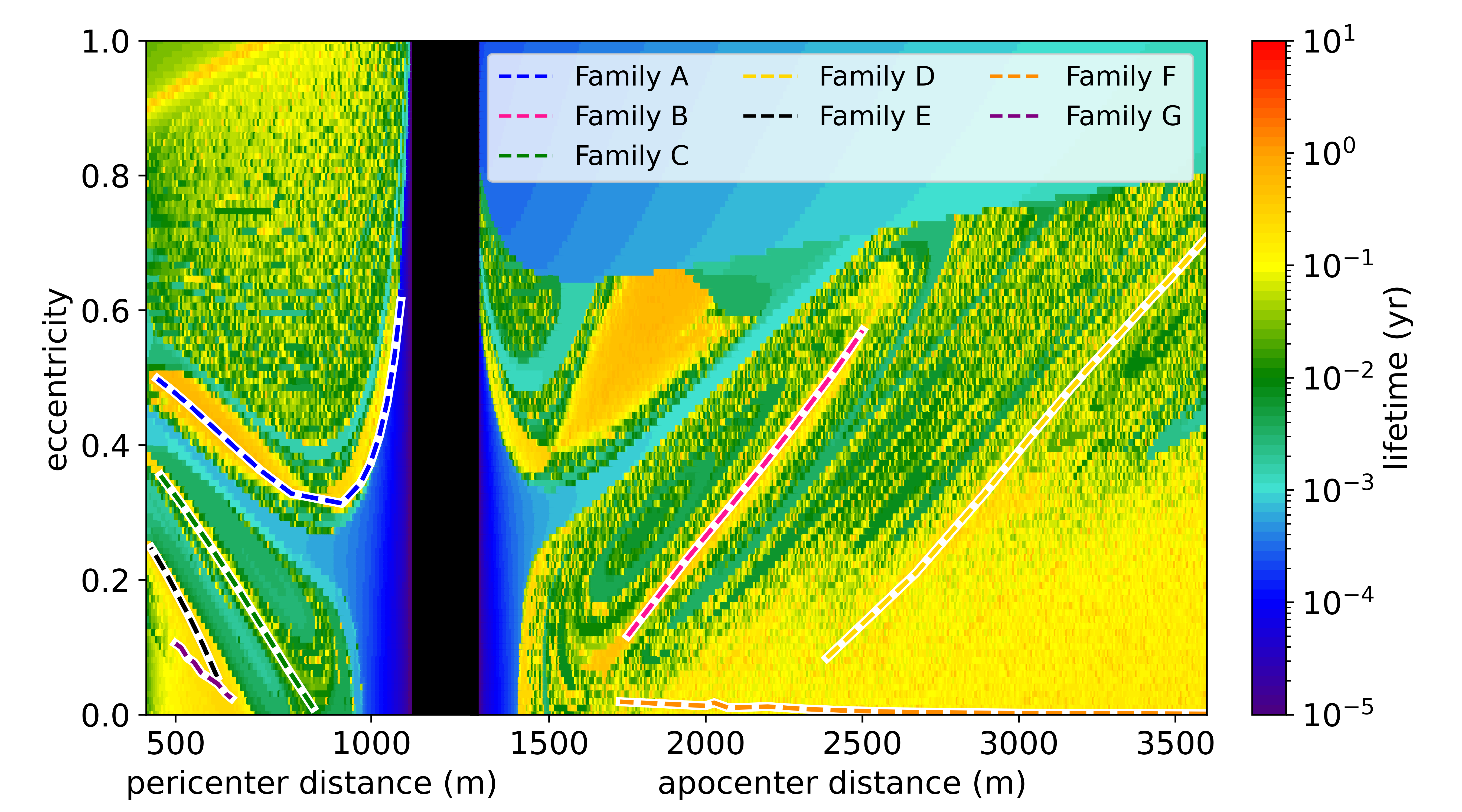}\label{mapsrfc}}
\subfloat[$\beta=10^{-6}$]{\includegraphics[width=0.48\columnwidth,trim={0 0 0 0},clip]{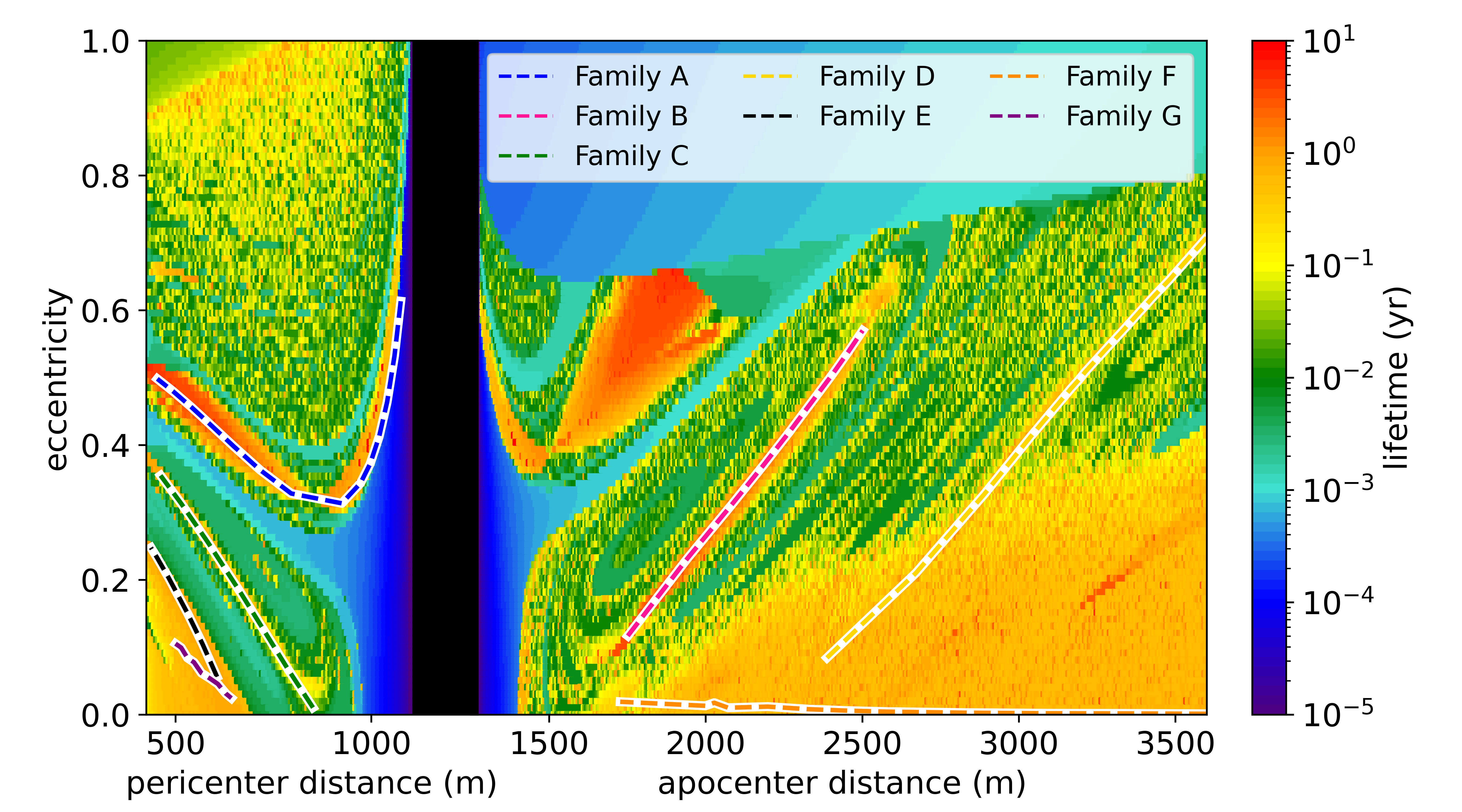}\label{mapsrfd}}\\
\centering
\subfloat[$\beta=10^{-7}$]{\includegraphics[width=0.48\columnwidth,trim={0 0 0 0},clip]{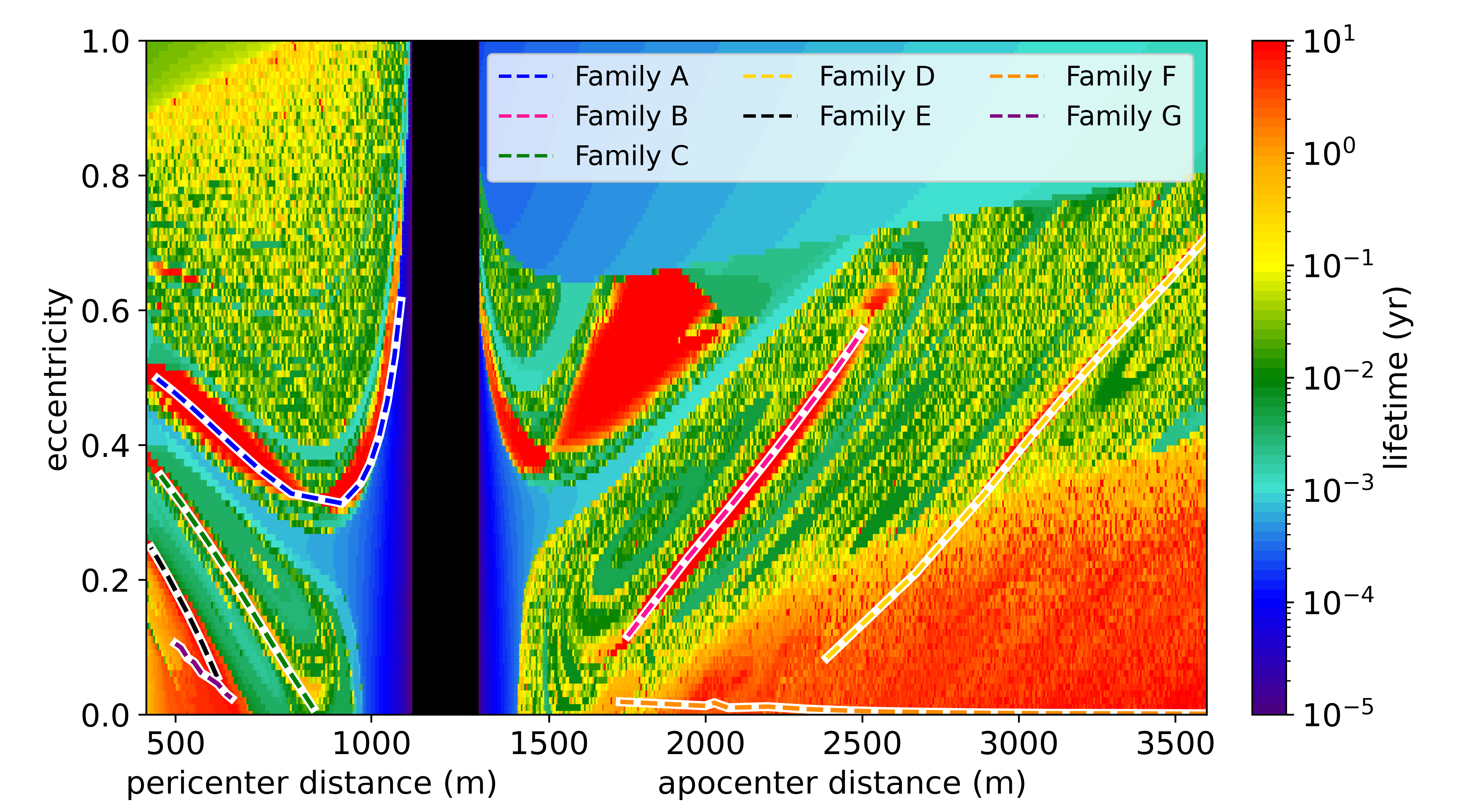}\label{mapsrfe}}
\subfloat[$\beta=0.0$]{\includegraphics[width=0.48\columnwidth,trim={0 0 0 0},clip]{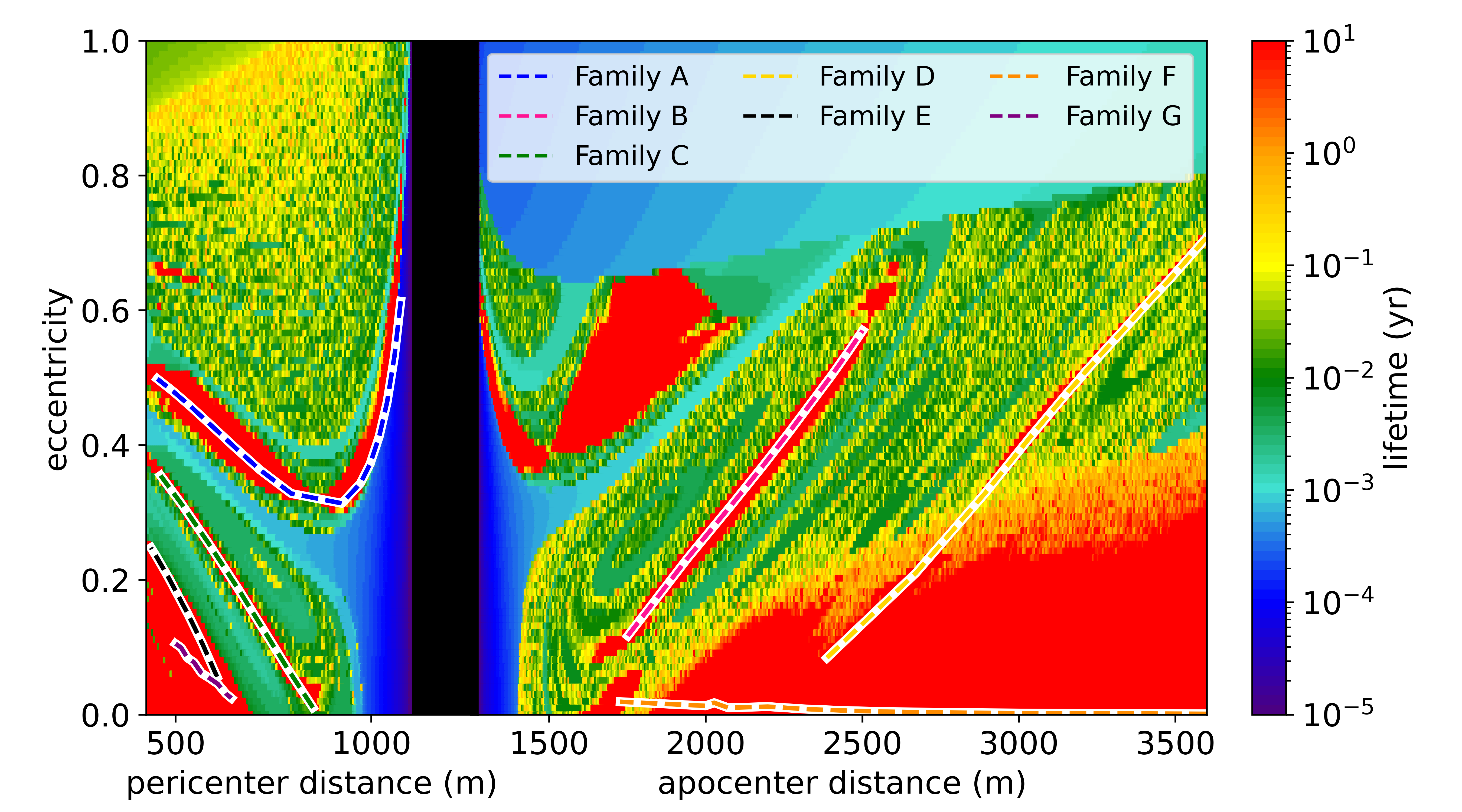}\label{mapsrff}}
\caption{Lifetimes of particles in the vicinity of Didymos and Dimorphos, under the solar gravity and radiation effects, with value of $\beta$ provided in the label of each panel. Each point corresponds to a particle with a different initial distance (pericenter for $X<R_{12}$ and apocenter for $X>R_{12}$) and eccentricity. The dashed lines give the periodic orbits of the stable families, and the black region places the region within Dimorphos.}
\label{mapsrf}
\end{figure}

\begin{table}{}
\caption{Lifetimes of the stability families, for different values of $\beta$. The last line gives the lifetime of the particles outside the families. See the text for the definition of lifetime.}
\label{tab_lifetime}
\centering
\begin{tabular}{ccccccc}
\hline\hline
 $\beta$ &  $10^{-3}$ & $10^{-4}$ & $10^{-5}$ & $10^{-6}$ & $10^{-7}$ & $0.0$ \\ \hline
Family A  &  0.003~yr & 0.092~yr & 0.546~yr & 3.186~yr & $>13.5$~yr & $>13.5$~yr \\
Family B & 0.004~yr & 0.037~yr & 0.417~yr & 2.721~yr & $>13.5$~yr & $>13.5$~yr \\
Family C &  0.002~yr & 0.019~yr & 0.171~yr & 0.640~yr & 3.859~yr & $>13.5$~yr \\
Family D  &  0.004~yr & 0.039~yr & 0.269~yr & 0.951~yr & 7.534~yr & $>13.5$~yr \\
Family E  &  0.001~yr & 0.028~yr & 0.277~yr & 0.796~yr & 6.877~yr & $>13.5$~yr \\
Family F  &  0.004~yr & 0.034~yr & 0.230~yr & 1.009~yr & 7.362~yr  & $>13.5$~yr \\
Family G  &  0.001~yr & 0.029~yr & 0.256~yr & 0.893~yr & 6.699~yr & $>13.5$~yr \\
Non-family & 0.003~yr & 0.027~yr & 0.091~yr & 0.193~yr & 0.229~yr & 0.229~yr \\
\hline
\end{tabular}
\end{table}

For the highest value of $\beta$, associated with sub-millimeter-sized particles, we find that the SRF is responsible for removing particles from stable motion early in the simulations (Figure~\ref{mapsrfa}). All particles have similar lifetimes of tens of hours, regardless of whether they are initially in a stable family or not. By decreasing $\beta$ by one order (Figure~\ref{mapsrfb}), we obtain a longer lifetime of particles in Family A, the most stable family in the system. For even smaller values of $\beta$ (larger particles), the effects of the SRF become weaker, and a clear distinction is seen between the lifetimes of particles initially inside or outside the stable regions. For centimeter-sized particles and larger ($\beta\geq10^{-6}$), particles in stable families last for more than a few months in the system, with a fraction of metric particles surviving for more than 10~years.

In general, families outside the orbit of Dimorphos live longer than families inside its orbit. This is the case for families B, D, and F, which are mostly removed from the system due to impacts with Dimorphos. Of these families, Family F is the only one that is not associated with an MMR and has the longest lifetime. This is because particles further away from Dimorphos naturally live longer, which increases the overall lifetime of the family. The most stable family is certainly Family A, which covers a wide region and is associated with highly eccentric orbits. For this reason, particles remain in this family even under the SRF and are removed only after several close encounters with Dimorphos. We note that part of the orbits of the other families are catapulted into Family A trajectories because of the SRF.

In summary, our results show that the SRF is responsible for breaking the stability of small particles, removing particles with millimeter radii and smaller within a few weeks. Only particles with radii of meters can reside in the vicinity of Didymos in stable orbits for years. For this size range, the SRF does not dominate the particle dynamics, but acts as a perturbation \citep[also see ][]{Ferrari2021trajectory,Raffa2023}. Seen this, it is possible that material formed due to the impacts of IDPs could reside, at least for while, in the vicinity of Didymos. Such a possibility is analysed in the next section.

\section{Production and ejection of material by Didymos and Dimorphos} \label{sec_ejected}

\subsection{Production of material due to IDP impacts }

The interplanetary environment is populated by dust originating from different objects, such as asteroids, Jupiter family comets, and Halley-type comets \citep{Divine1993}. Studies based on ground observations and spacecraft data show that the NEAs region is dominated by dust originated from Jupiter family comets \citep{Levison1997,Nesvorny2010,Nesvorny2011a,Nesvorny2011b,Pokorny2014}, and this material is responsible for impacting and eroding airless bodies, including Didymos and Dimorphos. 

The mass production rate due to impacts of IDPs in Didymos and Dimorphos is given by \citep{Krivov2003,Szalay2019}
\begin{equation}
{\rm M^{+}_i}=\pi\overline{R}_i^2\mathcal{F}Y_i   
\end{equation}
where $\overline{R}_i$ is the average radius of the object $i$, $\mathcal{F}$ is the flux of IDPs in the region, and $Y_i$ is the ejecta yield that depends on the physics of the impact and the composition of impactor and target. The ejecta yield is defined as the ratio of the total ejecta mass to the mass of impactors. Since spectroscopic data indicate a large fraction of silicates in Didymos \citep{deLeon2010,Pravec2022}, we compute the yield of Didymos and Dimorphos using the empirical prescription obtained by \cite{Koschny2001} for pure silicate objects:
\begin{equation}
Y_i=1.19\times 10^{-6}\left(\frac{m_{\rm imp}}{\rm kg}\right)^{0.23}\left(\frac{v_{\rm imp}}{\rm m/s}\right)^{2.46} \label{yield}
\end{equation}
where $v_{\rm imp}$ is the impactor velocity and $m_{\rm imp}$ its mass, assumed as $10^{-8}$~kg \citep{Poppe2016}.

Also as an approximation, we assume the same average $\mathcal{F}$ and $v_{\rm imp}$ for an object located at 1 AU: $\mathcal{F}=5\times 10^{-16}~{\rm kg\cdot m^{-2}s^{-1}}$ \citep{Plane2018} and $v_{\rm imp}=10$~km/s \citep{Janches2021}. We do not consider the effect of gravitational focusing onto the flux and velocity due to the small mass of Didymos and Dimorphos \citep[see][]{Szalay2019}. 

As result, we estimate the mass production rate by Didymos and Dimorphos, respectively, as:
\begin{equation}
{\rm M^+_{Did}}=8.6\times 10^{-1}~{\rm kg/year} \label{mplusdd}
\end{equation}
and
\begin{equation}
{\rm M^+_{Dim}}=3.4\times 10^{-2}~{\rm kg/year} \label{mplusdm}
\end{equation}

We take the reader's attention to the fact that ${\rm M^+_{Did}}$ and ${\rm M^+_{Dim}}$ should be regarded as rough estimates due to the uncertainties concerning the quantities $\mathcal{F}$, $m_{\rm imp}$, $v_{\rm imp}$, and specially $Y$. Equation~\ref{yield} is obtained for pure-silicate objects, while Didymos and Dimorphos, which are classified as S-type objects \citep{Lin2023}, might contain other species in their composition. Just as a comparison, the yield of a Didymos made of ice would be more than 6 times that given by Equation~\ref{yield} \citep{Koschny2001}, demonstrating that the yield can be affected up to an order of magnitude depending on the chemical composition of the objects. Didymos and Dimorphos are expected to have high porosity \citep[$>10\%$,]{Pajola2022}, which can also affect the yield. Another detail concerns Didymos' rotation. As the asteroid rotates near its critical rotation, it is imagined that it should be easier to remove material from its surface \citep{Yu2018,Yu2019,Trogolo2023}, and therefore it is possible that Didymos is more efficient at producing material.

\subsection{Fate of ejected material}

Having determined the mass production in the system, we turned our attention to the fate of this material, conducting a set of numerical simulations with particles ejected from the equator of Didymos and Dimorphos. For each body, we ran a simulation with a fixed $\beta$ (ranging from $\beta=10^{-7}$ to $10^{-3}$) and 200,000 particles ejected from the surface of the object. The initial conditions of the particles for each body are described below.

\subsubsection{Ejection from Didymos}
As initial conditions, we assume that the particles are ejected from Didymos' equatorial surface, with position-vector $\vec{r}_{\rm X_1Y_1}$ in the reference frame $X_1Y_1$:
\begin{equation}
\vec{r}_{\rm X_1Y_1}=a_1r_0\cos\Phi\hat{X}_1+a_1r_0\sin\Phi\hat{Y}_1
\end{equation}
where $\Phi$ is randomly and uniformly selected between $0^\circ$ and $360^\circ$.

In relation to the ejection velocity, it is composed of two distinct components: the tangential velocity $v_{\rm spin}$ and the velocity $v_{\rm ejec}$. The tangential velocity corresponds to the rotational speed at Didymos equator, with the asteroid rotating with a period of $T_{\rm spin}=2.26$~hours \citep{Naidu2020}. The $v_{\rm ejec}$ velocity, on the other hand, represents the velocity imparted to particles as a result of the impact in the reference frame that rotates with Didymos. The combined velocity-vector $\vec{v}_{\rm X_1Y_1}$ in the reference frame $X_1Y_1$ can be described as follows:
\begin{equation}
\vec{v}_{\rm X_1Y_1}=(v_{\rm ejec}\cos(\Phi+\Theta)-v_{\rm spin}\sin\Phi)\hat{X}_1+(v_{\rm ejec}\sin(\Phi+\Theta)+v_{\rm spin}\cos\Phi)\hat{Y}_1,
\end{equation}
where the tangential velocity $v_{\rm spin}$ is:
\begin{equation}
v_{\rm spin}=\left(\frac{2\pi}{T_{\rm spin}}-\omega\right)a_1r_0,
\end{equation}
while $\Theta$ is randomly and uniformly selected between $-90^\circ$ and $90^\circ$, and $v_{\rm ejec}$ is randomly distributed in the interval $v_{\rm min}$ to $0.45$ m/s, following the classical function for the distribution of ejecta velocities \citep{Krivov2003}:
\begin{equation}
f(v_{\rm ejec})=\frac{1.2}{v_{\rm min}}\left(\frac{v_{\rm ejec}}{v_{\rm min}}\right)^{-2.2}{\rm \mathcal{H}(v_{\rm ejec}-v_{\rm min})}. \label{distribution}
\end{equation}
Here, ${\rm \mathcal{H}(v_{\rm ejec}-v_{\rm min})}$ is the Heaviside function, and $v_{\rm min} = 0.04$~m/s is the minimum velocity required for the launched particles to to be placed in orbit instead of being directly recreated by the source object. This value was determined through test simulations of particles ejected from Didymos at different velocities. The upper limit of $v_{\rm ejec}$ (0.45 m/s) corresponds to the escape velocity of the Didymos-Dimorphos system. For higher velocities, the particles are expected to acquire a hyperbolic orbit.

\subsubsection{Ejection from Dimorphos}

Since Dimorphos is in a spin-locked orbit, it doesn't rotate in the rotating frame, implying that $v_{\rm spin} = 0$. Therefore, the positions $\vec{r}_{\rm X_2Y_2}$ and velocities $\vec{v}_{\rm X_2Y_2}$ of the particles in the reference frame $X_2Y_2$ are given by:
\begin{equation}
\vec{r}_{\rm X_2Y_2}=a_2r_0\cos\Phi\hat{X}_2+b_2r_0\sin\Phi\hat{Y}_2
\end{equation}
and
\begin{equation}
\vec{v}_{\rm X_2Y_2}=v_{\rm ejec}\cos(\Phi+\Theta)\hat{X}_2+v_{\rm ejec}\sin(\Phi+\Theta)\hat{Y}_2
\end{equation}
where $\Phi$ is randomly and uniformly given in the interval $0^{\circ}$ to $360^{\circ}$, $\Theta$ is randomly and uniformly selected between $-90^\circ$ and $90^\circ$, and $v_{\rm ejec}$ is randomly distributed following the distribution given by Eq.~\ref{distribution}, in the interval $v_{\rm min}=0.05$~m/s to 0.45~m/s. The minimum velocity was found using the same methodology as for Didymos.

\subsubsection{Results}
\begin{figure}
\centering
\subfloat[]{\includegraphics[width=0.48\columnwidth,trim={0 0 0 0},clip]{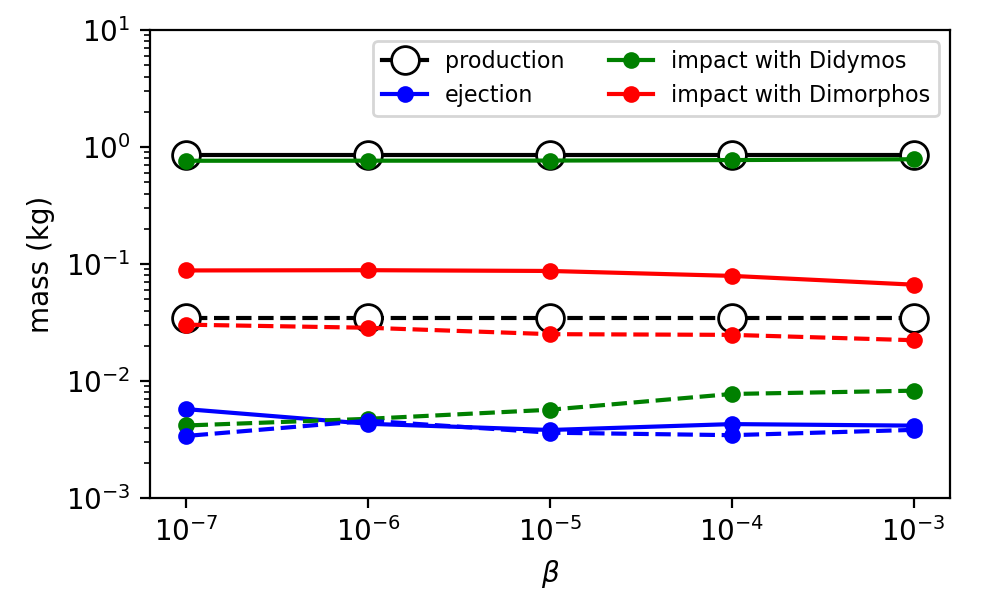}\label{productiona}}
\quad
\subfloat[]{\includegraphics[width=0.48\columnwidth,trim={0 0 0 0},clip]{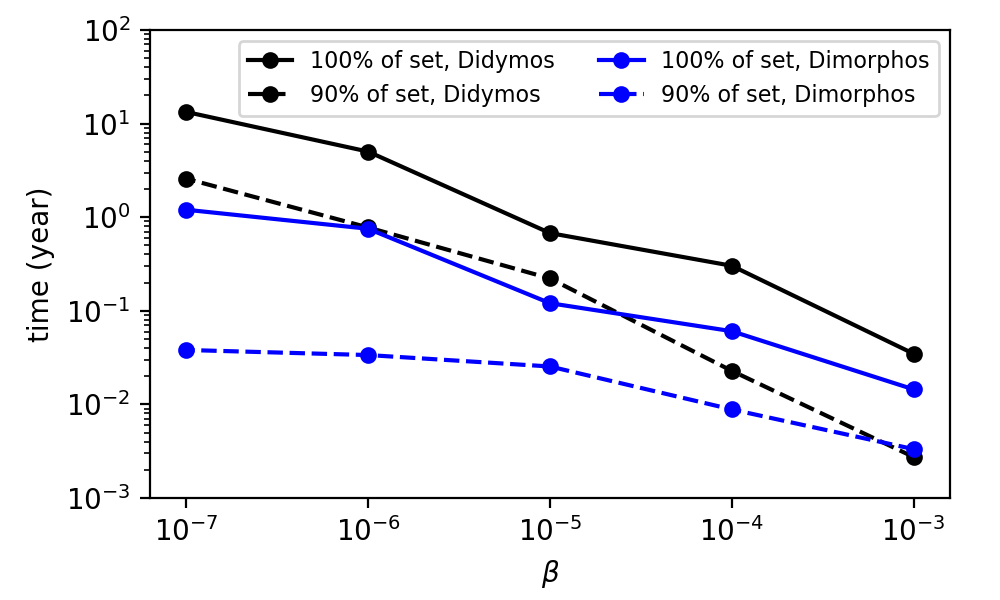}\label{productionb}}
\caption{a) Amount of mass launched from the asteroids in one year (black lines) and lost by impacts or ejection (colored lines), for Didymos  (solid lines) and Dimorphos (dashed lines). b) lifetime of the particles. The time for $90\%$ and $100\%$ of the particle set to be removed is shown by dashed and solid lines, respectively.}
\label{productiondd}
\end{figure}
The figure~\ref{productiona} shows the amount of mass released by the objects in one year (black line) and the amount of mass lost by collision with Didymos (green line), Dimorphos (red line), and ejection (blue line). Figure~\ref{productionb} shows the lifetime of the released material, as a function of $\beta$, with the dashed curves corresponding to the time required for 90\% of the particles to be lost from the system, while the solid curves give the time for the entire assembly to be removed. The black and blue lines correspond to the case with Didymos and Dimorphos as source objects, respectively.

The majority of particles ejected from Didymos are ejected into orbits inside Dimorphos' orbit, following trajectories associated with Families E and G. Meanwhile, a smaller fraction of particles is ejected into more distant orbits, in circumbinary trajectories. Although particles in circumbinary trajectories can persist for a fraction of years in the system, particles associated with Family G exhibit the longest lifetimes. The longest-lived particles are centimeter-sized or larger, located in the quasi-periodic orbits of the Family G (first sort orbits around Didymos). In the case of Dimorphos, particles are predominantly ejected into orbits associated with Families G and F. More than 90\% of the particles ejected from Dimorphos are removed within a few tens of days, and none survive for more than two years (Fig.~\ref{productionb}).

Most of the material ejected from an object is always re-accreted by the same object, and more than 88\% of the material ejected from Didymos collides with it in all simulations. For Dimorphos, the re-accretion fraction is between 65\%-85\%. Smaller particles (larger $\beta$) show greater variations in eccentricities, increasing the probability of collisions. Because of this, the overall lifetime of the assembly is reduced as $\beta$ increases. The particles tend to collide with Didymos due to its larger cross-section, and we observe that the increase of $\beta$ is accompanied by an increase in impacts with Didymos for both objects as sources.

It is also observed an exchange of material between the pair: $\sim$7-10\% of particles ejected from Didymos are accreted by Dimorphos, while $\sim$10\%-25\% of particles ejected from Dimorphos collide with Didymos. Taking into account the mass rates, we obtain a clear trend in the exchange of mass between the objects in the Didymos-Dimorphos direction. The mass supplied by the primary to the secondary is more than ten times greater than in the other direction.

\begin{figure}
\centering
\subfloat[Didymos, $\beta=10^{-3}$]{\includegraphics[width=0.37\columnwidth,trim={0 0 0 0},clip]{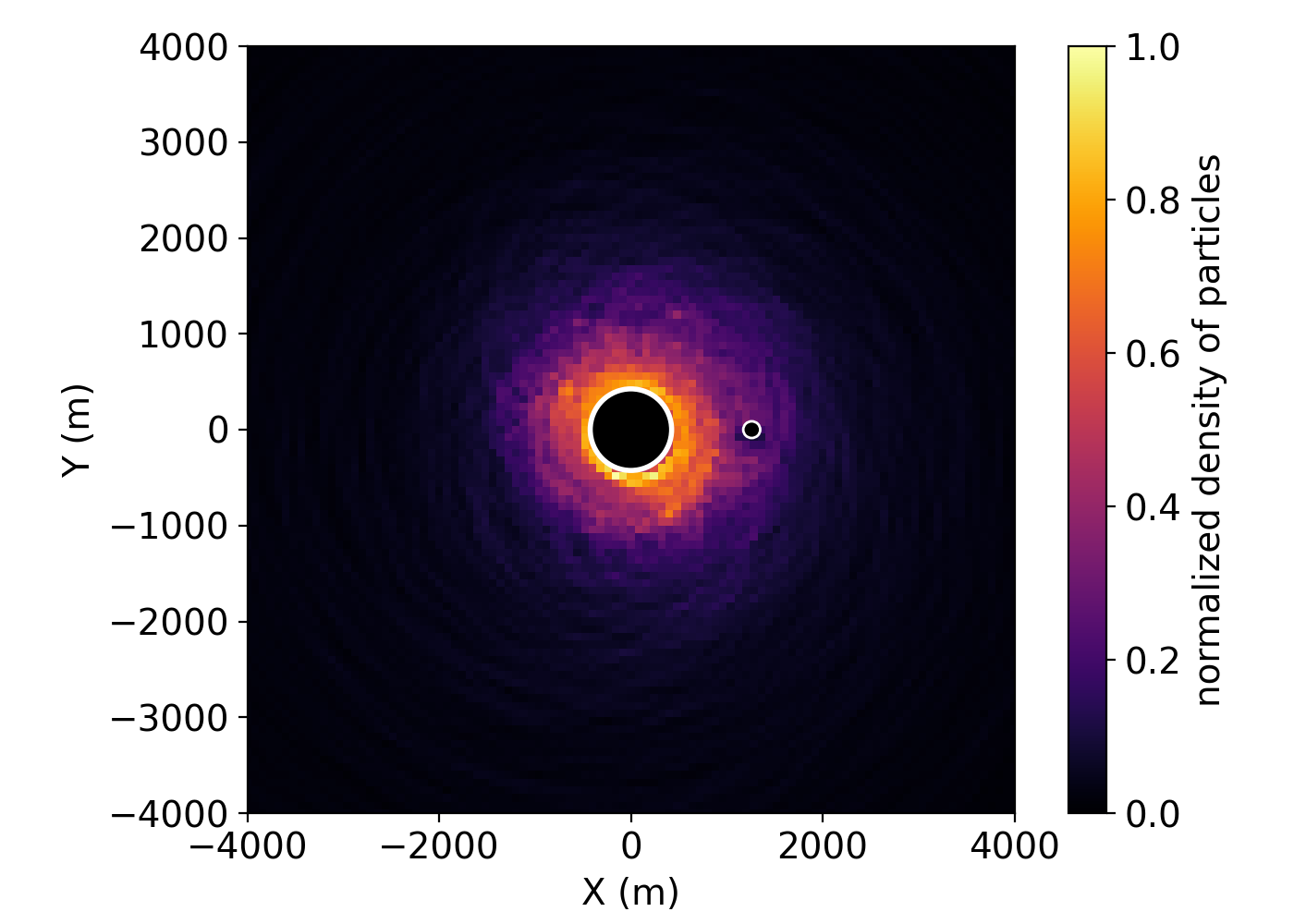}\label{densa}}\quad
\subfloat[Dimorphos, $\beta=10^{-3}$]{\includegraphics[width=0.37\columnwidth,trim={0 0 0 0},clip]{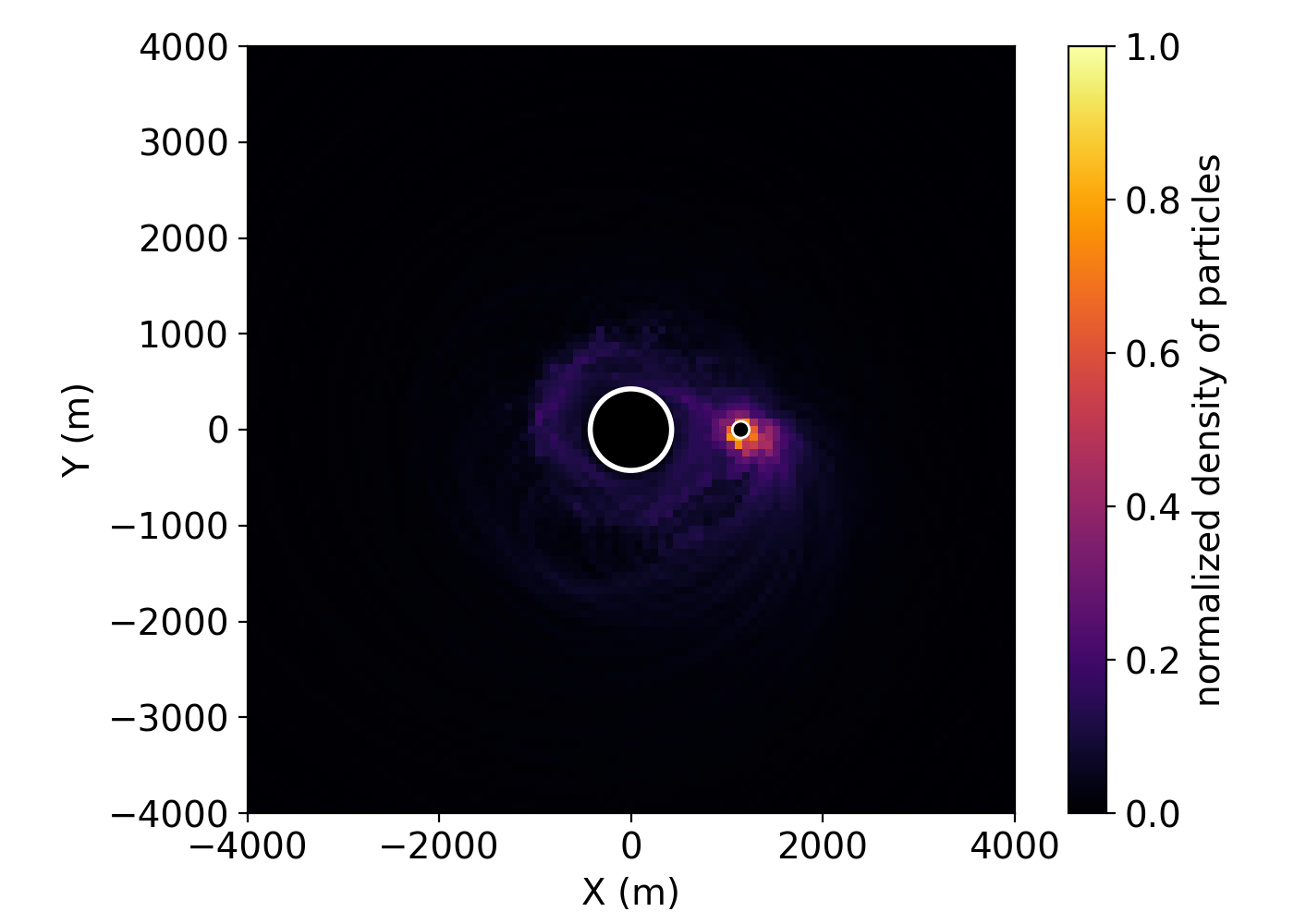}\label{densa}}
\\
\subfloat[Didymos, $\beta=10^{-5}$]{\includegraphics[width=0.37\columnwidth,trim={0 0 0 0},clip]{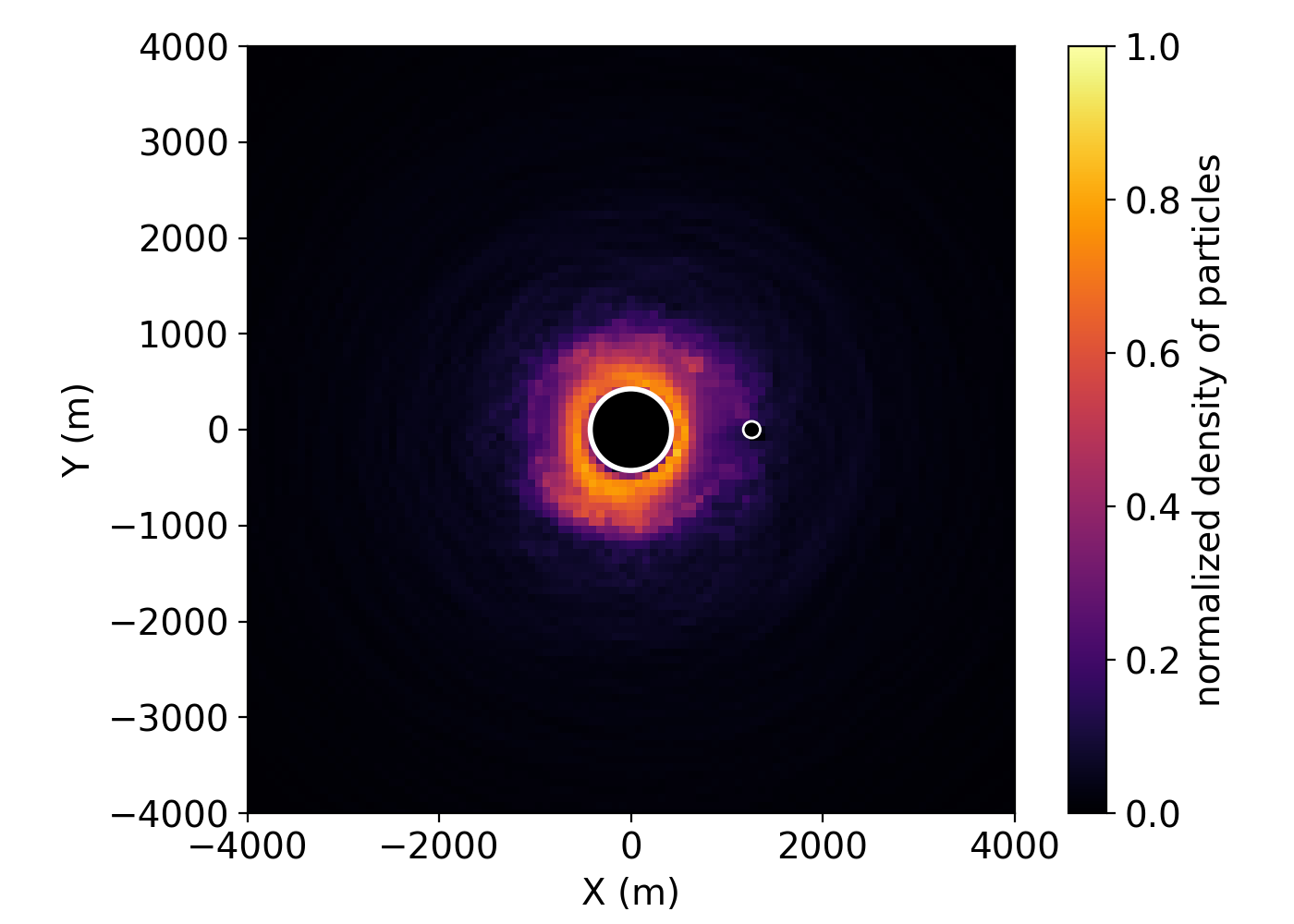}\label{densb}}\quad
\subfloat[Dimorphos, $\beta=10^{-5}$]{\includegraphics[width=0.37\columnwidth,trim={0 0 0 0},clip]{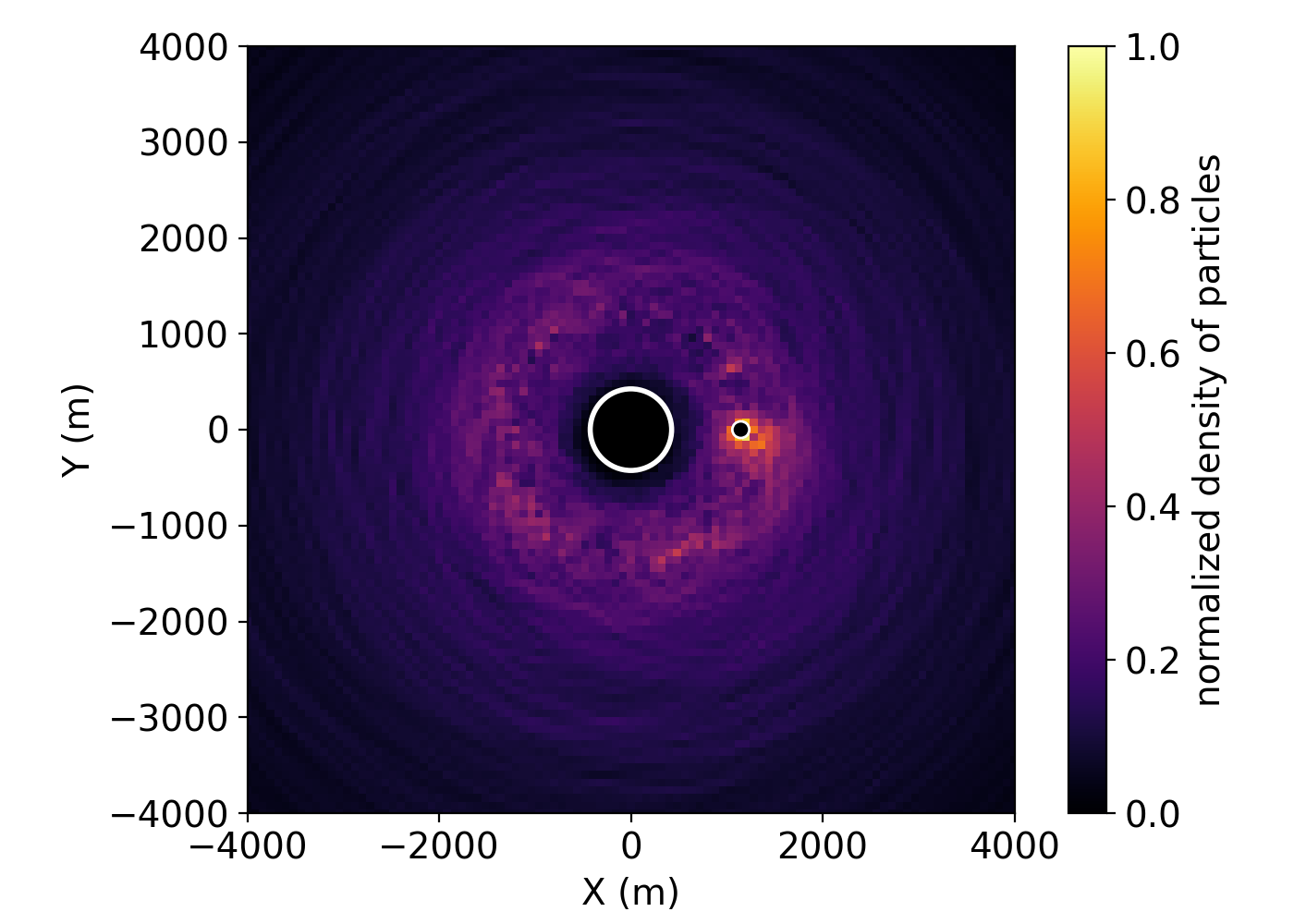}\label{densb}}\\
\subfloat[Didymos, $\beta=10^{-7}$]{\includegraphics[width=0.37\columnwidth,trim={0 0 0 0},clip]{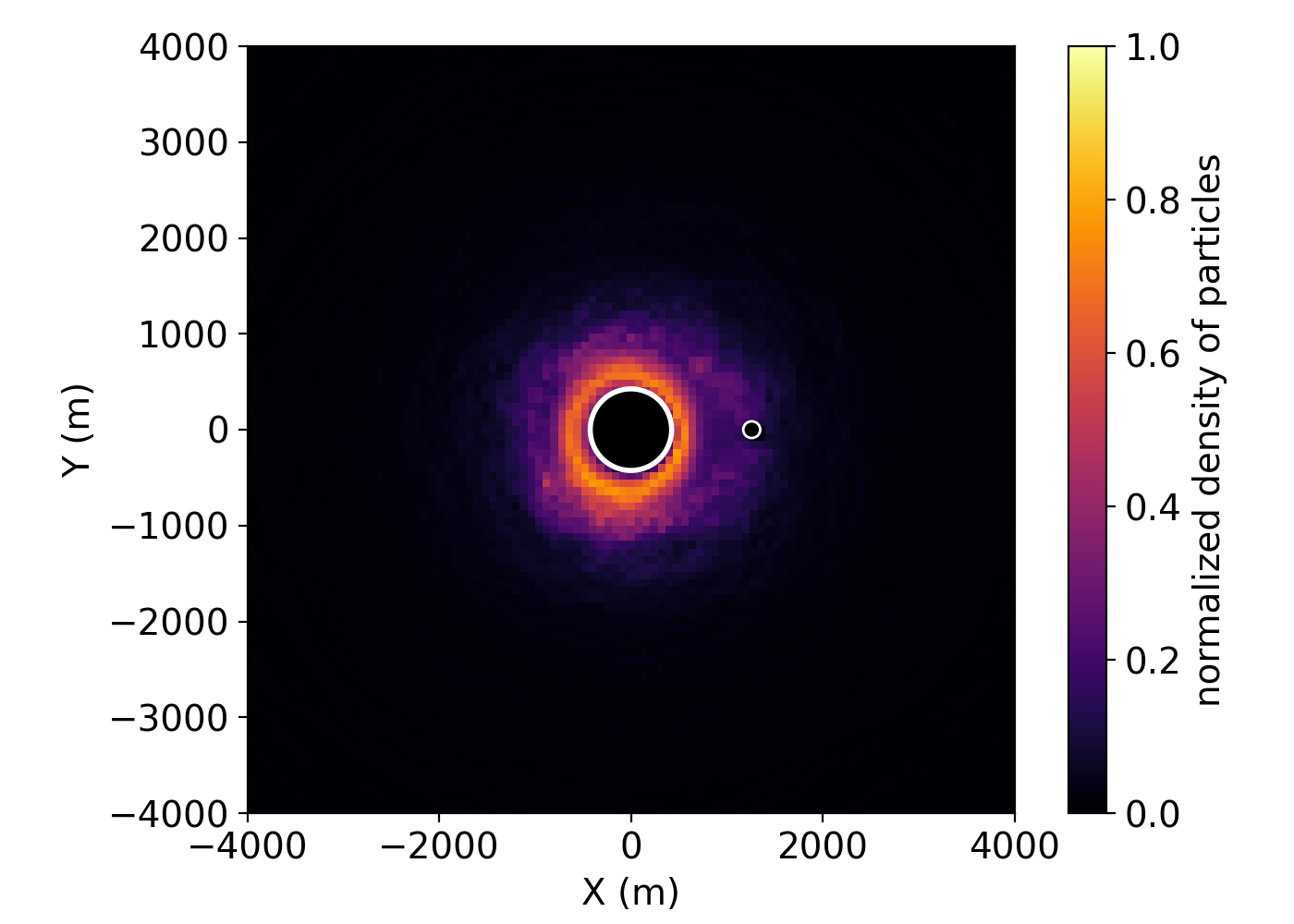}\label{densc}}\quad
\subfloat[Dimorphos, $\beta=10^{-7}$]{\includegraphics[width=0.37\columnwidth,trim={0 0 0 0},clip]{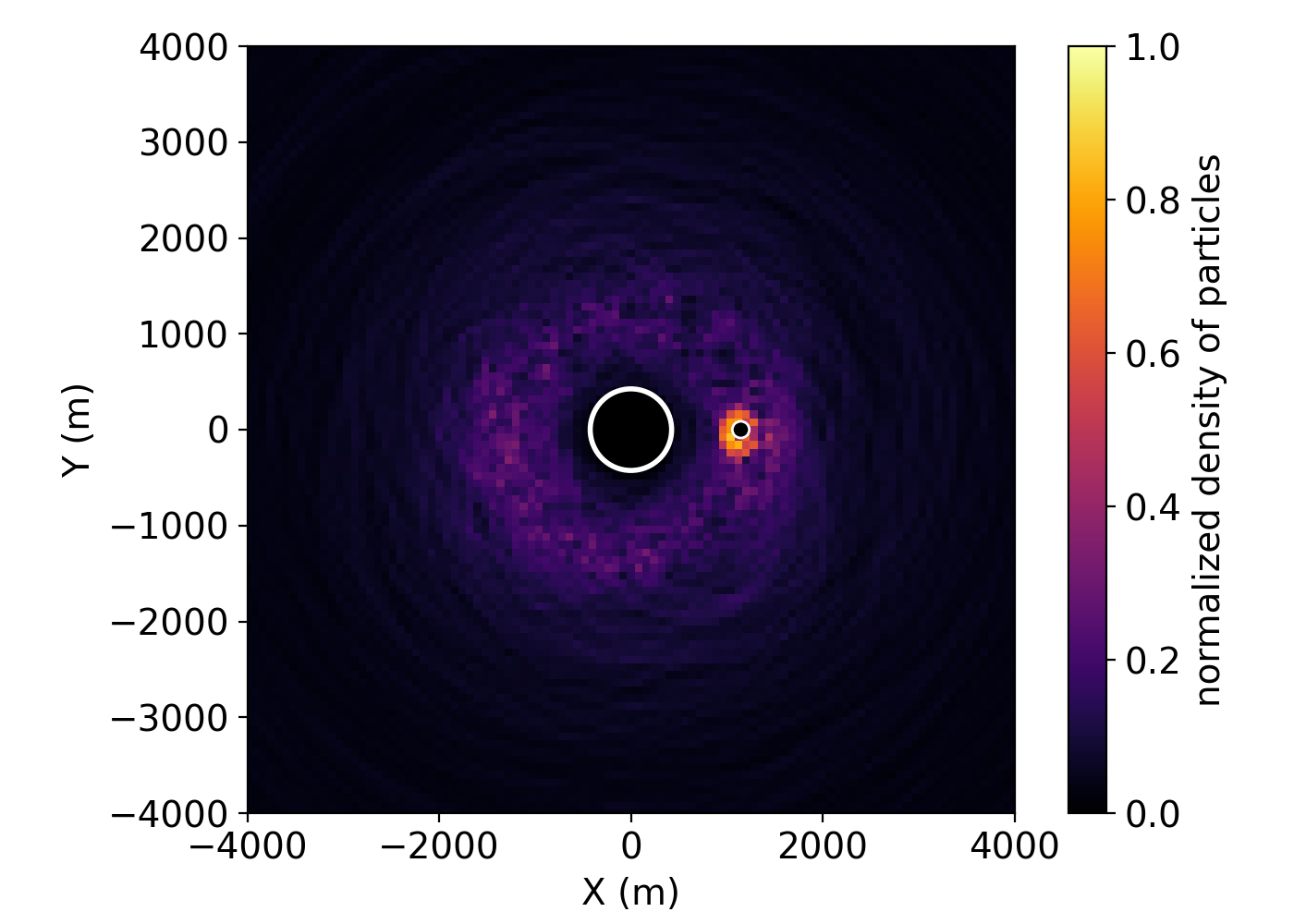}\label{densc}}\\
\caption{Surface density of dust material in the vicinity of Didymos and Dimorphos for a,b) $\beta=10^{-3}$, c,d) $\beta=10^{-5}$ and e,f) $\beta=10^{-7}$. The panels on the left correspond to the case with Didymos as source object and on the right, the case with Dimorphos as source. Didymos and Dimorphos correspond to the black dots with white border in the figures.}
\label{dens}
\end{figure}

An overview of the dust environment in the vicinity of Didymos and Dimorphos is shown in Figure~\ref{dens} for different values of $\beta$. Particles launched from Didymos and Dimorphos are shown in the left and right panels, respectively. We construct this figure by dividing the $XY$ plane into a set of regular bins and plotting the trajectories of all the launched particles. Then, the density of particles in each bin is computed. The material released from the equator of Didymos primarily populates the immediate vicinity of the asteroid, leading to the formation of the bright structure extending up to 1500 meters, shown in the left column of Figure~\ref{dens} (left column). On the other hand, particles released from the equator of Dimorphos spread over wider regions, populating both the inner and outer regions of the secondary. The smallest particles ($\beta=10^{-3}$) launched from Dimorphos are lost before spreading around the binary. For the lowest values of $\beta$, a fraction of the particles remains in trajectories associated with quasi-satellite motion and we obtain higher concentrations of particles around and within the orbit of the secondary. In the next section, we will discuss the implications of our results.

\section{Discussion} \label{sec_discussion}
In this work, we look at the possibility that the Didymos and Dimorphos environment is populated with material. Our analysis on the stability of the system considering only the gravitational effects of Didymos, Dimorphos and Sun showed that the Didymos equatorial neighborhood is generally unstable, existing only a few regions where particles can survive in stable motion for more than ten years. When considering the solar radiation force, we obtain that this perturbation dominates the evolution of millimeter-sized particles, reducing their lifetime to just a few days at most. 

Metric grain pieces can survive in the system for years under the SRF effects, preferably in simple-periodic symmetrical retrograde satellites orbits (Family A), the most stable orbits of the system. This family is associated with retrograde orbits with high eccentricity, covering the entire region between Didymos and Dimorphos. Although the solar radiation force is negligible for larger pieces, some processes can still destabilize their movement, perhaps removing them from the system. Some of these effects are erosion, collisional grinding, Yarkovsky effects, and tidal migration \citep{Stern1986,Cuk2005,Walsh2015,Caudal2023}. On the other hand, it is possible that particles in quasi-satellite orbits (Family A) or in Families B, C, D, and E can be kept in stable motion due to their resonant configuration, even under dissipative effects. This should be investigated in future publications. 

We are aware that our treatment of the shape of the objects is very simplistic, and therefore our system does not reproduce the full complexity of the real system. However it allows us to lead long term simulations, with shape much closer to reality than point-like objects. The physical characteristics of the system -- mass ratio and asteroids distance -- are responsible for generating large instabilities in the system and we find that the shape does play a minor role in the stability of the particles. Solar tides, on the other hand, only act by destabilizing the particles in the close region of Didymos that are in orbits with very high eccentricities, when the binary is at its perihelion.

Asymmetries in the shape of an object give rise to equilibrium points very close to the object's surface \citep{Yu2018}, but do not create stable regions in their vicinity \citep{Jiang2015,Jiang2016}. Rather, these shape irregularities tend to destabilize the regions. Asymmetries in the shape also induce spin-orbit resonances \citep{Sicardy2020}, which can lead to the emergence of stable regions when analysing a rotating Didymos. However, it has been observed that the overlap of spin-orbit resonances generates chaotic regions in vicinity of asymmetric objects \citep{Sicardy2019,Sicardy2020,Madeira2022a}, which may interfere with the stability of Families E and G. Therefore, the case explored here corresponds to a limit scenario, and the real Didymos system is expected to be even more unstable than what we observe.

The DART impact significantly altered the Didymos system, resulting in changes in the orbital period and eccentricity of Dimorphos \citep{Thomas2023,Meyer2023}. In addition, the impact may have induced a chaotic rotation state in the secondary \citep{Agrusa2021}. These changes are expected to affect the stability of any material that may have been populating the system during the impact, particularly the particles of Family A, which comprises orbits closest to Dimorphos, and Families B, C, D, and E, which are in mean motion resonance with the secondary.

The impact may also have changed the shape of Dimorphos \citep{Raducan2022,Nakano2022}. Considering the relatively small influence of Dimorphos shape on our results, we expect that changes in Dimorphos shape would have a less significant effect on particles stability compared to changes in its orbital and rotational elements. In light of all discussed, we conclude that the existence of a stable population of material in the equatorial neighborhood of Didymos is extremely unlikely.

We draw the reader's attention that our study was carried out exclusively in the equatorial plane of Didymos. Additional stable regions are also known in non-equatorial regions of Didymos \citep{Fodde2023,Raffa2023}. Moreover, there is evidence to suggest that the DART impact caused in the ejection of boulders, which may currently be orbiting the system \citep{Jewitt2023}. Studying the stability of such objects is beyond the scope of this study.

While it is unlikely that material resides in stable orbits in the vicinity of Didymos, material originating from IDP impacts on Didymos and Dimorphos does indeed populate this region. The amount of material released by Didymos exceeds that released by Dimorphos by more than 20 times, populating the region around the primary. Most of this material ejected from the equator of Didymos is re-accreted by the object within fractions of years. According to DART team's observations, although there are boulder tracks in the higher altitudes, Didymos' equatorial region appears to be smooth \citep{Barnouin2023}. While this could potentially be attributed to image resolution, it is also possible that recent dust deposition and potential surface motion contributed to this appearance. Our findings appear to support the last possibility.

Most of the material released from the equator of Dimorphos is re-accreted by the same object over a period of tens of days. Nonetheless, some particles with radius of centimeters or larger, ejected from both objects, survive for a few years in circumbinary orbits. A mass transfer from Didymos to Dimorphos due to IDP impacts is observed in the system, which could suggest a cumulative growth of the secondary body over time. However, our estimates reveal that such growth would be only $\sim 10^{-4}$ Dimorphos masses over 1~Myr, showing that the cumulative growth of Dimorphos due to IDP impacts is inefficient.

We conclude that if there is material produced by IDP impacts in the vicinity of Didymos and Dimorphos, it must be young and originate mainly from Didymos. Higher concentrations of material are expected to populate the region extending up to a radial distance of 1500 meters. Meanwhile, larger boulders that may be generated by this process could remain within the system for years, in first sort circum-Didymos orbits. The two CubeSats onboard the Hera spacecraft will allow us to constrain the spectroscopy and evolutionary history of Didymos and Dimorphos, which will give us insights about the evolutionary history of the near-Earth asteroids.

\section*{Acknowledgements}
This work was supported by Institut de Physique du Globe de Paris and European Research Council (101001282, METAL). Numerical computations were performed on the S-CAPAD/DANTE platform, IPGP, France. We thank Harrison Agrusa and Gonzalo Tancredi for the comments that helped us to significantly improve the article.



\end{document}